\documentclass[apj]{emulateapj}
\pdfoutput=1
\usepackage{epstopdf}
\usepackage{longtable}
\usepackage{fancyvrb}
\usepackage{multirow}
\usepackage{verbatim}
\usepackage{longtable}
\usepackage[caption=false]{subfig}
\usepackage{float}
\newcommand{\oiii}{[\ion{O}{3}]$\lambda5007$}

\newcommand{\nii}{[\ion{N}{2}]}
\newcommand{\niia}{[\ion{N}{2}]$\lambda$6583}

\newcommand{\deltaVproj}{$|\Delta V_{\rm{proj.}}|$}
\newcommand{\deltaVsig}{$|\Delta V/\sigma_{V}|$}
\newcommand{\deltaVforb}{$\Delta V_{\rm{f}}$}
\newcommand{\deltaVbalm}{$\Delta V_{\rm{B}}$}
\newcommand{\deltaVsigforb}{$|\Delta V/\sigma_{V}|_{\rm{f}}$}
\newcommand{\deltaVsigbalm}{$|\Delta V/\sigma_{V}|_{\rm{B}}$}
\newcommand{\NsigHbeta}{$N\sigma_{H\beta}$}
\newcommand{\NsigOIII}{$N\sigma_{[OIII]}$}
\newcommand{\NsigHalphaNII}{$N\sigma_{H\alpha,[NII]}$}
\newcommand{\NsigCont}{$N\sigma_{\rm{cont.}}$}
\newcommand{\Nsig}{$N\sigma$}
\newcommand{\dvdiff}{$|\Delta V_{\rm{f-B}}|/\sigma_{V}$}
\newcommand{\skewHbeta}{$\gamma_{H\beta}$}
\newcommand{\skewOIII}{$\gamma_{\rm{[OIII]}}$}
\newcommand{\dpeak}{$DPeak$}

\newcommand{\mass}{M$_{\rm{gal.}}$}
\newcommand{\BHmass}{M$_{\bullet}$}
\newcommand{\SUNmass}{M$_{\odot}$}

\newcommand{\sigxfinastr}{$\sigma_{astr.,X}$}
\newcommand{\sigyfinastr}{$\sigma_{astr.,Y}$}
\newcommand{\sigfinastr}{$\sigma_{astr.}$}
\newcommand{\sigxfinpos}{$\sigma_{pos.,X}$}
\newcommand{\sigyfinpos}{$\sigma_{pos.,Y}$}
\newcommand{\sigfinpos}{$\sigma_{pos.}$}
\newcommand{\sepphys}{$\Delta S_{3D}$}
\newcommand{\sepproj}{$\Delta S_{proj.}$}
\newcommand{\septheta}{$|\Delta \theta|$}

\newcommand{\sigmatrans}{$\sigma_{trans.}$}
\newcommand{\sigmarej}{$\sigma_{rej.}$}

\newcommand{\cata}{In-Fiber}
\newcommand{\catb}{Out-Fiber}

\newcommand{\FXH}{F$_{\rm{2-10keV,unabs.}}$}
\newcommand{\FXS}{F$_{\rm{0.5-2keV,unabs.}}$}
\newcommand{\LX}{$L_{X}$}
\newcommand{\LXH}{L$_{\rm{2-10keV}}$}
\newcommand{\LXHAGN}{L$_{\rm{2-10keV,AGN}}$}
\newcommand{\LXHAGNErr}{$\sigma_{\rm{LX,AGN}}$}
\newcommand{\LXHSFR}{L$_{\rm{2-10keV,SFR}}$}
\newcommand{\nHexgal}{n$_{\rm{H,exgal.}}$}
\newcommand{\nHgal}{$n_{\rm{H,Gal.}}$}

\newcommand{\XChandra}{$X_{\rm{AGN}}$}
\newcommand{\XChandraErr}{$\sigma_{\rm{X,AGN}}$}
\newcommand{\YChandra}{$Y_{\rm{AGN}}$}
\newcommand{\YChandraErr}{$\sigma_{\rm{Y,AGN}}$}

\newcommand{\XSDSS}{$X_{\rm{Gal.}}$}
\newcommand{\XSDSSErr}{$\sigma_{\rm{X,Gal.}}$}
\newcommand{\YSDSS}{$Y_{\rm{Gal.}}$}
\newcommand{\YSDSSErr}{$\sigma_{\rm{Y,Gal.}}$}

\newcommand{\zossy}{z$_{\rm{OSSY}}$}
\newcommand{\wvd}{\texttt{wavdetect}}
\newcommand{\sdssra}{$\alpha_{SDSS,J2000}$}
\newcommand{\sdssdec}{$\delta_{SDSS,J2000}$}
\newcommand{\sdssfib}{$r_{\rm{fib.}}$}
\newcommand{\sdssfibrad}{$1\farcs5$}

\newcommand{\brinch}{927,552}
\newcommand{\brinchagn}{23,279}

\newcommand{\parentAGNsdsssz}{20,098}
\newcommand{\sdsschandrasz}{2,292}
\newcommand{\parentsz}{48}
\newcommand{\catasz}{9}
\newcommand{\catbsz}{11}
\newcommand{\catabovrlpsz}{2}
\newcommand{\catabAGNsz}{18}
\newcommand{\catabiassz}{2}

\newcommand{\parentnobiassz}{46}
\newcommand{\catabbiassz}{8}
\newcommand{\catabbiasoffsz}{8}
\newcommand{\catabiasoffsz}{2}
\newcommand{\cataspecsigsz}{4}
\newcommand{\catanobiasoffsz}{7}

\newcommand{\seyfertsz}{42}
\newcommand{\linersz}{6}
\newcommand{\seyfertfrac}{$88_{-6}^{+4}\%$}

\newcommand{\catadeltaVforb}{6}
\newcommand{\catadeltaVbalm}{8}
\newcommand{\cataNHalphaNII}{3}
\newcommand{\catadvdiff}{1}
\newcommand{\cataskewHbeta}{4}
\newcommand{\cataskewOIII}{2}
\newcommand{\catadp}{2}
\newcommand{\excludevala}{$33$}
\newcommand{\excludevalb}{$67_{-17}^{+14}$}
\newcommand{\excludevalc}{$89_{-16}^{+7}$}

\newcommand{\twocompN}{$2.81\sigma$}
\newcommand{\twocompS}{$1.96\sigma$}
\newcommand{\sepmeansimpar}{$0\farcs98$}
\newcommand{\sepmeansimoff}{$1\farcs48$}
\newcommand{\velmeansimpar}{$34.19$ km s$^{-1}$}
\newcommand{\velmeansimoff}{$28.51$ km s$^{-1}$}
\newcommand{\ovrlp}{1}
\newcommand{\litsz}{9}
\newcommand{\simitersz}{100,000}
\newcommand{\ovrlpparenetsz}{$2.2_{-1.6}^{+3.7}\%$}

\shorttitle{Spatially Offset AGN I}
\shortauthors{Barrows et al.}

\begin{document}

\submitted{Accepted for publication in ApJ}

\title{Spatially Offset Active Galactic Nuclei I: Selection and Spectroscopic Properties}
\author{R. Scott Barrows,\altaffilmark{1} Julia M. Comerford,\altaffilmark{1} Jenny E. Greene,\altaffilmark{2} David Pooley\altaffilmark{3}}

\altaffiltext{1}{Department of Astrophysical and Planetary Sciences, University of Colorado Boulder, Boulder, CO 80309, USA \newline \indent
\indent \emph{author email: Robert.Barrows@Colorado.edu}}
\altaffiltext{2}{Department of Astrophysical Sciences, Princeton University, Princeton, NJ 08544, USA and }
\altaffiltext{3}{Department of Physics and Astronomy, Trinity University, San Antonio, TX 78212, USA}
\bibliographystyle{apj}

\begin{abstract}

We present a sample of \catabAGNsz~optically-selected and X-ray detected spatially offset active galactic nuclei (AGN) from the Sloan Digital Sky Survey (SDSS). In \catasz~systems, the X-ray AGN is spatially offset from the galactic stellar core that is located within the $3''$ diameter SDSS spectroscopic fiber.  In \catbsz~systems, the X-ray AGN is spatially offset from a stellar core that is located outside the fiber, with an overlap of \catabovrlpsz.  To build the sample, we cross-matched Type II AGN selected from the SDSS galaxy catalogue with archival $Chandra$ imaging and employed our custom astrometric and registration procedure.  The projected angular (physical) offsets span a range of $0\farcs6$ (0.8 kpc) to $17\farcs4$ (19.4 kpc), with a median value of $2\farcs7$ (4.6 kpc).  The offset nature of an AGN is an unambiguous signature of a galaxy merger, and these systems can be used to study the properties of AGN in galaxy mergers without the biases introduced by morphological merger selection techniques.  In this paper (Paper I), we use our sample to assess the kinematics of AGN photoionized gas in galaxy mergers.  We find that spectroscopic offset AGN selection may be up to \excludevalc$\%$ incomplete due to small projected velocity offsets.  We also find that the magnitude of the velocity offsets are generally larger than expected if our spatial selection introduces a bias toward face-on orbits, suggesting the presence of complex kinematics in the emission line gas of AGN in galaxy mergers.

\end{abstract}


\section{Introduction}
\label{sec:intro}

Overwhelming observational evidence suggests that most galaxies host a nuclear supermassive black hole (SMBH) of mass $\sim10^{6}-10^{9}M_{\odot}$.  Despite the ubiquitous presence of SMBHs in galaxies, however, the primary mechanisms for their growth are largely uncertain.  The discovery that many global properties of the host galaxy appear to be correlated with the central SMBH mass, most notably properties of the central bulge - stellar velocity dispersion \citep{Gebhardt00,Ferrarese2000,Gultekin:2009}, mass \citep{McLure:2002,Haring:2004} and luminosity \citep{Marconi:Hunt:2003,Bentz:2009c} - suggests that the mechanisms responsible for SMBH growth are linked to the formation of the host galaxy.  
 
 Due to the vastly different physical size scales of SMBHs and galaxies, they are not directly interacting and therefore must be affected by processes which influence both scales at a significant level.  In particular, one potential route for the correlated growth of SMBHs and galaxies is through mergers of gas-rich galaxies.  These events are capable of both feeding SMBHs - by removing sufficient angular momentum from the gas in the merging galaxy system such that it accumulates in the nuclear regions - and of growing and evolving the stellar populations of galaxies - through merger-triggered star-formation \citep{Barnes:1991,Mihos:1996,Hopkins05,Springel2005}.  This scenario is attractive within the hierarchical paradigm of galaxy formation which asserts that mergers are the primary mechanism for evolving galaxies toward the populations observed at both high cosmological redshifts and in the local Universe.
 
To obtain direct evidence for a connection between SMBH growth and galaxy evolution, a clean and unambiguous sample of galaxy mergers hosting AGN is necessary.  Traditional attempts to build samples of galaxy mergers have relied upon visual morphology, either selecting mergers `by eye,' which is inherently qualitative and can lead to significant numbers of false merger classifications \citep{Lotz:2011}, or by galaxy asymmetry, which is biased by dust obscuration and redshift dependent surface brightness dimming \citep{Lotz:2011,Conselice:2003}.  

Galaxy mergers may also be identified via AGN catalogues.  In particular, if an AGN is observed to have a kinematic offset from the systemic velocity of the host galaxy, the hypothesis is that a galaxy merger has taken place.  Specifically, in a galaxy merger, the two SMBHs from each galaxy will form a dual SMBH system (separated by $<10$kpc), each with a non-zero motion relative to the center of mass of the merging system.  If either of the SMBHs is accreting as an AGN, and the orbital radial velocity is non-negligible, their motion will be manifested spectroscopically as an AGN-photoionized emission line that is offset in velocity-space from the stellar absorption line system (presumed to represent the systemic velocity of the host galaxy).  If only one emission line system is observed (\emph{single}-peaked offset emission line), the hypothesis is that only one of the SMBHs is an AGN (\emph{offset AGN}).  If two kinematically offset emission lines are observed (\emph{double}-peaked offset emission lines), the hypothesis is that both SMBHs are AGN (\emph{dual AGN}). The original spectroscopic searches targeted dual AGN candidates by looking for double-peaks in the bright AGN \oiii~emission line accessible in optical spectra at $z<0.8$ from spectroscopic surveys such as DEEP2 \citep{Comerford2009a}, AGES \citep{Comerford:2013}, and the Sloan Digital Sky Survey (SDSS) \citep{Wang2009,Liu2010a,Smith:2010,Ge:2012}, while other emission lines from the SDSS ([NeV]$\lambda3426$ and [NeIII]$\lambda3869$) have probed out to higher redshifts ($0.8<z<1.6$\citealp[; ][]{Barrows:2013}).  However, kinematically offset emission lines can be produced by mechanisms other than orbital motion of a dual SMBH system.  In particular, kinematics associated with the narrow line region (NLR) gas such as small scale or large-scale outflows (AGN and/or star-formation driven) and rotation within the host galaxy potential can drive the observed line-splitting.  Therefore, follow-up spatially-resolved observations are necessary to determine if two spatially offset AGN are present.

Long-slit spectroscopy provides a means to spatially locate the peaks of ionized gas emission, either in one dimension along the slit length \citep{Rosario:2010,Shen:2011b,McGurk:2015}, or in two dimensions from orthogonal slit position angles \citep{Comerford:2012}.  Adaptive optics (AO) imaging in the near-infrared has been used to locate the peaks of stellar continuum emission in the host galaxies to determine if multiple stellar nuclei are present, as expected in a galaxy merger \citep{Fu:2011a,Rosario:2011,Barrows:2012}.  A combination of the two above methods can be used to identify systems with dual stellar cores in which a peak in the ionized gas is spatially coincident with one or both of the stellar cores.  A more direct means of achieving this result is through integral field unit (IFU) spectroscopy which can spatially resolve the two-dimensional position of ionized gas emission \citep{McGurk:2011,Fu:2012,McGurk:2015}.  From these follow-up observatations of dual AGN candidates, $\sim15\%$ of double-peaked AGN are systems in which a galaxy merger hosts two actively accreting SMBHs.  However, these methods can not uniquely distinguish between a scenario in which two AGN are present, each associated with a stellar core, or one in which only a single AGN is present and ionizing the NLR gas associated with two SMBHs or an extended NLR such as an outflow.    

\indent To spatially constrain the AGN location decisively, observations at wavelengths that directly probe near the SMBH accretion disk are necessary, rather than extended regions of nebular gas.  For example, compact, flat-spectrum radio sources are considered strong evidence for synchrotron emission from magnetic field lines generated by an AGN accretion disk \citep{Readhead:1979,Pushkarev:2012}, and the excellent spatial resolution at these frequencies has been used to look for dual AGN in double-peaked spectroscopic candidates \citep{Fu:2011,Mueller-Sanchez:2015,Fu:2015a,Fu:2015b}.  AGN can also be spatially located with $Chandra$ X-ray imaging (0.5-10keV) since emission at energies above $\sim2$ keV is produced by the inverse Compton scattering of far-ultraviolet to soft X-ray continuum emission from the AGN accretion disk off of a nearby corona of hot electrons.  The spatial resolution of $Chandra$ makes this method effective at confirming dual AGN through two distinct hard X-ray sources.  Indeed, some of the original dual AGN confirmations were serendipitously achieved this way through $Chandra$ observations of ultra-luminous infrared galaxies (heavily obscured galaxy mergers) because hard X-rays suffer little attenuation below neutral Hydrogen columns of $\sim10^{24}$ cm$^{-2}$ \citep{Komossa2003,Guainazzi2005,Hudson2006,Bianchi2008,Piconcelli2010,Koss:2011,Mazzarella:2012}.
  
\indent By registering $Chandra$ imaging (which spatially constrains the AGN positions) with $HST$ imaging (which spatially locates galaxy stellar cores, e.g. \citealt{Lackner:2014}), dual AGN can be identified unambiguously as systems in which two X-ray AGN are spatially coincident with two galaxy stellar cores.  \citet{Liu:2013} obtained $Chandra$ and $HST$ imaging of four double-peaked AGN which had previously been identified as strong dual AGN candidates from combined long-slit and near-infrared (NIR) AO imaging as described above \citep{Liu2010b}.  They conclude that two of the four systems are unambiguously dual AGN, while the remaining two may be offset AGN. Similarly, \citet{Comerford:2015} present $Chandra$ and $HST$ imaging of ten double-peaked AGN with long-slit spectroscopy that suggests they are strong dual AGN candidates.  The $HST$ imaging reveals dual stellar cores (i.e. galaxy mergers) in six of the systems, and using a detailed astrometric procedure, \citet{Comerford:2015} confirm the dual AGN nature of one system.  The remaining five systems may be offset AGN.  This procedure of registering high-resolution X-ray and optical imaging is the currently optimal method for distinguishing between the dual and offset AGN scenarios.     

While offset and dual AGN are both observational tracers of galaxy mergers, offset AGN may exist under different merger conditions and comparisons of the two populations will reveal the parameters that control activation of AGN in galaxy mergers.  Besides the potential offset AGN identified in \citet{Liu:2013} and \citet{Comerford:2015}, only one other offset AGN has been explicitly identified, and its discovery was serendipitous due to an SDSS fiber placed at an off-nuclear location in a galaxy merger \citep{Barth:2008}.  To systematically target offset AGN candidates, \citet{Comerford:Greene:2014}, hereafter CG14, used the SDSS to select AGN with kinematically offset single-peaked emission line systems.  Several features of the CG14 sample suggest that spectroscopic selection may indeed target some authentic offset AGN.  However, as with double-peaked AGN, NLR kinematics can mimic this kinematic offset feature, and high-resolution X-ray imaging is necessary to confirm the offset nature of an AGN relative to the host galaxy.  To this end, here we systematically identify offset AGN directly using archival $Chandra$ X-ray imaging and corresponding optical SDSS imaging of AGN.  In addition to building a clean sample of galaxy mergers hosting AGN, this sample bypasses the initial spectroscopic selection of previously identified offset AGN candidates and therefore does not reflect the biases introduced by the kinematics of emission line selection parameters.  Furthermore, the offset X-ray AGN are spatially located within the SDSS fiber, allowing for a comparison with the spectroscopic selection technique. 

The present paper is Part I of our spatially offset AGN analysis, in which our aim is to describe the novel selection process for our sample and to compare the spectroscopic properties of our spatially selected offset AGN sample to the those of the spectroscopic sample of candidates from CG14.  This paper is structured as follows: in Section \ref{sec:initial} we describe the initial AGN sample, in Section \ref{sec:astrometry} we describe the astrometric procedure, in Section \ref{sec:image_modeling} we describe the image modeling, in Section \ref{sec:Xray_AGN} we define the X-ray criteria for AGN status, in Section \ref{subsec:parent_agn} we describe the parent sample of AGN, in Section \ref{subsec:offset_agn} we describe the final sample of offset AGN, in Section \ref{sec:analysis} we present the analysis of the spectroscopic properties, in Section \ref{sec:discussion} we discuss the results of the analysis, and Section \ref{sec:conclusions} contains our conclusions.  
Throughout the paper we adopt the cosmological parameters $\Omega_{\Lambda}=0.728$, $\Omega_{b}=0.0455$, $\Omega_{m}h^{2}=0.1347$, and $H_{0}=70.4$ km s$^{-1}$ Mpc$^{-1}$. This corresponds to the maximum likelihood cosmology from the combined WMAP+BAO+H0 results from the WMAP 7 data release of \citet{Komatsu:2011}.


\section{The Optically-Selected AGN Sample}
\label{sec:initial}
Our intention is to construct a sample of galaxies hosting accreting SMBHs that appear as AGN, from which we will select AGN that are spatially offset from a galaxy stellar core.  To do so, we started with the \brinch~galaxies in the MPA-JHU value-added catalogue of spectroscopic properties from Data Release 7 of the SDSS \citep{Abazajian09}.  We then selected those with emission lines indicating a substantial contribution from AGN photo-ionization based on the criteria set forth in \citet{Brinchmann:2004}: the four emission lines \oiii, H$\beta$, \niia~ and H$\alpha$ are spectroscopically accessible, have signal-to-noise ratios (S/N) greater than 3.0, and lie above the theoretical limit for pure starbursts \citep{Kewley:2001} in the space defined by the flux ratios $F_{[OIII]}\lambda 5007$/$F_{H\beta}$ and $F_{[NII]}\lambda 6583$/$F_{H\alpha}$, also known as a Baldwin-Phillips-Terlevich (BPT) diagram \citep{Baldwin1981}. This produces a sample of \brinchagn~optically-selected AGN.  While this cut may eliminate AGN in galaxies with a significant contribution from star-formation, it guarantees that an AGN is present.  Finally, we cross-matched the sample of \brinchagn~AGN with the $z<0.21$ OSSY catalogue \citep{Oh:2011}, providing a final, optically-selected sample of \parentAGNsdsssz~AGN.  This final step is made to arrive at the same initial sample from which CG14 identified spectroscopic candidate offset AGN.  All subsequent spectroscopic parameters adopted in this work are obtained from the OSSY catalogue, including the galaxy redshifts (\zossy) obtained from stellar absorption lines.  Since these AGN are selected from the SDSS sources spectroscopically classified as galaxies, they do not have broad emission lines and thus are restricted to Type II AGN.

\begin{figure*}
\hspace*{-0.in} \includegraphics[width=7.0in]{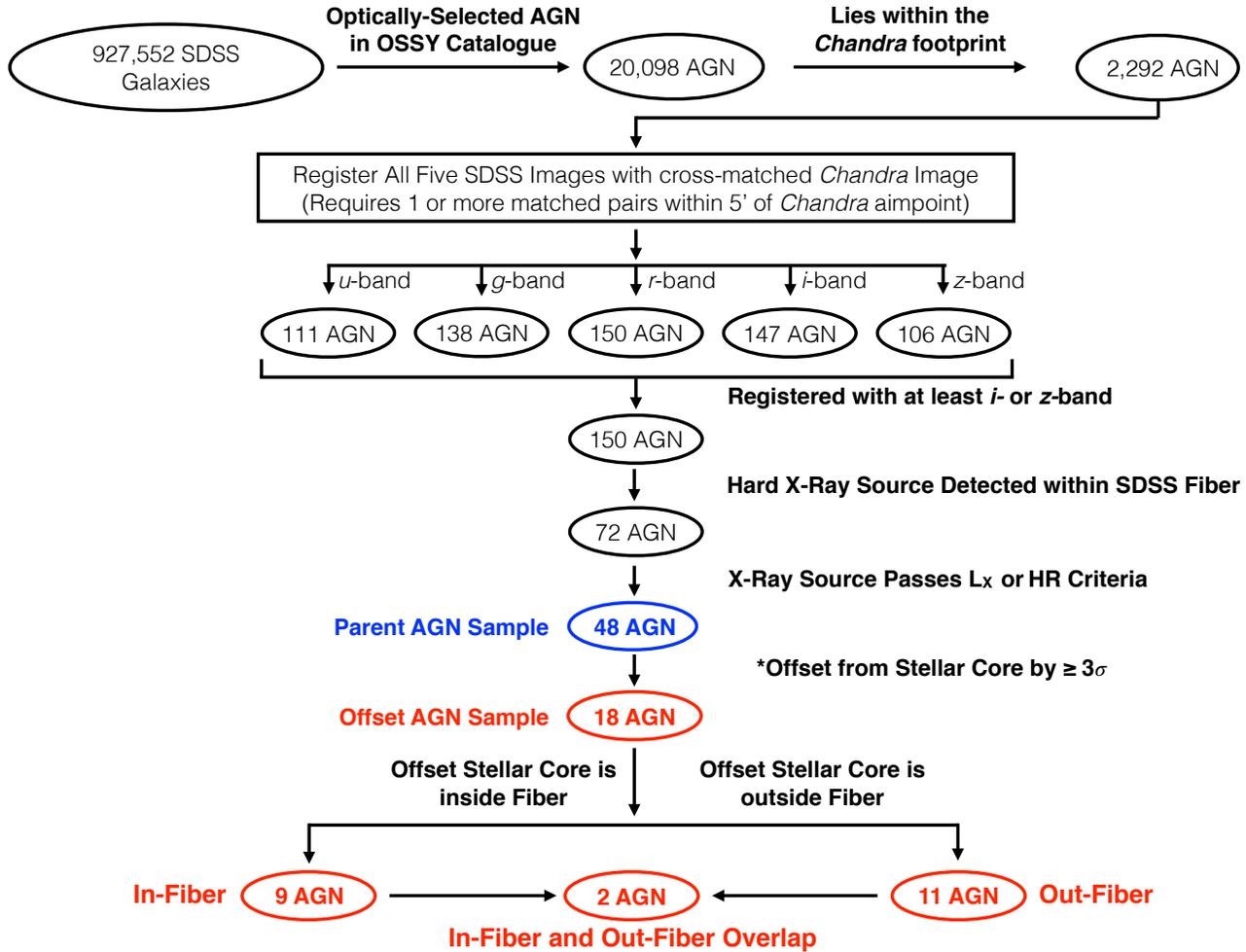}
\caption{\footnotesize{Flow chart showing the progression of the offset AGN sample selection from $Chandra$ and the SDSS, as detailed in Sections \ref{sec:initial}, \ref{sec:astrometry}, \ref{sec:image_modeling}, \ref{sec:Xray_AGN}, \ref{subsec:parent_agn} and \ref{subsec:offset_agn}. *Valid offsets correspond to offsets measured in either the i- or z-bands, or offsets measured in any of the other three bands that are consistent, within the uncertainties, with the i- or z-band offsets (See Section \ref{sec:waveband} for details).}
}
\label{fig:flow_chart}
\end{figure*}

\section{Astrometry and Registration}
\label{sec:astrometry}

Our goal is to identify a subset of the \parentAGNsdsssz~optically-selected AGN that are X-ray detected and have significant spatial offsets from a galaxy stellar core.  To do so, we will find overlapping SDSS and $Chandra$ coverage (Section \ref{subsec:cross_corr}), register the images (Section \ref{subsec:register}) and obtain estimates of the relative astrometric uncertainty (Section \ref{subsec:accuracy}).  These steps are described in detail below.

\subsection{Cross-Correlating the SDSS AGN with $Chandra$ Observations}
\label{subsec:cross_corr}

The coordinates of each optically-selected SDSS AGN (Section \ref{sec:initial}) are the J2000.0 right ascension (\sdssra) and declination (\sdssdec) in the International Celestial Reference Frame \citep[ICRF;][]{Arias:1995} which is tied to the Hipparcos catalogue \citep{Perryman:1997} and stored in the FITS standard World Coordinate System (WCS) as described in \citet{Calabretta:2002}.  The actual values of \sdssra~and \sdssdec~depend on the astrometric calibrations applied to SDSS CCD frames \citep{Pier:2003}.  In short, absolute astrometry is determined from stars in the SDSS $r$-band CCD frames, producing absolute astrometric root-mean-square (rms) uncertainties of 65-75 mili-arcseconds (mas) when matched to the U.S. Naval Observatory Astrograph Catalog \citep{Zacharias:2000} and 95-105 mas when matched to Tycho-2 \citep{Hog:2000}.  The astrometry for the other four SDSS bands ($u$, $g$, $i$, and $z$) is determined by matching them to the calibrated $r$-band frames, with relative astrometric uncertainties of 25-35 mas.  The catalogued values of \sdssra~and \sdssdec~for a SDSS object come from the $r$-band calibrations.

We cross-correlated the SDSS coordinates (\sdssra, \sdssdec) of the 20,098 AGN with the coordinates of $Chandra$ non-proprietary (as of October 27th, 2014) archival coverage using the \texttt{find\char`_chandra\char`_obsid} tool in the \emph{Chandra Interactive Analysis of Observations} (CIAO 4.7) package.  The coordinates of $Chandra$ data are also stored as FITS World Coordinates in the ICRF, and the absolute astrometric 90\% (99\%) uncertainty in $Chandra$ observations is $0\farcs6$ (0\farcs8) within $3'$ of the observation aim point.  Our search criteria required that the positions defined by \sdssra~and \sdssdec~lie within the footprint of a $Chandra$ frame (\texttt{radius}=0) to be considered a match.  This cross-correlation resulted in \sdsschandrasz~unique SDSS optically-selected AGN that are covered by one or more $Chandra$ observations. 

\subsection{Registering the Chandra and SDSS Images}
\label{subsec:register}

While the World Coordinates of the SDSS and $Chandra$ images are defined by the same reference frame (ICRF), there is uncertainty in the accuracy of the WCS values (i.e the absolute astrometric uncertainties described in Section \ref{subsec:cross_corr}) which limits the extent to which we can determine the relative locations on the sky between sources in two different images.  Therefore, to reduce the uncertainties on the relative astrometry between the SDSS and $Chandra$ images, we have registered each pair of SDSS and $Chandra$ images returned by the cross-correlation.  Our registration procedure involves detecting suitably strong sources in both images and computing the coordinate transformations between them.  These steps are described in detail below.

\subsubsection{Detecting Sources for Registration}
\label{sec:detect}

We used Source Extractor \citep{Bertin:Arnouts:1996} to find SDSS sources suitable for registration, requiring a detection threshold of $3\sigma$.  We ran Source Extractor on all five SDSS bands ($u$,$g$,$r$,$i$ and $z$) rejecting any sources with flags generated by Source Extractor that indicate a source either has bad pixels, saturated pixels, or is close to the image boundary.  Separately, we applied the bad pixel masks generated from the SDSS pipeline and omitted any detections that are contaminated by bad pixels.

To find suitable objects for registration in the \emph{Chandra} images, we used the \wvd~algorithm in CIAO.  Detections that are significantly off-axis will have large uncertainties due to deterioration of the PSF as a function of off-axis distance \citep{Gaetz:2004}.  We find that, while the \wvd~centroid errors generally increase with distance from the aimpoint, the errors are relatively stable out to $\gtrsim5^{\prime}$, beyond which they drastically increase.  Therefore, we restricted our $Chandra$ source detection radius to no greater than $5'$ from the observation aimpoint and provided  \wvd~with a PSF map (generated using \texttt{mkpsfmap}) to account for a variable PSF across the frame.  For a frame of $4096\times 4096$ pixels (several times larger than the $5'$ circular fields used in our analysis), a threshold of $1/(4096\times 4096)$, or $\sim6\times 10^{-8}$, corresponds to 1 random background fluctuation detected as a source.  Therefore, we adopt \texttt{sigthresh}=$10^{-8}$, which corresponds to less than one false detection per search region.  

Due to the relatively low counts of X-ray sources compared to optical sources, the accuracy of our registration procedure is limited by the positional uncertainties of $Chandra$ detections.  While the counts detected by $Chandra$ are binned to the native pixel size of $0\farcs492$ by default, the spacecraft dithering and aspect corrections allow the relative positions of counts to be  positionally accurate to better than $0\farcs492$.  Therefore, to achieve optimal spatial resolution, we have applied sub-pixel binning to the $Chandra$ images.  Studies have shown that sub-pixel binning does not worsen the positional uncertainties and can reduce them by $\sim20\%$ for sources close to the aimpoint \citep{Anderson:2011}.  Therefore, before running \wvd, we used \texttt{dmcopy}, a CIAO tool for writing tables of $Chandra$ data to an image, to bin $Chandra$ counts into pixels 1/4th the native pixel size.  We choose a binning factor of 1/4th to improve the resolution and accuracy of the centroids while also limiting computational expense.   For a subset of the $Chandra$ images, we compared the results using the ``$\sqrt{2}$ sequence" of wavelet scales: 1, $\sqrt{2}$, 2, 2$\sqrt{2}$, 4, 4$\sqrt{2}$, 8, 8$\sqrt{2}$ and 16 pixels \citep{Xue:2011}, and a subset of that sequence: 2 and 4 pixels.  We find that the number and specific sources detected remains unchanged between between the two sequences for the $Chandra$ images tested.  The changes in specific values of the native $Chandra$ coordinates are much less than the errors on each source detection.  Therefore, we use the \wvd~results from the 2, 4 pixel sequence to allow all $Chandra$ images to be analyzed uniformly in a reasonable amount of time.

\begin{deluxetable*}{ccccccccc}
\tabletypesize{\footnotesize}
\tablecolumns{9}
\tablecaption{Spatially Offset AGN: X-ray Properties and Best-fit Spectral Model Parameters.}
\tablehead{
\colhead{SDSS Name \vspace*{0.05in}} &
\colhead{T$_{\rm{exp.}}$} &
\colhead{S} &
\colhead{H} &
\colhead{$\Gamma$} &
\colhead{\nHexgal} &
\colhead{\FXS} &
\colhead{\FXH} &
\colhead{D$_{aim}$} \\
\colhead{~ \vspace*{0.05in}} &
\colhead{(s)} &
\colhead{(counts)} &
\colhead{(counts)} &
\colhead{~} &
\colhead{($10^{20}$ cm$^{-2}$)} &
\colhead{{\tiny ($10^{-14}$ erg s$^{-1}$ cm$^{-2}$)}} &
\colhead{{\tiny ($10^{-14}$ erg s$^{-1}$ cm$^{-2}$)}} &
\colhead{($^{\prime}$)} \\
\colhead{(1)} &
\colhead{(2)} &
\colhead{(3)} &
\colhead{(4)} &
\colhead{(5)} &
\colhead{(6)} &
\colhead{(7)} &
\colhead{(8)} &
\colhead{(9)}
}
\startdata 
SDSSJ080523.29$+$281815.84 & 17382 & $23_{-5.8}^{+5.9}$ & $17_{-5.1}^{+5.2}$ & $1.1_{-0.6}^{+1.4}$ & $0.0$\tablenotemark{b} & $0.62_{-0.36}^{+0.77}$ & $2.90_{-1.67}^{+3.61}$ & 0.16 \\
SDSSJ081330.15$+$541844.44 & 33208 & $16_{-5.0}^{+5.1}$ & $16_{-5.0}^{+5.1}$ & $1.4_{-0.7}^{+1.6}$ & $0.6_{-0.3}^{+0.7}$ & $0.06_{-0.03}^{+0.07}$ & $0.19_{-0.10}^{+0.23}$ & 3.43 \\
SDSSJ084135.09$+$010156.31 & 22417 & $106_{-10.3}^{+10.3}$ & $95_{-9.7}^{+9.7}$ & $1.7$\tablenotemark{a} & $1.2_{-0.7}^{+1.5}$ & $3.56_{-1.99}^{+4.27}$ & $6.55_{-3.66}^{+7.87}$ & 0.19 \\
SDSSJ090714.45$+$520343.42 & 16634 & $6_{-3.4}^{+3.6}$ & $37_{-6.1}^{+6.1}$ & $1.7$\tablenotemark{a} & $591.4_{-326.3}^{+707.7}$ & $11.37_{-6.28}^{+13.61}$ & $14.27_{-7.87}^{+17.08}$ & 0.17 \\
SDSSJ104232.05$+$050241.94 & 51536 & $24_{-5.9}^{+6.0}$ & $583_{-24.1}^{+24.1}$ & $1.7$\tablenotemark{a} & $988.5_{-542.9}^{+1177.1}$ & $66.39_{-36.46}^{+79.05}$ & $92.38_{-50.73}^{+110.01}$ & 0.26 \\
SDSSJ105842.58$+$314459.76 & 17088 & $10_{-4.1}^{+4.3}$ & $93_{-9.6}^{+9.6}$ & $1.7$\tablenotemark{a} & $756.7_{-416.5}^{+903.1}$ & $41.78_{-23.00}^{+49.86}$ & $42.63_{-23.47}^{+50.87}$ & 0.16 \\
SDSSJ110851.04$+$065901.45 & 21731 & $39_{-6.2}^{+6.2}$ & $22_{-5.7}^{+5.8}$ & $1.1_{-0.6}^{+1.3}$ & $1.0_{-0.5}^{+1.2}$ & $0.64_{-0.36}^{+0.78}$ & $2.97_{-1.68}^{+3.60}$ & 0.17 \\
SDSSJ111458.02$+$403611.41 & 31623 & $9_{-4.0}^{+4.1}$ & $9_{-4.0}^{+4.1}$ & $1.7$\tablenotemark{a} & $40.3_{-23.4}^{+50.9}$ & $0.44_{-0.26}^{+0.56}$ & $0.74_{-0.43}^{+0.93}$ & 4.95 \\
SDSSJ111519.98$+$542316.65 & 21485 & $104_{-10.2}^{+10.2}$ & $1342_{-36.6}^{+36.6}$ & $1.7$\tablenotemark{a} & $490.1_{-269.1}^{+583.7}$ & $289.13_{-158.76}^{+344.38}$ & $350.45_{-192.42}^{+417.41}$ & 0.14 \\
SDSSJ114528.80$+$494510.81 & 19653 & $4_{-2.9}^{+3.2}$ & $9_{-4.0}^{+4.1}$ & $1.7$\tablenotemark{a} & $38.1_{-22.0}^{+47.6}$ & $0.42_{-0.24}^{+0.52}$ & $0.65_{-0.37}^{+0.81}$ & 2.29 \\
SDSSJ123420.14$+$475155.86 & 6798 & $<1.9$ & $11_{-4.3}^{+4.4}$ & $1.7$\tablenotemark{a} & $371.7_{-213.3}^{+460.0}$ & $1.33_{-0.76}^{+1.65}$ & $1.14_{-0.66}^{+1.41}$ & 2.71 \\
SDSSJ131739.20$+$411545.61 & 4994 & $<1.9$ & $8_{-3.8}^{+4.0}$ & $1.7$\tablenotemark{a} & $480.9_{-270.0}^{+589.3}$ & $8.65_{-4.86}^{+10.60}$ & $10.86_{-6.10}^{+13.31}$ & 0.59 \\
SDSSJ134442.17$+$555313.59 & 46725 & $773_{-27.8}^{+27.8}$ & $733_{-27.1}^{+27.1}$ & $1.7$\tablenotemark{a} & $89.5_{-52.0}^{+111.4}$ & $26.58_{-15.44}^{+33.10}$ & $45.26_{-26.29}^{+56.36}$ & 1.06 \\
SDSSJ145210.53$+$292409.00 & 5414 & $1_{-1.9}^{+2.3}$ & $60_{-7.7}^{+7.7}$ & $1.7$\tablenotemark{a} & $1043.1_{-569.8}^{+1241.2}$ & $121.10_{-66.14}^{+144.09}$ & $118.36_{-64.65}^{+140.84}$ & 0.60 \\
SDSSJ151004.01$+$074037.17 & 6456 & $7_{-3.6}^{+3.8}$ & $9_{-4.0}^{+4.1}$ & $1.7$\tablenotemark{a} & $3.9_{-2.1}^{+4.7}$ & $2.58_{-1.43}^{+3.11}$ & $4.74_{-2.62}^{+5.73}$ & 4.70 \\
SDSSJ153924.57$+$032453.71 & 6463 & $1_{-1.9}^{+2.3}$ & $7_{-3.6}^{+3.8}$ & $1.7$\tablenotemark{a} & $2.5_{-1.8}^{+3.8}$ & $0.37_{-0.26}^{+0.55}$ & $0.70_{-0.48}^{+1.04}$ & 2.44 \\
SDSSJ160003.55$+$412845.43 & 12037 & $9_{-4.0}^{+4.1}$ & $9_{-4.0}^{+4.1}$ & $1.7$\tablenotemark{a} & $0.0$\tablenotemark{b} & $0.43_{-0.39}^{+0.85}$ & $0.78_{-0.72}^{+1.56}$ & 1.86 \\
SDSSJ212512.48$-$071329.90 & 8669 & $197_{-14.0}^{+14.0}$ & $98_{-9.9}^{+9.9}$ & $1.6_{-0.9}^{+1.9}$ & $3.5_{-1.9}^{+4.2}$ & $13.67_{-7.50}^{+16.33}$ & $29.92_{-16.40}^{+35.72}$ & 0.29
\enddata
\tablecomments{Column 1: SDSS galaxy name; Column 2: exposure time; Column 3: counts in the soft band; Column 4: counts in the hard band; Column 5: adopted best-fit power-law photon index; Column 6: adopted best-fit intrinsic absorbing column density; Column 7: unabsorbed, rest-frame soft band flux; Column 8: unabsorbed, rest-frame hard band flux; Column 9: angular distance from the observation aimpoint.}
\tablenotetext{a}{fixed}
\tablenotetext{b}{frozen at 0.0 in \texttt{Sherpa}}
\label{tab:offset_xray}
\end{deluxetable*}

\subsubsection{Computing the Coordinate Transformations}
\label{sec:trans}

We supplied the \texttt{dmcoords} tool in \texttt{CIAO} with \sdssra~and \sdssdec~to make astrometric projections of the SDSS sources onto the $Chandra$ native coordinate system.  We then found matching pairs between the coordinates of the $Chandra$ source lists and SDSS source lists using the \texttt{xyxymatch} task in IRAF.  Given the typically small number of matches between images in our sample, we used the basic \emph{Tolerance} algorithm in \texttt{xyxymatch} that finds the input coordinate nearest the reference coordinate but within a user-supplied threshold radius (the `tolerance' parameter).  In our analysis we used a `tolerance' of $2''$ which is similar to the $2.''5$ matching radii used for the $Chandra$ Deep Field North \citep{Alexander:2003} and Extended Deep Field South \citep{Lehmer:2005} and the $1.''2$ matching radius used for the $Chandra$ Deep Field South \citep{Xue:2011}.  To investigate the dependency of the matches on the `tolerance' parameter, we re-ran \texttt{xyxymatch} using $1.''2$ and $2.''5$ radii and found that there was no change in the number or specific sources matched.

Then, we used the \texttt{geomap} task in IRAF to calculate X and Y linear transformations for every $ith$ matched pair ($\Delta X_{i}$,$\Delta Y_{i}$).  As with the CIAO program \texttt{\texttt{wcs\char`_match}}, this method allows us to compute transformations with only one matched pair since both source lists have been placed in the $Chandra$ native coordinate system so that a rotation or scale factor is not necessary. To eliminate potential false matches, in \texttt{geomap} we reject matched pairs that produce transformations that are outliers by more than a specified number of standard deviations, \sigmatrans, from the mean of the transformations.  We allowed the rejection procedure to iterate up to ten times for convergence.  Typically, a rejection threshold of \sigmarej=$1.5-3.0$\sigmatrans~is used for rejecting outliers \citep{Hog:1994,Souchay:2009}.  However, due to the small number of matched pairs in our images, the astrometric solutions are sensitive to individual pairs and a single mis-matched pair can affect our conclusions.  To investigate the dependency of our astrometry on \sigmarej, we re-calculated the linear transformations for $1.0$\sigmatrans$\leq$\sigmarej$\leq3.0$\sigmatrans~and found that the transformations are stable between the values of \sigmarej$=1.5-3.0$\sigmatrans, even with the loss of matched pairs within this range.  For all registered image pairs, we find that the weighted average transformation between \sigmarej$=1.5-3.0$\sigmatrans~is consistent (within the standard errors) with the transformation calculated at \sigmarej$=1.5$\sigmatrans.  Below \sigmarej$=1.5$\sigmatrans, the number of matches pairs drops significantly and the transformations become unstable, indicating that a threshold below \sigmarej$=1.5$\sigmatrans~is too strict.  Therefore, we have adopted \sigmarej$=1.5$\sigmatrans~as the rejection threshold.  If no matched pairs pass these steps, the image pair is not used.  Of the sources that are not rejected, the final linear transformations between the two images in X and Y are taken to be the error-weighted averages of the individual transformations between each matched pair, $\overline{\Delta X}={\sum_{i=1}^{n} \Delta X_{i}\times w_{i,X}}/\sum_{i=1}^{n}w_{i,X}$ and $\overline{\Delta Y}={\sum_{i=1}^{n} \Delta Y_{i}\times w_{i,Y}}/\sum_{i=1}^{n}w_{i,Y}$, respectively (see Section \ref{subsec:accuracy} for a discussion of the weights $w_{i,X}$ and $w_{i,Y}$).

Since we require an un-biased knowledge of the spatial offset between the X-ray AGN and optical host galaxy nuclei, these sources must be excluded from the final transformations.  Therefore, we applied the original transformations, as described above, to \sdssra~and \sdssdec, and repeated the transformation calculations \emph{excluding} any detections within $2''$ of the transformed SDSS AGN coordinates.

\subsubsection{Choice of SDSS Filter}
\label{sec:waveband}
Each $Chandra$ image may potentially be registered with up to all five SDSS bands.  The average wavelengths of the SDSS filters are 3551\AA~($u$), 4686\AA~($g$), 6165\AA~($r$), 7481\AA~($i$) and 8931\AA~($z$), with each sampling emission from a different region of the optical electromagnetic spectrum and thus mapping different mass distributions and subject to different radiative transfer effects.  Here we mention the most pertinent of these differences:  

\noindent \emph{Obscuration:} the level of extinction increases from the reddest ($z$) to the bluest ($u$) filter, such that the observed galaxy morphology in bluer filters is more affected by the presence of dust that can complicate locating the galaxy nucleus. 

\noindent \emph{Line Emission:} AGN and star-formation produce several intense emission lines at optical wavelengths, most notably the H$\beta$/\oiii~and H$\alpha$/\niia~complexes.  At a redshift of $0$, H$\beta$/\oiii~falls in the $g-$band and H$\alpha$/\niia~falls in the $r-$band.  Gas producing these lines is subject to outflow-driven kinematics that do not trace the potential of the galactic nucleus. 

\noindent \emph{Star-formation:} regions of actively forming stars contain larger populations of young stars that contribute to emission in the bluer filters.  These star-forming regions are not necessarily located near the galaxy's nucleus.

For each SDSS band, we have visually inspected the galaxy image and GALFIT model, omitting any sources for which there are clearly dust lanes and/or multiple peaks of emission that suggest kinematics in the ionized gas or result in false centroids.  If these features are present but on scales smaller than the SDSS resolution then they are unlikely to effect our measurement of the host galaxy nucleus above the level of uncertainty in our astrometry.  However, to safeguard against this possibility, we make the following requirement: an AGN is included in the sample if at least either the $i-$ or $z-$band SDSS image is registered with $Chandra$.  This requirement eliminates reliance on the bluer filters that detect emission subject to obscuration, emission line contamination and star-formation.

\subsection{Astrometric Accuracy}
\label{subsec:accuracy}
To measure real (statistically significant) physical offsets between $Chandra$ X-ray AGN detections and SDSS galaxy stellar cores, we need to quantify the level of accuracy to which the images have been registered.  For each dimension, X and Y, of two registered images, we combined in quadrature the errors of source positions in the $Chandra$ image ($\sigma_{C,i,X}$,$\sigma_{C,i,Y}$) and the SDSS image ($\sigma_{S,i,X}$,$\sigma_{S,i,Y}$) to calculate the relative astrometric error for an individual pair of matched sources, $\sigma_{i,X}=\sqrt{\sigma_{C,i,X}^{2}+\sigma_{S,i,X}^{2}}$ and $\sigma_{i,Y}=\sqrt{\sigma_{C,i,Y}^{2}+\sigma_{S,i,Y}^{2}}$.  These errors were used to determine the relative weights of each matched pair ($w_{i,X}=1/\sigma_{i,X}^{2}$ and $w_{i,Y}=1/\sigma_{i,Y}^{2}$), and the final uncertainty on the relative astrometry for this pair of images, (\sigxfinastr,\sigyfinastr), is the inverse of the square root of the sum of the weights of all the sources, $n$, used for registration: \sigxfinastr$=1/\sqrt{\sum_{i=1}^{n} w_{i,X}}$ and \sigyfinastr$=1/\sqrt{\sum_{i=1}^{n} w_{i,Y}}$.  Note that this section describes how we are determining the relative \emph{astrometric} uncertainties and does not include the uncertainties in the X-ray AGN and optical nuclear \emph{detections}, which are discussed in Section \ref{sec:image_modeling}.

\section{Image Modeling}
\label{sec:image_modeling}

Once the $Chandra$ and SDSS images have been registered, measuring significant offsets between X-ray AGN and optical stellar cores requires estimates of the source coordinates, which we obtain by fitting two-dimensional parametric models.  In this section we describe the procedure for modeling the $Chandra$ and SDSS images of the AGN.

\subsection{X-ray Image Modeling}
\label{subsec:Xray_img_mod}

To detect X-ray sources associated with the AGN, we fit models to the rest-frame `soft' (0.5-2keV), `hard' (2-10keV) and `full' (0.5-10 keV) Chandra images to measure their detection significance and position.  To achieve the optimal spatial resolution, we have run the models on $Chandra$ images binned to 1/10th the $Chandra$ native pixel size as in \citet{Harris:2004,Comerford:2015}.  As with pixel binning, the energy filtering was done with \texttt{dmcopy}.  Using the modeling facilities in \texttt{Sherpa}, we simultaneously modeled the sources as two-dimensional Lorenztian functions (\texttt{beta2d}: $f(r)=A(1+[r/r_{0}]^{2})-\alpha$) and the background as a fixed count rate estimated using a source-free adjacent circular region of $30''$ radius.  The initial positions of the \texttt{beta2d} component were set to the location of the transformed SDSS AGN coordinates.  The amplitude $A$ was constrained to have a minimum value of 0.0, and as in \citet{Pooley:2009} we tied $r_{0}$ and $\alpha$ to each other.  While running the fit in \texttt{Sherpa}, we allowed the model to fit a region around the AGN coordinates of 3 times the PSF size (estimated using \texttt{psfSize}) at that location on the chip.  The best-fit model parameters, including the positions and uncertainties (\XChandra$\pm$\XChandraErr~and (\YChandra$\pm$\YChandraErr) were obtained by minimizing the Cash statistic using \texttt{Sherpa}'s implementation of the `Simplex' minimization algorithm \citep{Lagarias:1998}.  We required that a significant detection have an amplitude of $\ge2$ times the amplitude error.  To account for the possibility of multiple X-ray sources, we iteratively increased the number of \texttt{beta2d} components until not all components had significant amplitudes.  In all cases, we found that a single \texttt{beta2d} component provided the best fit.  The detections are only considered to be within the SDSS fiber if the final X and Y solutions are within \sdssfib$\pm$\sigxfinastr~and \sdssfib$\pm$\sigyfinastr, respecitvely (where \sdssfib$=$\sdssfibrad~is the SDSS fiber radius) of the transformed SDSS AGN coordinates.  
Since emission not directly from the AGN corona, such as from outflow-induced shocks or an inter-cluster medium, can produce substantial amounts of soft X-ray photons, we only consider an X-ray detection to be a potential AGN if it passes the above detection criteria in the hard X-ray images.

\begin{figure}
\hspace*{-0.1in} \includegraphics[width=3.5in]{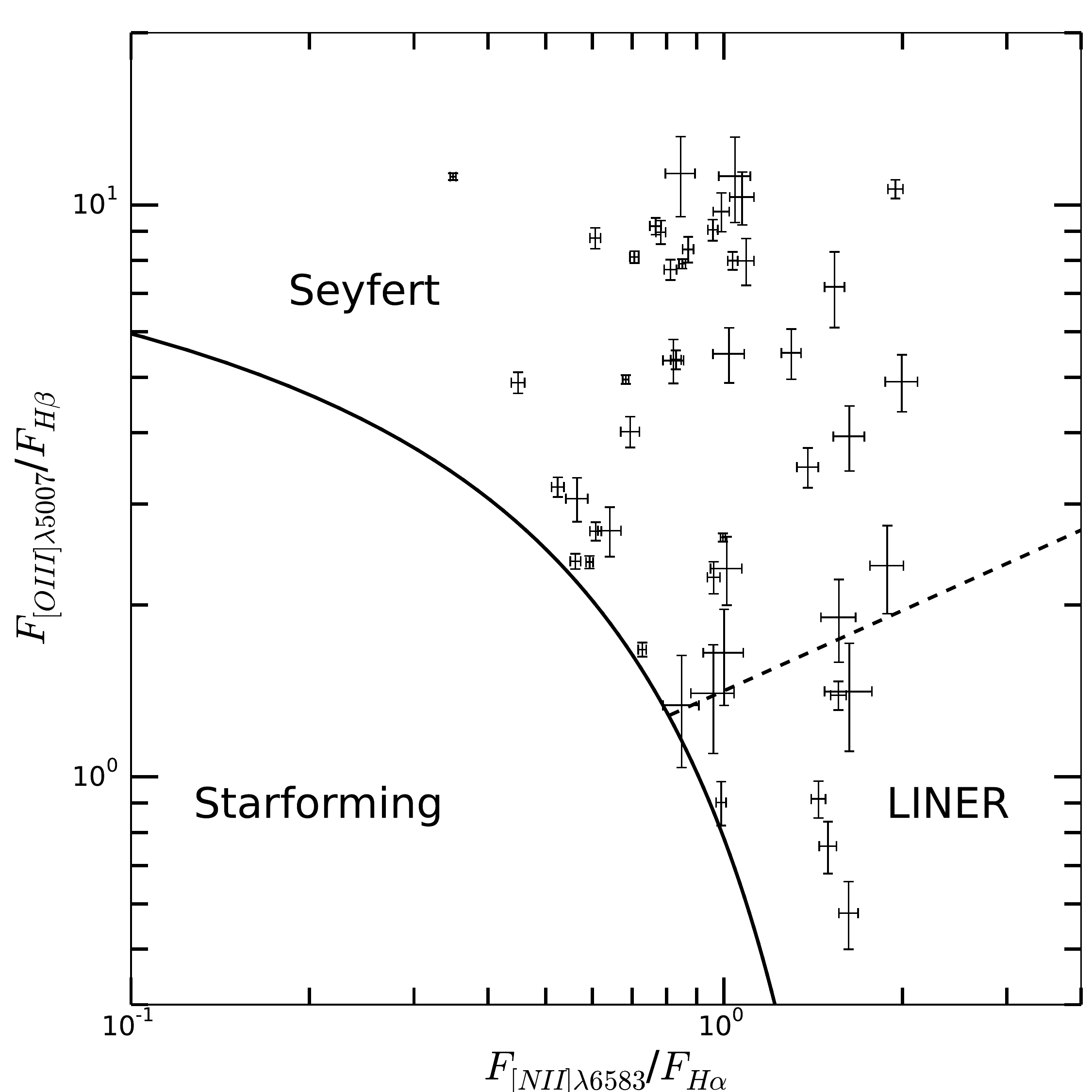}
\caption{\footnotesize{The parent AGN sample plotted on a standard BPT diagram, with the ratio of \oiii~to H$\beta$ flux plotted against the ratio of \niia~to H$\alpha$ flux.  
The solid, curved line marks the theoretical upper limit for pure starburst models from \citet{Kewley:2001}, separating composite star-forming and AGN host galaxies from AGN host galaxies.  The dashed line is the empirical separation between Seyerts and LINERS within the AGN regime from \citet{Kauffmann:2003}.}
}
\label{fig:BPT}
\end{figure}

\subsection{Optical Image Modeling}
\label{subsec:opt_img_mod}

To detect and measure the positions of optical stellar cores in the host galaxy of each SDSS AGN (primary galaxy), we fit the SDSS images with two-dimensional models using GALFIT \citep{Peng:2010}, which is capable of decomposing images of galaxies into multiple components.  We are interested in locating the positions and uncertainties of galactic stellar cores (\XSDSS$\pm$\XSDSSErr~and \YSDSS$\pm$\YSDSSErr), in which case the most relevant component to use is the Sersic profile, which has been empirically shown to be a good approximation of the light profiles of stellar bulges \citep{Graham:2005}.  Therefore, we fit single-Sersic profile fits (plus a uniform sky component) to each galaxy.  The fits were run on square regions of projected physical size 40 kpc on each side (to explore within 20 kpc of the AGN position and thereby include a wide range of merger stages) with the angular size scale calculated from the \zossy~and assuming the cosmology stated in Section \ref{sec:intro}.  

The errors returned by GALFIT are purely statistical in that they are computed directly from the variance of the input images.  Practically, the true radial profiles may deviate significantly from the model components used in GALFIT, particularly at large radii, and this will introduce additional uncertainties in the best-fit parameters.  \citet{Haussler:2007} have examined the uncertainties in GALFIT parameters due to non-statistical effects such as neighbors within the fitting region and non-uniform sky.  They found that the errors are dominated by these effects, rather than the variance in the data.  However, for our purposes of locating the position of optical nuclei, we are only concerned with the location of the Sersic profile peak (\XSDSS~and \YSDSS), and radial profiles of the residuals confirm that these parameters are effectively insensitive to the poor fitting at larger radii.  We note that in \citet{Comerford:2015} we used Sersic profiles for the same purpose of locating the positions of galactic nuclei in galaxy mergers, finding that these profiles are effective at locating the global peak of emission, regardless of significant residuals at large radii.

We also attempted a two-Sersic component fit (over the same fitting region) to test for the presence of secondary stellar cores in either satellite galaxies or close interacting neighbors.  In these cases, we either rejected or adopted the two-component model based on visual inspection.  To ensure that a secondary nucleus is associated with the primary galaxy, we require that it be at a similar redshift (implying they are gravitationally interacting), based on either spectroscopic or photometric redshifts, or that the two nuclei are visibly connected via stars or tidal streams.  In some cases, redshifts are unavailable for the secondary galaxy and we are forced to rely upon visual inspection of the image for connecting features.  We therefore acknowledge the possibility of chance projections.  To estimate this probability, we use the SDSS DR7 galaxy catalogue to calculate the galaxy surface density in the redshift range that covers our entire sample (\zossy$<0.21$).  Assuming an angular separation of $30''$, we estimate a frequency of $\sim3.6\times 10^{-6}$ for such a chance projection.  Given this low probability, combined with the morphologies suggesting interactions, we consider the chance projection possibility to be unlikely.

\begin{deluxetable*}{cccccccccc}
\tabletypesize{\footnotesize}
\tablecolumns{10}
\tablecaption{Spatially Offset AGN: Redshifts, X-ray and Offset Properties, SDSS Filters, Pair matches, and Fiber Coverage.}
\tablehead{
\colhead{SDSS Name  \vspace{0.05in}} &
\colhead{\zossy} &
\colhead{\LXHAGN} &
\colhead{HR} &
\colhead{\septheta} &
\colhead{\sepproj} &
\colhead{P.A.} &
\colhead{Filt.} &
\colhead{n} &
\colhead{Fiber Coverage} \\
\colhead{~ \vspace{0.05in}} &
\colhead{~} &
\colhead{($10^{40}$ erg s$^{-1}$)} &
\colhead{~} &
\colhead{($^{\prime \prime}$)} &
\colhead{(kpc)} &
\colhead{($^{\circ}$ E of N)} &
\colhead{~} &
\colhead{~} &
\colhead{~} \\
\colhead{(1)} &
\colhead{(2)} &
\colhead{(3)} &
\colhead{(4)} &
\colhead{(5)} &
\colhead{(6)} &
\colhead{(7)} &
\colhead{(8)} &
\colhead{(9)} &
\colhead{(10)}
}
\startdata 
J0805$+$2818 & $0.13$ & $114.3_{-65.9}^{+142.0}$ & $-0.15_{-0.19}^{+0.20}$ & $2.32 \pm 0.20$ & $5.2 \pm 0.5$ & $317.5 \pm 28.0$ & $i$ & 1 & Out-Fiber \\
J0813$+$5418 & $0.04$ & $0.7_{-0.4}^{+0.8}$ & $0.00_{-0.22}^{+0.23}$ & $1.73 \pm 0.68$ & $1.3 \pm 0.5$ & $123.2 \pm 48.2$ & $z$,$i$,$r$,$g$,$u$ & 1,2,2,2,1 & In-Fiber \\
J0841$+$0101 & $0.11$ & $193.6_{-108.0}^{+232.5}$ & $-0.05_{-0.07}^{+0.07}$ & $3.75 \pm 0.28$ & $7.4 \pm 0.6$ & $55.1 \pm 4.2$ & $i$,$r$,$g$,$u$ & 2,3,1,1 & Out-Fiber \\
J0907$+$5203 & $0.06$ & $114.1_{-62.9}^{+136.5}$ & $0.72_{-0.16}^{+0.16}$ & $7.62 \pm 0.43$ & $8.5 \pm 0.5$ & $191.3 \pm 10.7$ & $z$,$i$,$r$,$g$,$u$ & 1,1,1,1,1 & Out-Fiber \\
J1042$+$0502 & $0.03$ & $147.4_{-81.0}^{+175.5}$ & $0.92_{-0.04}^{+0.04}$ & $9.44 \pm 0.19$ & $5.0 \pm 0.1$ & $64.5 \pm 1.3$ & $i$,$r$,$g$,$u$ & 2,2,2,1 & Out-Fiber \\
J1058$+$3144 & $0.07$ & $512.5_{-282.1}^{+611.6}$ & $0.81_{-0.10}^{+0.10}$ & $2.66 \pm 0.56$ & $3.6 \pm 0.7$ & $39.0 \pm 8.1$ & $g$,$u$ & 1,1 & Out-Fiber \\
J1108$+$0659 & $0.18$ & $256.5_{-144.6}^{+310.9}$ & $-0.28_{-0.14}^{+0.14}$ & $0.66 \pm 0.24$ & $2.0 \pm 0.7$ & $161.9 \pm 58.1$ & $r$ & 1 & In-Fiber \\
J1114$+$4036 & $0.08$ & $9.3_{-5.4}^{+11.7}$ & $0.00_{-0.31}^{+0.32}$ & $1.86 \pm 0.57$ & $2.6 \pm 0.8$ & $14.6 \pm 4.5$ & $z$,$u$ & 3,1 & In-Fiber \\
J1115$+$5423 & $0.07$ & $3997.3_{-2194.8}^{+4761.1}$ & $0.86_{-0.03}^{+0.03}$ & $0.64 \pm 0.05$ & $0.8 \pm 0.1$ & $29.9 \pm 2.3$ & $i$ & 1 & In-Fiber \\
$-$ & $-$ & $-$ & $-$ & $9.18 \pm 0.05$ & $12.0 \pm 0.1$ & $47.3 \pm 0.3$ & $i$,$r$ & 1,1 & Out-Fiber \\
J1145$+$4945 & $0.16$ & $41.8_{-24.1}^{+52.3}$ & $0.38_{-0.38}^{+0.40}$ & $3.16 \pm 0.52$ & $8.6 \pm 1.4$ & $247.4 \pm 40.8$ & $i$,$r$ & 2,2 & Out-Fiber \\
J1234$+$4751 & $0.18$ & $98.5_{-56.5}^{+121.9}$ & $1.00_{-0.39}^{+0.44}$ & $1.34 \pm 0.57$ & $4.0 \pm 1.7$ & $98.1 \pm 42.0$ & $g$ & 1 & In-Fiber \\
J1317$+$4115 & $0.07$ & $105.3_{-59.1}^{+129.0}$ & $1.00_{-0.48}^{+0.55}$ & $1.35 \pm 0.35$ & $1.7 \pm 0.4$ & $274.9 \pm 72.1$ & $z$,$i$,$r$,$g$,$u$ & 2,2,2,2,2 & In-Fiber \\
J1344$+$5553 & $0.04$ & $132.8_{-77.1}^{+165.4}$ & $-0.03_{-0.03}^{+0.03}$ & $1.18 \pm 0.28$ & $0.8 \pm 0.2$ & $224.4 \pm 52.8$ & $z$,$i$,$r$,$g$ & 3,3,4,4 & In-Fiber \\
J1452$+$2924 & $0.06$ & $940.5_{-513.7}^{+1119.0}$ & $0.97_{-0.13}^{+0.13}$ & $17.39 \pm 0.31$ & $19.4 \pm 0.3$ & $298.6 \pm 5.4$ & $i$,$r$ & 1,2 & Out-Fiber \\
J1510$+$0740 & $0.05$ & $21.8_{-12.1}^{+26.3}$ & $0.12_{-0.33}^{+0.35}$ & $10.02 \pm 0.69$ & $8.8 \pm 0.6$ & $273.5 \pm 18.9$ & $z$,$u$ & 2,1 & Out-Fiber \\
J1539$+$0324 & $0.06$ & $5.2_{-3.6}^{+7.8}$ & $0.75_{-0.51}^{+0.56}$ & $0.92 \pm 0.39$ & $1.1 \pm 0.5$ & $110.0 \pm 46.6$ & $i$ & 2 & In-Fiber \\
J1600$+$4128 & $0.03$ & $1.1_{-1.0}^{+2.3}$ & $0.00_{-0.31}^{+0.32}$ & $9.11 \pm 0.57$ & $5.9 \pm 0.4$ & $59.2 \pm 3.7$ & $z$,$i$,$g$,$u$ & 1,1,1,1 & Out-Fiber \\
J2125$-$0713 & $0.06$ & $278.5_{-152.7}^{+332.5}$ & $-0.34_{-0.06}^{+0.06}$ & $1.34 \pm 0.32$ & $1.6 \pm 0.4$ & $269.8 \pm 64.0$ & $z$,$i$,$r$,$g$,$u$ & 2,2,2,2,2 & In-Fiber \\
$-$ & $-$ & $-$ & $-$ & $7.30 \pm 0.32$ & $8.7 \pm 0.4$ & $13.3 \pm 0.6$ & $z$,$g$ & 2,2 & Out-Fiber
\enddata
\tablecomments{Column 1: abbreviated SDSS galaxy name; Column 2: redshift from the OSSY catalogue; Column 3: AGN 2-10 keV luminosity; Column 4: hardness ratio; Column 5: projected angular offset magnitude; Column 6: projected physically offset magnitude;  Column 7: position angle of the offset X-ray AGN; Column 8: filters of the SDSS image in which significant offsets are measured; Column 9: number of SDSS-$Chandra$ matches for each of the filters in Column 7; Column 10: whether or not the offset is contained entirely within the SDSS fiber.  The offset values in Columns 5 and 6 are from the reddest filter listed in Column 8.}
 \label{tab:offset}
\end{deluxetable*}

\section{X-Ray AGN Selection Criteria}
\label{sec:Xray_AGN}

From Section \ref{subsec:Xray_img_mod}, we know that the hard X-ray detections are coincident with the SDSS fiber and therefore are likely to be associated with the optical AGN detection.  However, in this section we describe several possible alternative sources of this hard X-ray emission and describe the X-ray spectral thresholds used to rule out a null-AGN contribution.  Then, we describe our procedure for X-ray spectral modeling.

\subsection{Sources of X-ray Emission} 

The observational category of ultra luminous X-ray sources (ULXs) consists of off-nuclear objects with X-ray luminosities in the range \LX$=$$10^{39}-10^{41}$ erg s$^{-1}$.  Common interpretations of ULX power include accretion onto compact stellar-mass objects ($<10^{2}$\SUNmass) such as in X-ray binaries (XRBs) or neutron stars \citep{Bachetti:2014}.  Contribution to the observed X-ray emission from these sources may be significant, particularly if vigorous star-formation is occurring.  To account for this effect, we have estimated the expected 2-10 keV luminosity produced by star-formation, \LXHSFR, within the SDSS fiber.  To do so, we have used the relation that describes \LXHSFR~as a function of both the star-formation rate (SFR) and the stellar mass ($M_{\star}$), where SFR and $M_{\star}$ are measured from the galaxy spectrum and available in the MPA-JHU catalogue.  \LXHSFR~is then subtracted from \LXH~to obtain the AGN contribution to the hard X-ray luminosity, \LXHAGN.  The uncertainties on \LXHAGN~(\LXHAGNErr) are the quadrature sums of the \LXH~and \LXHSFR~standard errors.  For AGN in our sample, we require \LXHAGN$\ge3\times$\LXHAGNErr~so that additional X-ray emission must be contributed beyond that from stellar-mass objects.  

Hyper luminous X-ray sources (HLXs) are bright (\LX$>$$10^{41}$ erg s$^{-1}$) off-nuclear objects, and theory suggests they are powered by accretion onto objects with masses $>$$10^{2}$\SUNmass~such as intermediate mass black holes (IMBHs, $10^{2}$\SUNmass$<$\BHmass$<$$10^{6}$)\SUNmass~or SMBHs (\BHmass$>$$10^{6}$\SUNmass).  Due to the often faint nature of their optical counter-parts relative to typical galactic nuclei, they may be tidally-stripped dwarf galaxies hosting IMBHs, though they may also be severely-stripped massive galaxies hosting SMBHs \citep{King:2005,Wolter:2006,Feng:Kaaret:2009,Jonker:2010}.  The brightest observed HLX luminosities are \LX$\sim$$10^{42}$ erg s$^{-1}$, with higher X-ray luminosities only observed in the nuclei of normal and massive galaxies where SMBHs reside.  Therefore, we consider an X-ray detection to be produced by an accreting SMBH (AGN) if \LXHAGN$\ge$$10^{42}$ erg s$^{-1}$.  While this cut will naturally exclude low-luminosity AGN, it is a commonly used conservative threshold for ruling out a null-AGN contribution to the X-ray emission.  

 \begin{figure*}[t!]
\hspace*{0.25in} \includegraphics[width=6.6in]{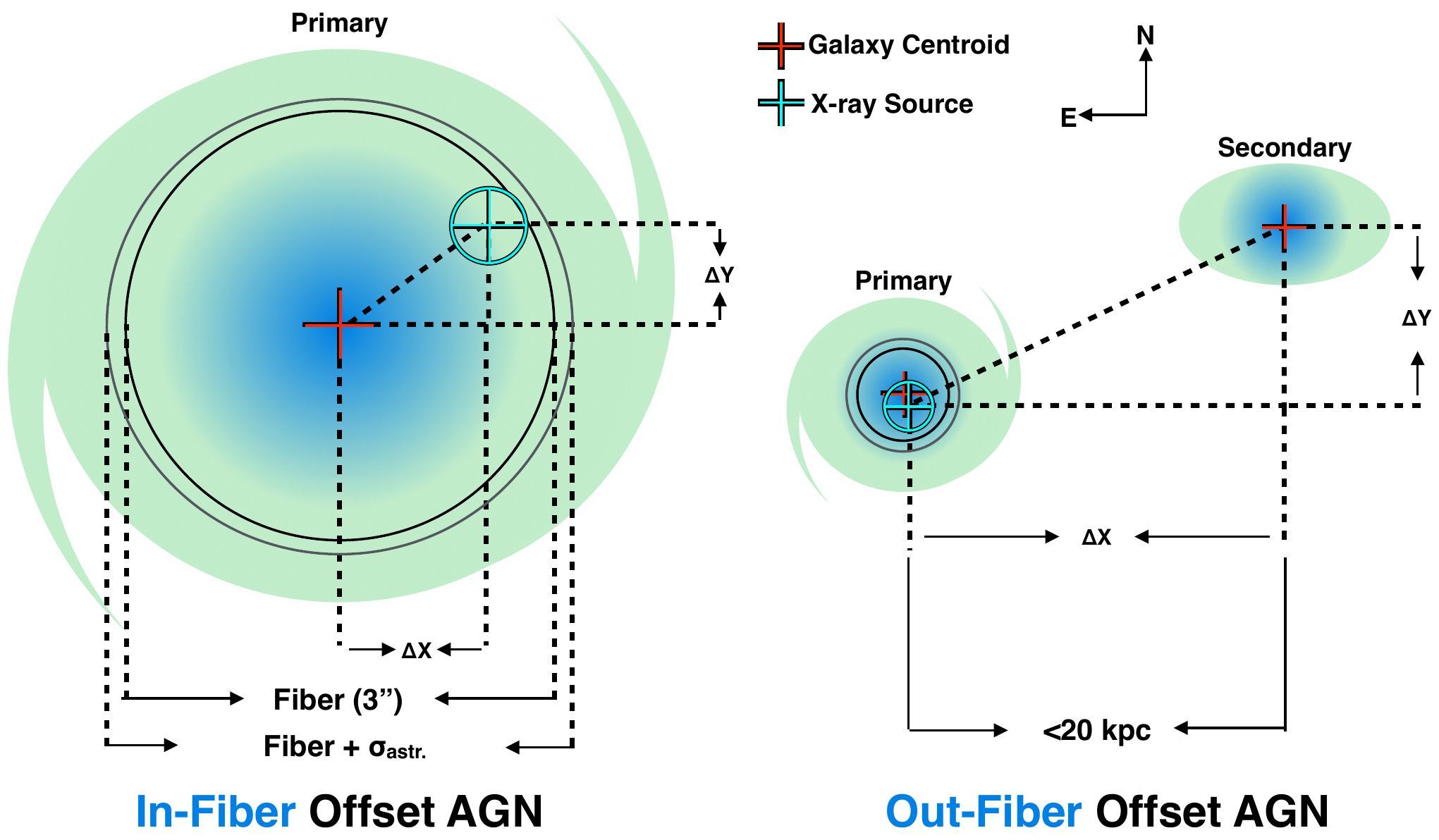}
\caption{\footnotesize{Schematic illustrating the distinctions between the \cata~spatially offset AGN (left) and \catb~spatially offset AGN (right).  The SDSS fiber size is shown by the black circle, with the larger grey circle showing the fiber radius including the astrometric uncertainty, \sigfinastr.  The red cross denotes the centroids of the galaxy stellar cores (\XSDSS,\YSDSS) with sizes adjusted for clarity.  The cyan crosses denote the X-ray detection centroid (\XChandra,\YChandra), and the surrounding ellipse X- and Y-axes denote the $3$\sigxfinpos~and $3$\sigxfinpos~uncertainties, respectively.  In all cases, the primary galaxy contains the X-ray AGN within the fiber by design (see the modeling constraints in Section \ref{subsec:Xray_img_mod}).  
Note that an AGN may in both samples (e.g. the overlap of \catabovrlpsz~shown in Figure \ref{fig:flow_chart}).}
}
\label{fig:schematic}
\end{figure*}

Accreting black holes, regardless of mass, are known to pass through a range of accretion states that can partially be described by the hardness ratio parameter: $HR=(H-S)/(H+S)$, where $H$ and $S$ are the numbers of hard and soft counts, respectively.  Black holes that fall into the ULX and HLX categories (\BHmass$<10^{6}$\SUNmass) typically can not achieve hardness ratios as high as AGN.  While AGN may pass through multiple accretion states in observable timescales, sometimes appearing in a `soft' state, a typical AGN hardness ratio is $HR=-0.6$ \citep{Wilkes:2005}, and a value above -0.1 will exclude stars and clusters \citep{Silverman:2005}.  Therefore, in our selection an X-ray source is considered to be associated with an AGN if it has a hardness ratio of $HR\ge-0.1$.

\subsection{X-ray Spectral Modeling}

Measurements of \LXHAGN~and $HR$ require extracted events files.  We used \texttt{dmextract} to extract events from a source region of two times the radius encircling $90\%$ of the PSF centered on the X-ray source detection and a background region of radius 15 $Chandra$ native pixels centered directly adjacent to and not including any of the source events.
 
From the extracted events files, we measured $H$ and $S$ using \texttt{calc\char`_data\char`_sum}, and calculated $HR$.  Errors on the counts were estimated assuming Poisson noise and propagated through the $HR$ calculations.  The un-binned energy spectra were modeled by minimizing the Cash statistic with \texttt{Sherpa}'s implementation of the Levenberg-Marquardt algorithm.  We are interested in the intrinsic fluxes integrated over rest-frame hard energy range.  Therefore, to each spectrum we fit a redshifted power law, $F\sim E^{-\Gamma}$ (intended to represent the intrinsic AGN X-ray emission at the redshift \zossy) that is multiplied by two absorbing column densities of neutral Hydrogen, one of which is assumed to be intrinsic to the source (\nHexgal), and the other which is fixed to the Galactic value (\nHgal).  We determined \nHgal~by taking the visual extinction along the line of sight to the source, estimated from the dust map of \citet{Schlegel98}, and converting to a column density using the gas-to-dust ratio from \citet{Maiolino:2001}.  Each spectrum was first fit with $\Gamma$ and \nHexgal~allowed to vary freely.  If the best-fit value of $\Gamma$ is not within the typical range of observed AGN power-law indices, i.e. $1\le \Gamma \le 3$ \citep{Nandra:1994,Reeves:2000,Piconcelli:2005,Ishibashi:2010} then $\Gamma$ is fixed at a value of 1.7 (a value typical for AGN, e.g. \citealt{Stern:2002,Middleton:2008}) and the fit is run again.  \LXHAGN~is then calculated from \FXH, \zossy~and the cosmology stated in Section \ref{sec:intro}.

If an X-ray detection satisfies either the \LXHAGN~or $HR$ threshold then it is considered to be an X-ray AGN detection \citep{Juneau:2011} and associated with the optical AGN detection.  For each of the final offset AGN (Section \ref{subsec:offset_agn}), the exposure times, $H$, $S$, best-fitting values of $\Gamma$ and \nHexgal, \FXS, \FXH~and distance of the detection from the aimpoint are reported in Table \ref{tab:offset_xray}, while \LXHAGN~and $HR$ are reported in Table \ref{tab:offset}.


\begin{figure*}
\subfloat {\hspace*{0.10in} \includegraphics[width=6.8in]{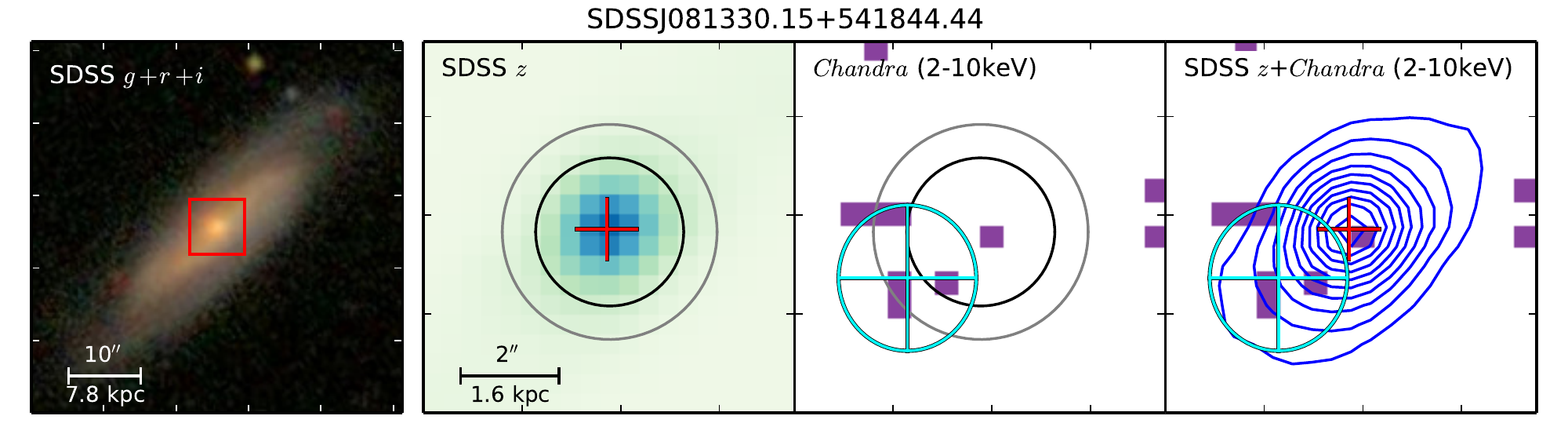}}
\vspace{-0.20in}
\subfloat {\hspace*{0.10in} \includegraphics[width=6.8in]{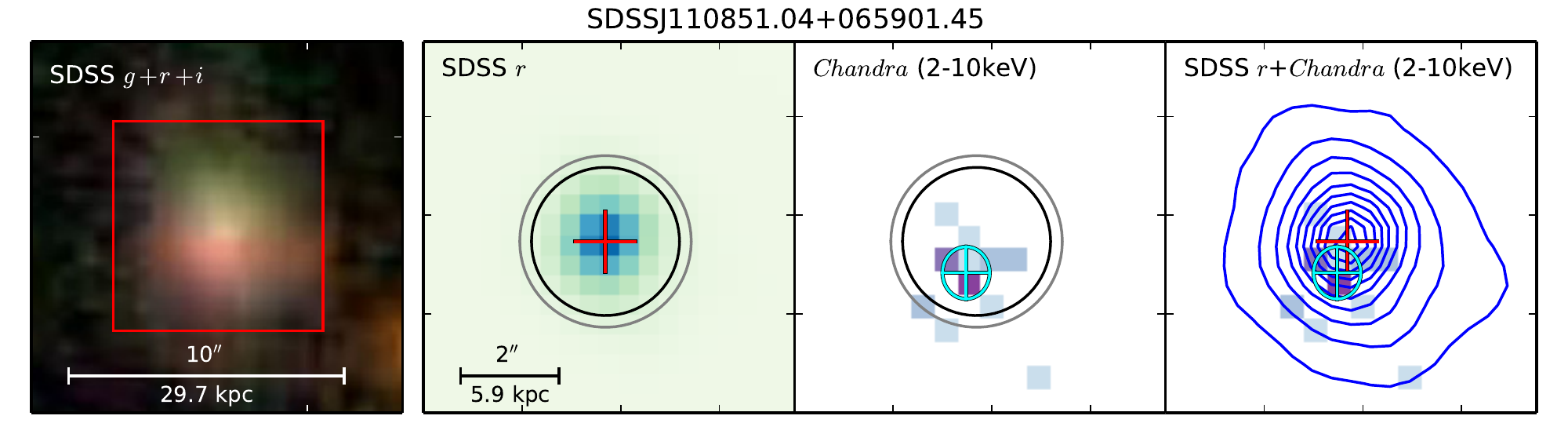}}
\vspace{-0.20in}
\subfloat {\hspace*{0.10in} \includegraphics[width=6.8in]{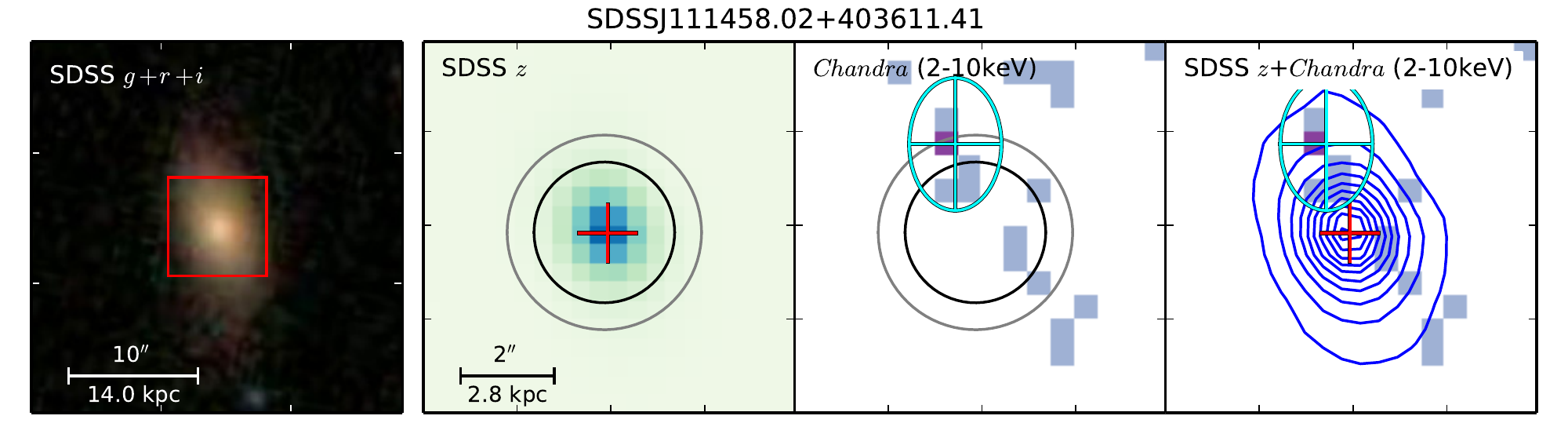}}
\vspace{-0.20in}
\subfloat {\hspace*{0.10in} \includegraphics[width=6.8in]{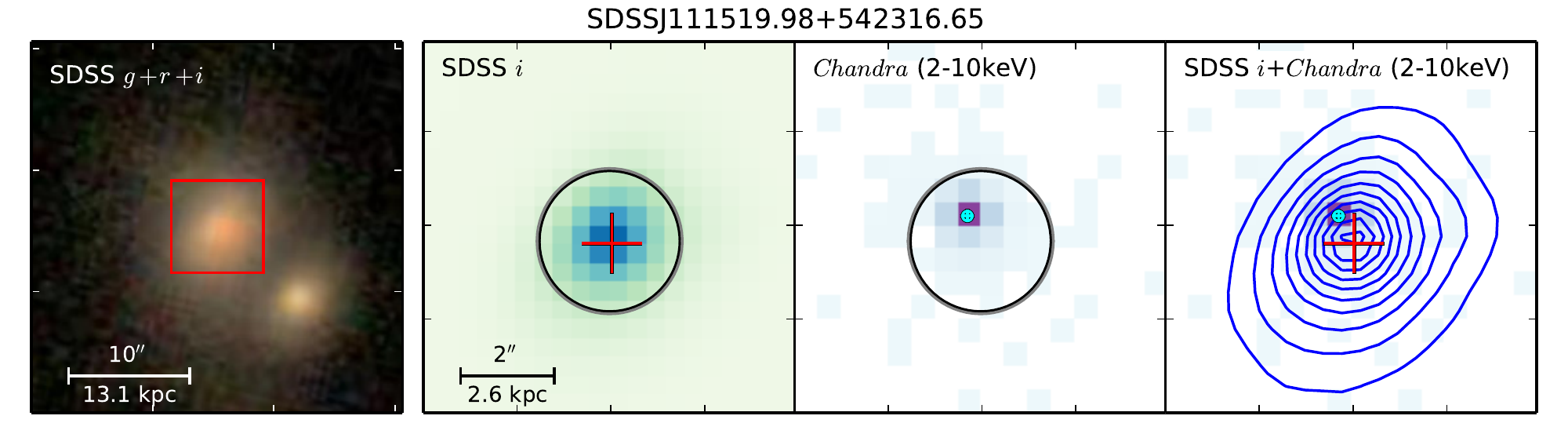}}
\caption{The sample of \cata~offset AGN.  In all panels, North is up and East is to the left.  For reference to the galactic environment, the far left panels show a $40\times40$ kpc field of the SDSS $g+r+i$ composite image centered on the SDSS AGN coordinates.  The red box is $8\times8^{\prime \prime}$ on a side, also centered on the SDSS AGN coordinates, and denotes the field-of-view of the three right-most panels.  From left-to-right, the three right-most panels show the SDSS image (reddest filter from Table \ref{tab:offset}), the rest-frame 2-10 keV image and the rest-frame 2-10 keV image with the SDSS image contours overlaid.  The black circle, grey circle, red crosses and cyan ellipses have the same meaning as described in Figure \ref{fig:schematic}.  For clarity, the X-ray images are shown in the native $Chandra$ pixel size.  All images and coordinates shown are in the SDSS reference frame.}
\label{fig:SDSS_offset_catA}
\end{figure*}
\begin{figure*}
\ContinuedFloat
\subfloat {\hspace*{0.10in} \includegraphics[width=6.8in]{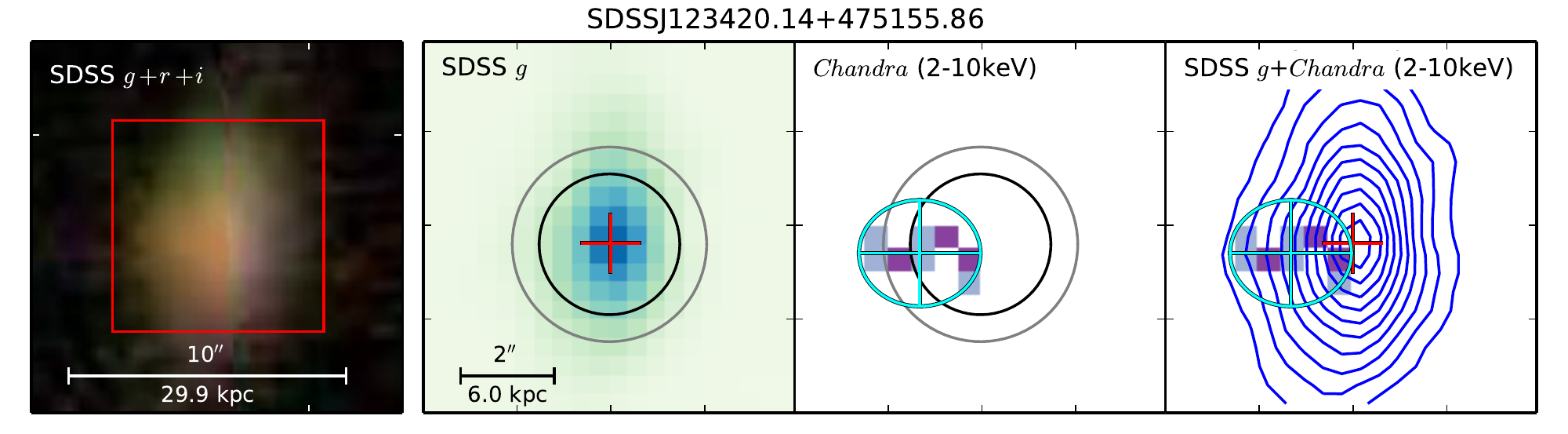}}
\vspace{-0.20in}
\subfloat {\hspace*{0.10in} \includegraphics[width=6.8in]{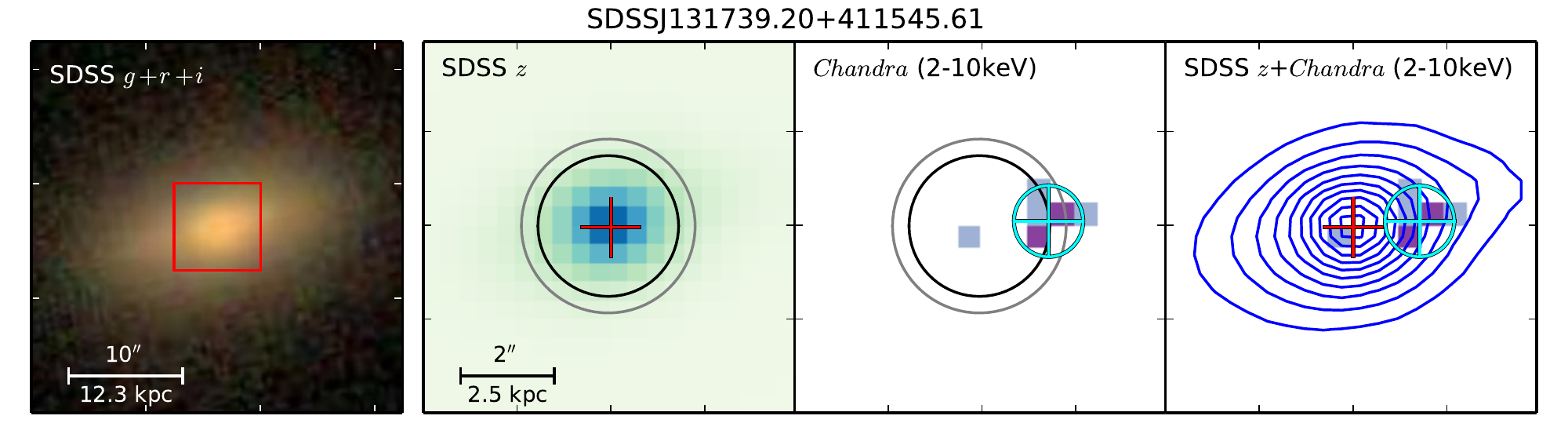}} 
\vspace{-0.20in}
\subfloat {\hspace*{0.10in} \includegraphics[width=6.8in]{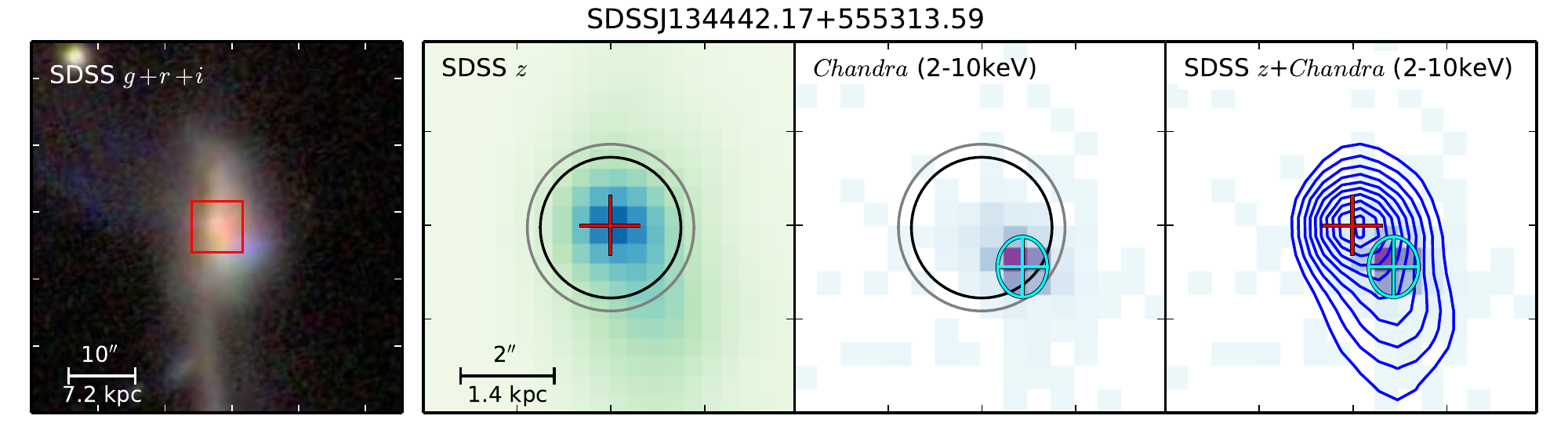}}
\vspace{-0.20in}
\subfloat {\hspace*{0.10in} \includegraphics[width=6.8in]{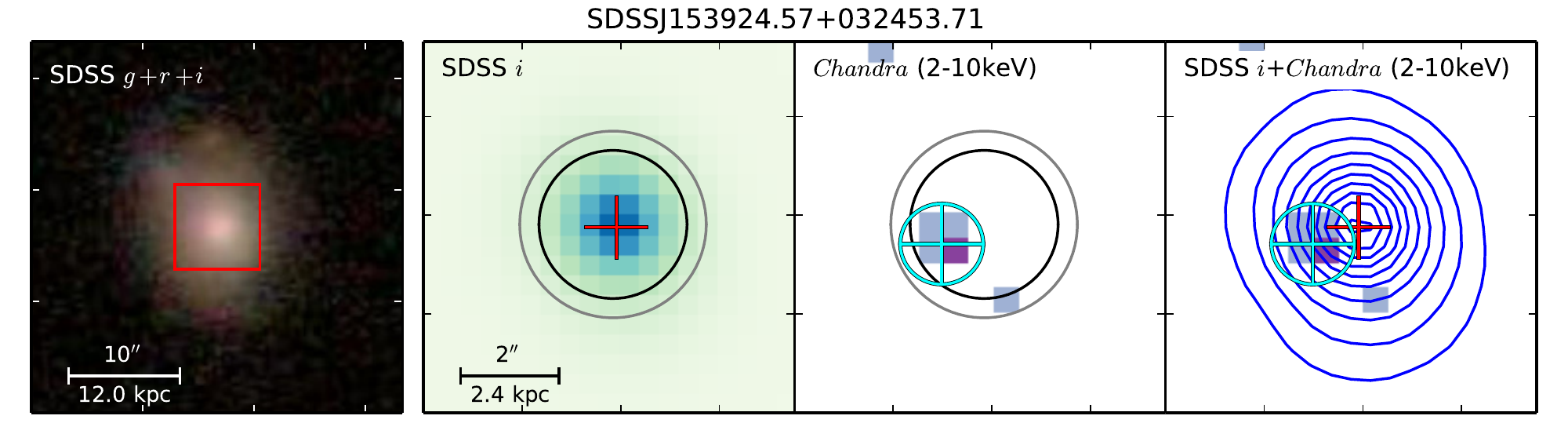}}
\vspace{-0.20in}
\subfloat {\hspace*{0.10in} \includegraphics[width=6.8in]{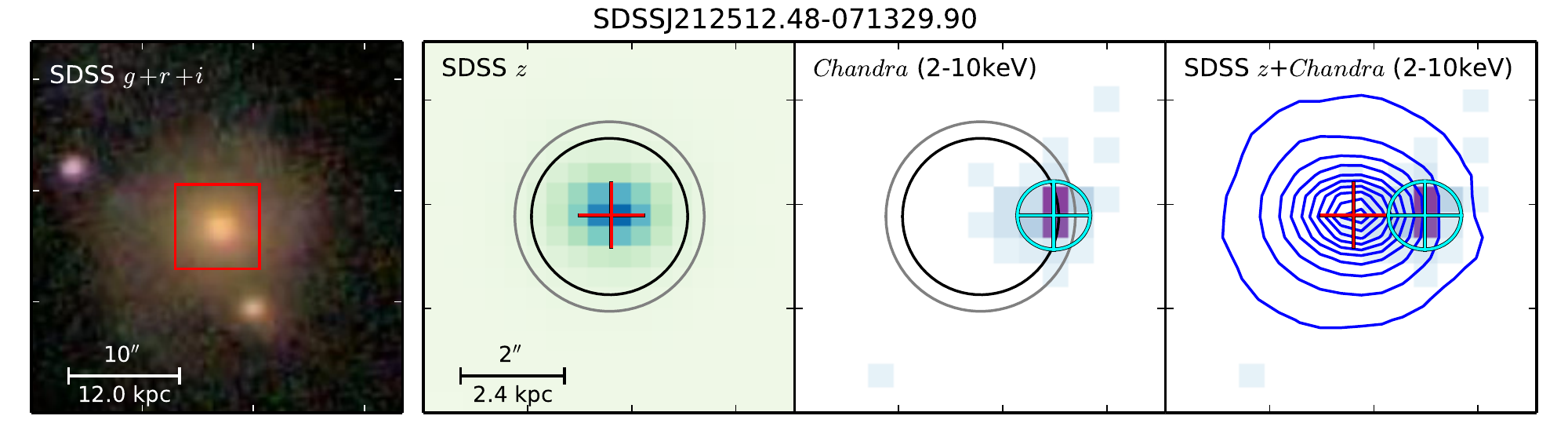}}
\caption{continued}
\end{figure*}



\section{The Parent AGN Sample}
\label{subsec:parent_agn}

Of the \parentAGNsdsssz~SDSS-selected AGN with spectroscopic measurements from the OSSY catalogue (Section \ref{sec:initial}), those that could be registered with over-lapping $Chandra$ coverage (Section \ref{sec:astrometry}), have models for both the X-ray and optical images (Section \ref{sec:image_modeling}) and pass the X-ray AGN criteria (Section \ref{sec:Xray_AGN}) define the parent AGN sample (\parentsz~AGN).  This procedure is displayed graphically, along with the sample size at each step, in Figure \ref{fig:flow_chart}.

Since our initial sample was drawn from the SDSS galaxy catalogue and defined to have emission line ratios satisfying the AGN criteria from \citet{Kewley:2001}, all sources in the final sample are optically classified as Type II AGN (Figure \ref{fig:BPT}).  The general population of Type II AGN on the BPT diagram can be divided into Seyfert II nuclei and low-ionization nuclear emission-line regions (LINERs), for which the nature of the photo-ionizing source may not be related to SMBH accretion.  While LINERs typically have lower luminosities compared to Seyferts, the emission line ratios tend to be similar except that LINERs have relatively strong lines of low-ionization potential (e.g. \niia) such that the flux ratio of $F_{[NII]\lambda 6583}$/$F_{H\alpha}$ is larger relative to $F_{[OIII]\lambda 5007}$/$F_{H\beta}$ and therefore generally occupy a different region of the BPT diagram.  To estimate the probability of LINER classifications for our parent AGN sample, we have shown in Figure \ref{fig:BPT} the empirical Seyfert-LINER division defined in \citet{Kauffmann:2003}.  From Figure \ref{fig:BPT} we see that \seyfertsz/\parentsz~(\seyfertfrac) of the parent sample are classified as Seyferts based on this division.  The fact that the parent sample is strongly skewed toward the Seyfert regime of the BPT diagram is a result of our X-ray AGN selection, indicating that this step is indeed selecting AGN with emission produced by accreting SMBHs.  This fact, combined with their proximity to the Seyfert-LINER division, strongly suggests that the photo-ionizing source of the \linersz~sources in the LINER regime is also accretion onto a SMBH.



\begin{figure*}
\subfloat{\includegraphics[width=7.in]{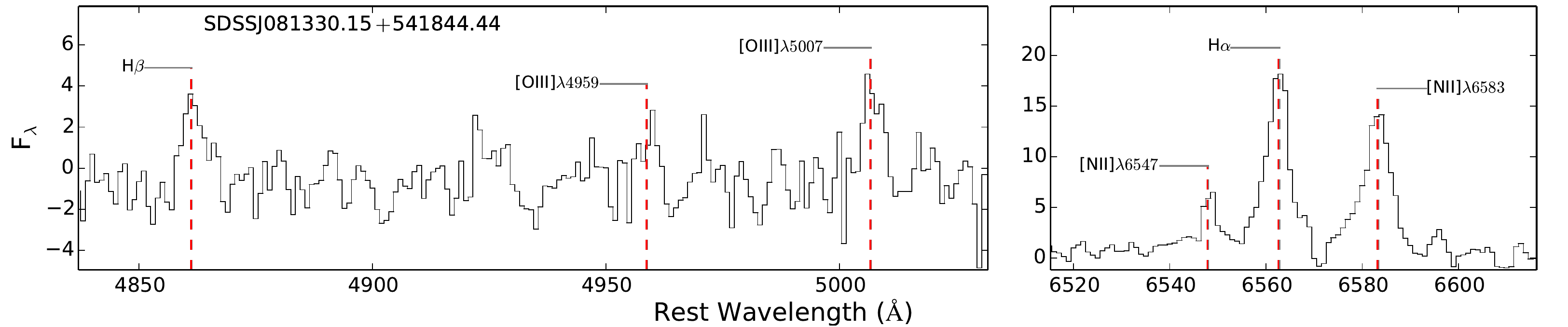}}
\vspace*{-0.38in}
\subfloat{\includegraphics[width=7.in]{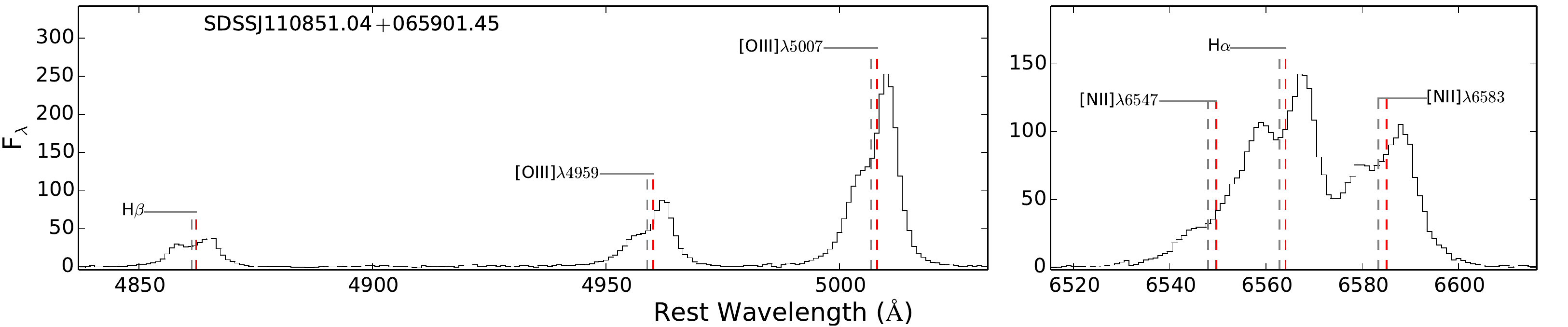}}
\vspace*{-0.38in}
\subfloat{\includegraphics[width=7.in]{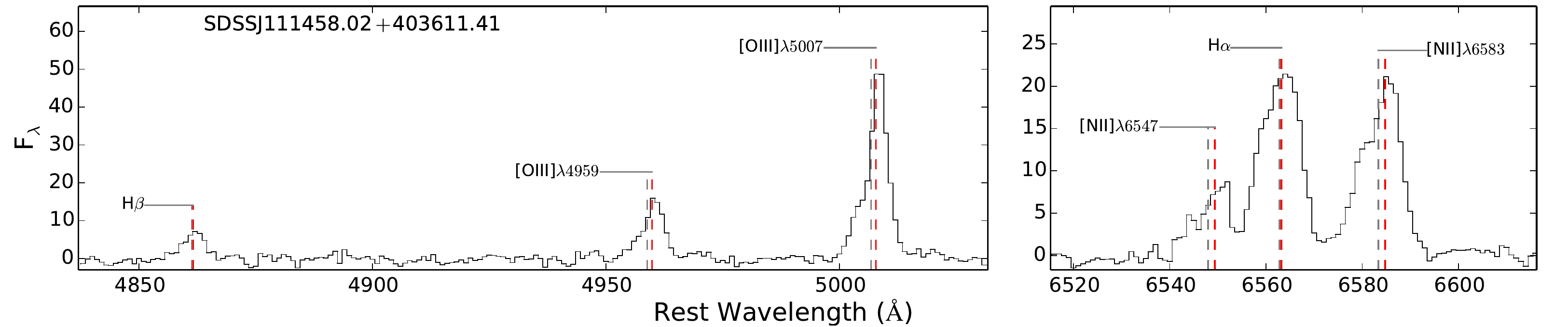}}
\vspace*{-0.38in}
\subfloat{\includegraphics[width=7.in]{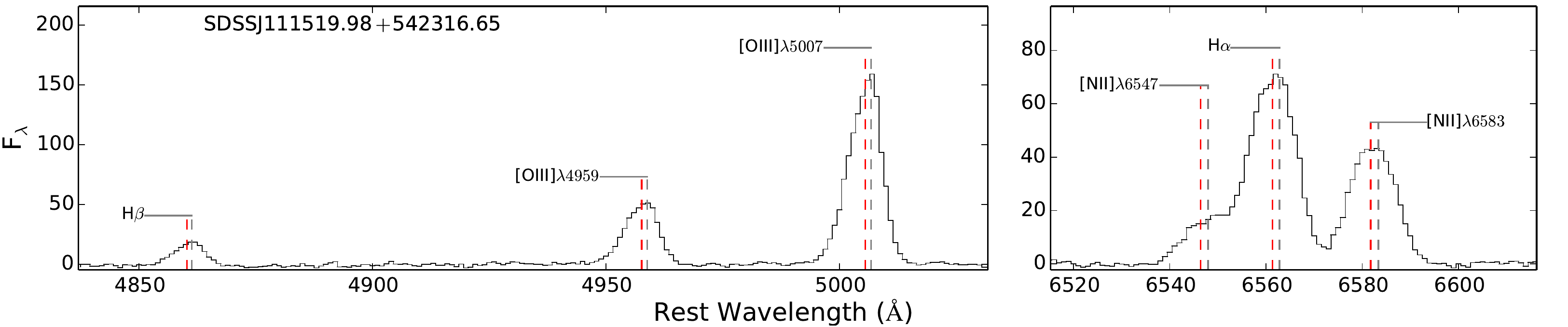}}
\vspace*{-0.38in}
\subfloat{\includegraphics[width=7.in]{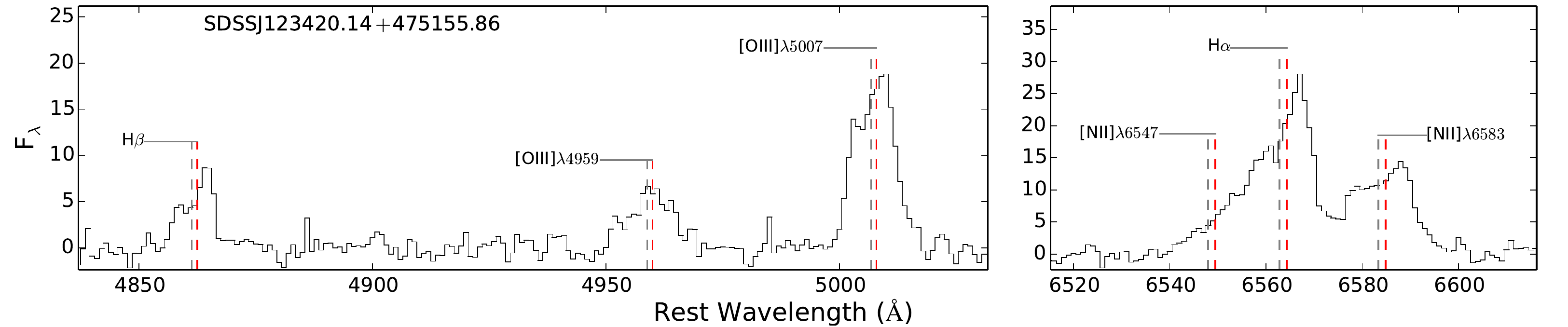}}
\caption{\footnotesize{SDSS spectra of the \catasz~spatially offset AGN from the \cata~sample shifted to the galaxy rest-frame based on \zossy.  Shown are the wavelength regions encompassing the H$\beta$ and [\ion{O}{3}]$\lambda 4959,5007$ emission lines (left) and the H$\alpha$ and [\ion{N}{2}]$\lambda 6547,6583$ emission lines (right).  The spectral flux densities, $F_{\lambda}$, are in units of $10^{-17}$ erg s$^{-1}$ cm$^{-2}$ \AA$^{-1}$.  The stellar continuum has been subtracted.  The grey, dashed lines show the expected wavelength of the emission lines if they were at the same redshift as the host galaxy stars.  The red, dashed lines show the measured wavelengths of the emission lines.}
}
\label{fig:spectra}
\end{figure*}

\begin{figure*}
\ContinuedFloat
\subfloat{\includegraphics[width=7.in]{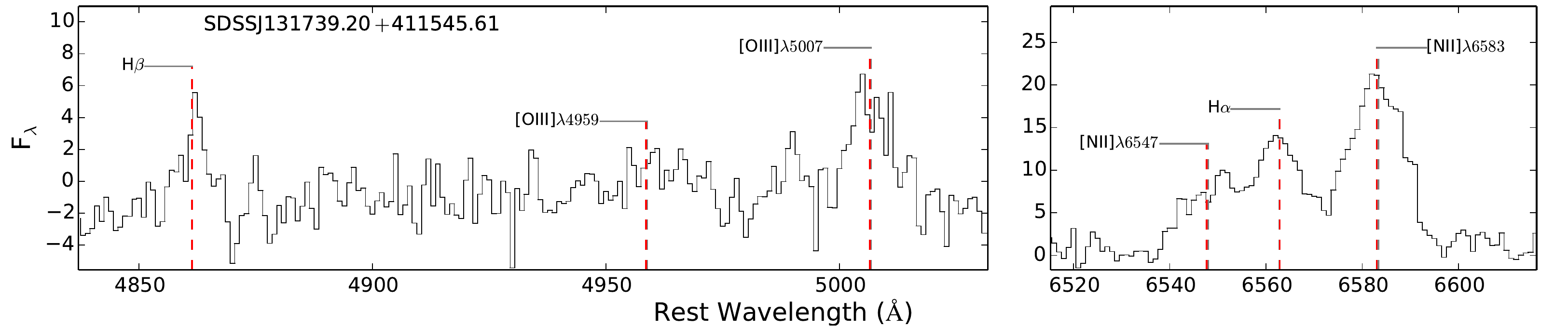}}
\vspace*{-0.38in}
\subfloat{\includegraphics[width=7.in]{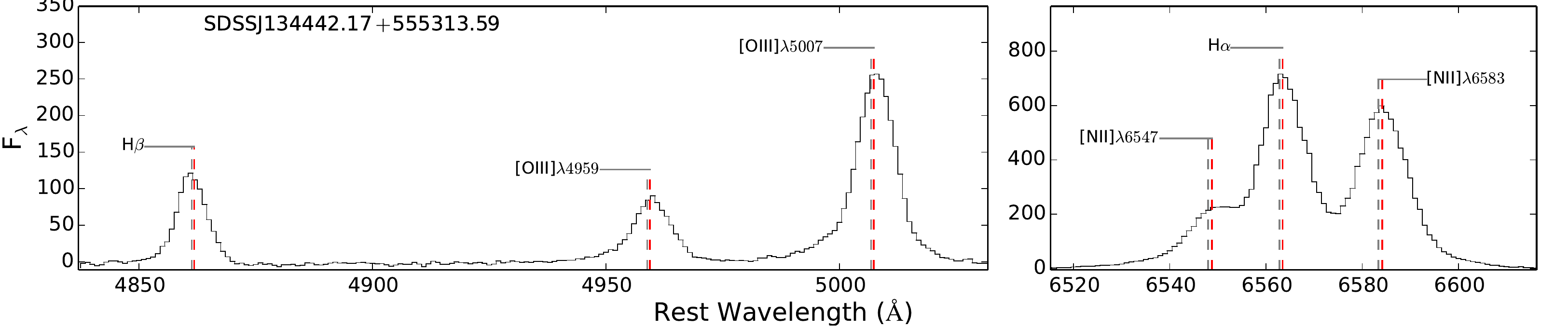}}
\vspace*{-0.38in}
\subfloat{\includegraphics[width=7.in]{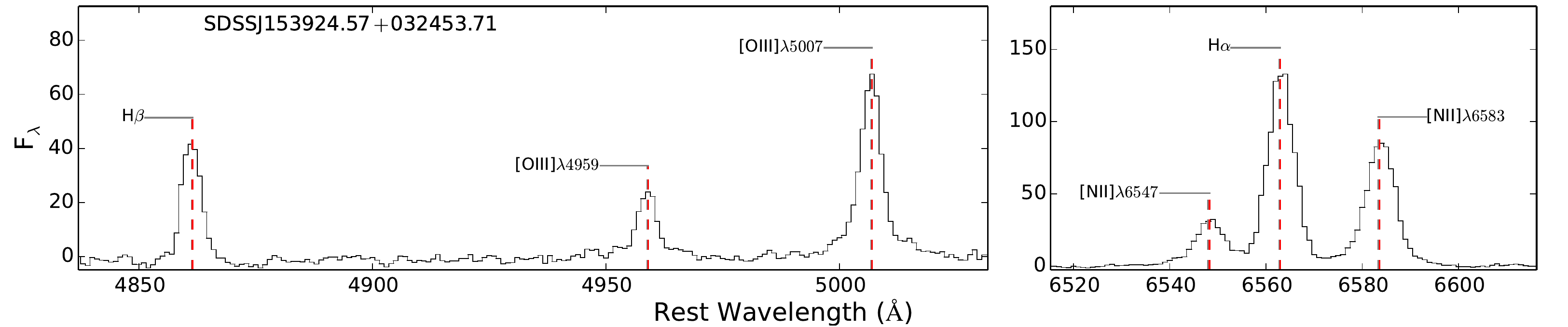}}
\vspace*{-0.38in}
\subfloat{\includegraphics[width=7.in]{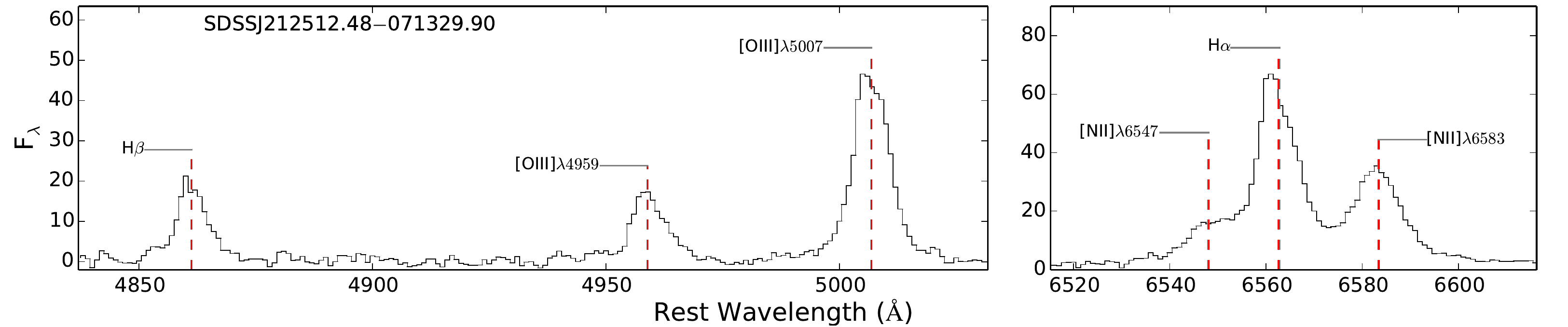}}
\caption{continued
}
\end{figure*}

\section{The Spatially Offset AGN Sample}
\label{subsec:offset_agn}

An offset AGN is any system from the parent sample in which the X-ray AGN (constrained to be within the SDSS fiber in Section \ref{subsec:Xray_img_mod}) has a significant projected offset - in either of the X or Y dimensions - from an optical galaxy stellar core (constrained to be within 20 kpc as defined in Section \ref{subsec:opt_img_mod}).  We define `significant' as $|\Delta X|\ge3.0$\sigxfinpos~and/or $|\Delta Y|\ge3.0$\sigyfinpos, where \sigxfinpos~and \sigyfinpos~are the quadrature sums of the astrometric uncertainty (\sigxfinastr,\sigyfinastr), X-ray image model position uncertainty (\XChandraErr,\YChandraErr) and optical image model position uncertainty (\XSDSSErr,\YSDSSErr).  Due to the filter criteria defined in Section \ref{sec:waveband}, we have offset measurements in at least one of the $i-$ or $z-$bands.  Therefore, we include sources in our offset sample if a significant spatial offset is measured in either the $i-$ or $z-$band, or alternatively if a significant spatial offset  is measured in one or more of the other three bands that is consistent, within the uncertainties \sigxfinpos~and \sigyfinpos, with offsets in the $i-$ or $z-$bands.  This requirement ensures that significant offsets are tied to the $i-$ and/or $z-$bands, which are the optimal SDSS filters for tracing stellar cores.  The final offset values adopted correspond to the reddest filter available that reaches this criterion.  Since the values of \sigxfinpos~and \sigyfinpos~are standard errors, offsets are at a confidence of $\ge99.73\%$ assuming the errors are distributed normally \citep{Lupton:1993}.  Implementing these requirements on the parent AGN sample results in \catabAGNsz~offset AGN.

\subsection{Physical Interpretation of the Spatially Offset AGN}
\label{subsec:physical}

SMBHs naturally exist in the nuclei of galaxies, and the pathways leading to the off-nuclear position of an AGN are limited.  In particular, two scenarios are envisioned, both of which are preceded by a galaxy merger.  In the first scenario, the merger is on-going or the merged system has relaxed, but the SMBHs of the two progenitor galaxies have not yet merged.  In this scenario, the two SMBHs, one of which is actively accreting as an AGN, are losing angular momentum via dynamical friction as their separation evolves toward coalescence.  In the second scenario, the two SMBHs have already coalesced, with the inevitable anisotropic emission of gravitational waves that imparts a non-zero net momentum to the merged SMBH.  The associated velocity will spatially displace the SMBH, with sufficiently large velocities resulting in a spatial offset resembling the offsets we have detected.  This `recoiling' SMBH is expected to maintain its accretion disk, potentially appearing as a spatially offset X-ray AGN.  While spatially offset AGN coincident with detected stellar cores are unlikely to be recoils since the ejected SMBH is not expected to maintain a significant number of bound stars, this constraint can not be put on the spatially offset AGN without such detections.  Therefore, we have used the results of simulations to estimate the relative probabilities of each scenario based on available data.  

\citet{Van_Wassenhove:2012} and \citet{Steinborn:2015} have simulated the occurrence of AGN in galaxy mergers before the SMBHs have coalesced.  Between the two studies, the fraction of AGN in mergers ranges from $1-5\%$.  Applying this frequency to our parent sample results in a prediction of $\sim1-2$ pre-coalescence spatially offset AGN in our sample.  \citet{Blecha:2016} have simulated the observability of spatially offset recoiling AGN in existing imaging surveys.  They predict up to 1 such system may be found in the $Chandra$ Deep Field North.  
Assuming each recoil event is associated with an $HST$ host galaxy implies a fraction of $4\times 10^{-5}$.  Applying this frequency to our parent sample corresponds to $2\times 10^{-3}$ recoiling offset AGN in our sample.  The probability of a recoil event is likely to be even smaller for our sample because the AGN are of the Type II class, implying obscuration of the broad line region (BLR).  Since the offset AGN in our sample (spatial offsets $\gtrsim0.8$ kpc) are well beyond typical radii of obscuring tori and recoiling SMBHs are not expected to carry significant quantities of obscuring material \citep{Komossa:2008,Blecha:2016}, a recoiling AGN would likely be viewed as a Type I AGN.  We note that a third scenario involves a triple SMBH interaction in which the two most massive SMBHs merge and the spatially offset AGN corresponds to the dynamically ejected lightest SMBH.  However, triple SMBH systems are intrinsically rarer than dual SMBH systems.  Therefore, since the probability of in-spiraling SMBH pairs are far more likely than observing a recoil event or triple SMBH interaction, we view our offset AGN sample as pre-coalescence dual SMBH systems in all subsequent analysis.

Finally, while we only identify a single offset AGN via X-ray detections, this does not preclude the possibility of a dual AGN system in the pre-coalescence dual SMBH scenario if the second AGN is fainter in hard X-rays.  One example of this scenario is SDSSJ110851.04+065901.45 (discussed individually in Section \ref{subsec:notes}).  The X-ray spectral and spatial profile of this source was analyzed in detail by \citet{Liu:2013}.  The counts were too few to resolve two sources, so \citet{Liu:2013} performed a one-dimensional re-projection of the counts along the axis connecting the two optical nuclei, finding that the model was marginally consistent with a dual AGN.  Since we do not use $HST$ resolution imaging, we do not attempt this re-projection in our analysis and thus do not detect two nuclei.  Instead, we detect the South-East AGN that is brightest in hard X-rays, consistent with the South-East detection from \citet{Liu:2013}.  We also see marginal evidence for two X-ray sources in SDSSJ111458.02+403611.41.  While a single \texttt{beta2d} component provided the best fit (centroiding on a location in the North-East of the fiber) the double \texttt{beta2d} component model provided a nearly satisfactory fit.  In particular, in addition to a significant detection in the North-East location (\twocompN), the two-component model yielded a \twocompS~detection in the South-West of the fiber.  This marginal detection can be seen in Figure \ref{fig:SDSS_offset_catA} and may represent a second source.

\subsection{Notes on Individual Spatially Offset AGN}
\label{subsec:notes}

We note that the spatially offset AGN fraction of \catabAGNsz/\parentsz~is artificially elevated since the X-ray AGN are not completely random samplings of offset AGN.  In particular, \catabbiassz~of them were targeted by $Chandra$ programs as potential offset AGN systems, and they are the targets of the $Chandra$ observations we use.  These \catabbiassz~objects all have previously reported X-ray AGN offsets from independent works that are consistent with our measurements.  Additionally, \cataspecsigsz~objects from our sample (three of which are included in the above \catabbiasoffsz~objects) have spectroscopic properties suggestive of a dual SMBH system.  In this section we describe these \litsz~objects individually. \\

\noindent SDSSJ080523.29+281815.84: This object was identified as being in an AGN pair by \citet{Liu:2011}, and it was the target of the $Chandra$ observation we use (OBSID: 14964, PI: Liu). \\ \\
SDSSJ084135.09+010156.31: The SDSS fiber spectrum of this object exhibits two kinematically offset AGN emission line systems \citep{Reyes:2008,Greene:2011}.  It was the target of the $Chandra$ observation we use (OBSID: 13950, PI: Comerford), imaged with the intention of spatially constraining the AGN position relative to the host galaxy \citep{Comerford:2015}. \\ \\
SDSSJ090714.45+520343.42: This object was identified as being in an AGN pair by \citet{Liu:2011}, and it was the target of the $Chandra$ observation we use (OBSID: 14965, PI: Liu). \\ \\
SDSSJ104232.05+050241.94: Also known as NGC3341, this object was identified as a spatially offset AGN in \citet{Barth:2008} from the off-nuclear location of an SDSS fiber, and it was the target of the $Chandra$ observation we use (OBSID: 13871, PI: Bianchi). \\ \\
SDSSJ105842.58+314459.76: This object was identified as being in an AGN pair by \citet{Liu:2011}, and it was the target of the $Chandra$ observation we use (OBSID: 14966, PI: Liu). \\ \\
SDSSJ110851.04+065901.45: The SDSS fiber spectrum of this object exhibits two kinematically offset AGN emission line systems \citep{Liu2010a}.  It was followed up with longslit spectroscopy \citep{Liu2010b}, NIR imaging \citep{Liu2010b,Shen:2011b}, AO imaging \citep{Fu:2011a}, IFU spectroscopy \citep{Fu:2012} and radio observations \citep{Bondi:2016}.  The combination of these observations confirms that it is a merging system that likely hosts two AGN.  It was the target of the $Chandra$ observation we use (OBSID: 12749, PI: Shen), imaged with the intention of spatially constraining the AGN position relative to the host galaxy \citep{Liu:2013}. \\ \\  
SDSSJ111519.98+542316.65:  Also known as MCG +09-19-015, the SDSS fiber spectrum of this object exhibits a single AGN emission line system that is kinematically offset from the host galaxy absorption line system (CG14). 
It is also in an AGN pair \citep{Liu:2011}, and it was the target of the $Chandra$ observation we use (OBSID: 13902, PI: Mushotzky). \\ \\  
SDSSJ123420.14+471555.86:  The SDSS fiber spectrum of this object exhibits two kinematically offset AGN emission line systems \citep{Ge:2012}. \\ \\
SDSSJ134442.17+553113.59:  Also known as Mrk 273, it is a well-known ULIRG for which AO imaging has confirmed the AGN location to be in the South-West nucleus \citep{U:2013}.  It was the target of the $Chandra$ observation we use (OBSID: 809, PI: Xia). 

\begin{deluxetable*}{ccccccc}
\tabletypesize{\footnotesize}
\tablecolumns{7}
\tablecaption{\cata~Spatially Offset AGN: Spatial Properties.}
\tablehead{
\colhead{SDSS Name \vspace*{0.05in}} &
\colhead{RA$_{\rm{Gal.}}$} &
\colhead{DEC$_{\rm{Gal.}}$} &
\colhead{RA$_{\rm{AGN}}$} &
\colhead{DEC$_{\rm{AGN}}$} &
\colhead{$\Delta X$} &
\colhead{$\Delta Y$} \\
\colhead{~ \vspace*{0.05in}} &
\colhead{(hh:mm:ss.s)} &
\colhead{(dd:mm:ss.s)} &
\colhead{((hh:mm:ss.s))} &
\colhead{(dd:mm:ss.s)} &
\colhead{($^{\prime \prime}$)} &
\colhead{($^{\prime \prime}$)} \\
\colhead{(1)} &
\colhead{(2)} &
\colhead{(3)} &
\colhead{(4)} &
\colhead{(5)} &
\colhead{(6)} &
\colhead{(7)}
}
\startdata 
J0813$+$5418 & 08:13:30.3253 & +54:18:43.5454 & 08:13:30.1599 & +54:18:44.4932 & $[-1.45\pm0.47]$ & $-0.95\pm0.49$ \\
J1108$+$0659 & 11:08:51.0493 & +06:59:00.8147 & 11:08:51.0355 & +06:59:01.4443 & $-0.21\pm0.16$ & $[-0.63\pm0.18]$ \\
J1114$+$4036 & 11:14:58.0541 & +40:36:13.1877 & 11:14:58.0128 & +40:36:11.3929 & $-0.47\pm0.33$ & $[+1.80\pm0.47]$ \\
J1115$+$5423 & 11:15:20.0078 & +54:23:17.1583 & 11:15:19.9714 & +54:23:16.6057 & $[-0.32\pm0.04]$ & $[+0.55\pm0.04]$ \\
J1234$+$4751 & 12:34:20.2603 & +47:51:55.6966 & 12:34:20.1283 & +47:51:55.8872 & $[-1.33\pm0.43]$ & $-0.19\pm0.38$ \\
J1317$+$4115 & 13:17:39.0751 & +41:15:45.6962 & 13:17:39.1943 & +41:15:45.5819 & $[+1.34\pm0.25]$ & $+0.11\pm0.25$ \\
J1344$+$5553 & 13:44:42.0682 & +55:53:12.7888 & 13:44:42.1663 & +55:53:13.6316 & $[+0.83\pm0.18]$ & $[-0.84\pm0.21]$ \\
J1539$+$0324 & 15:39:24.6262 & +03:24:53.3416 & 15:39:24.5683 & +03:24:53.6563 & $[-0.87\pm0.28]$ & $-0.31\pm0.27$ \\
J2125$-$0713 & 21:25:12.3879 & -07:13:29.8771 & 21:25:12.4782 & -07:13:29.8731 & $[+1.34\pm0.23]$ & $-0.00\pm0.22$
\enddata
\tablecomments{Column 1: abbreviated SDSS galaxy name; Columns 2-3: right ascension and declination of the optical galaxy stellar core; Columns 4-5: right ascension and declination of the X-ray AGN; Column 6-7: angular offsets of the X-ray AGN in the X-dimension and Y-dimensions.  All values of RA and DEC are in the SDSS reference frame.  The offset values in Columns 6-7 are from the reddest filter listed in Column 8 of Table \ref{tab:offset} and correspond to frames oriented with North up and East to the left.  Values in Columns 6-7 that satisfy the criteria for inclusion in the spatially offset AGN sample are bracketed.  A source with either value bracketed will be in the final spatially offset AGN sample of this work.}
 \label{tab:offset_gen}
\end{deluxetable*}

\subsection{Fiber Coverage of the Spatially Offset AGN}
\label{subsec:fiber}

The X-ray AGN can be offset from a stellar core that is either the nucleus of the primary galaxy (the galaxy on which the SDSS fiber was placed), or that of a secondary galaxy.  The combined effects of \sdssfib~and the $1\farcs6$ SDSS resolution mean that all detected secondary stellar cores are outside of the fiber.  This effect of the fiber is shown in the final steps of Figure \ref{fig:flow_chart} and schematically in Figure \ref{fig:schematic}.  Since the analysis (Section \ref{sec:analysis}) and discussion (Section \ref{sec:discussion}) in this first paper is focused on the spectroscopic properties of offset AGN, we have divided the sample into two categories based on whether or not the measured spatial offset is contained entirely within the fiber.  We note that this division has no inherent physical significance but is merely an artifact of the fiber size:  


\noindent {\bf \cata~Spatially Offset AGN:}  As shown in Figure \ref{fig:schematic} (left), these are systems in which the uncertainties are small enough to measure a significant offset between the X-ray source and the stellar core of the primary galaxy.  As denoted by the grey circle, the offset is contained entirely within the fiber when accounting for the relative astrometric uncertainties, and the emission line kinematics induced by the merger will be reflected in the spectroscopic signatures.  We find that \catasz~of the offset AGN are included in this category.  Though we do not detect a distinct optical component associated with the X-ray source, we do find evidence of morphological and/or photometric asymmetries that may be the result of a past galaxy merger.  These asymmetries may possibly account for some cases in which the SDSS fiber is slightly offset from the optical peak of emission, though this may also be due to differences between the SDSS pipeline and our GALFIT modeling.  For \catabiasoffsz~of these two cases (SDSSJ110851.04+065901.45 and SDSSJ134442.17+553113.59), our angular offsets are consistent with the independent measurements described in Section \ref{subsec:notes}, regardless of the different datasets and methods used, providing a powerful indication of the reliability of our procedure and leaving \catanobiasoffsz~previously unreported offset AGN from our \cata~sample.

\noindent {\bf \catb~Spatially Offset AGN:} As shown in Figure \ref{fig:schematic} (right), these are systems in which the X-ray source is spatially consistent with the primary stellar core, but offset from a detected secondary stellar core.  As denoted by the grey circle, the offset is not contained entirely within the fiber, and the spectroscopic signatures can not be used to infer kinematics of the merger.  
As denoted in Figure \ref{fig:schematic}, an AGN in the \catb~sample may also be included in the \cata~sample if the X-ray source is significantly offset from the primary stellar core.  We find that \catbsz~of the offset AGN are included in the \catb~sample.  Of these, \catabovrlpsz~are also included in the \cata~sample.

Table \ref{tab:offset} lists the offset AGN, along with their X-ray properties, offset values, position angles, the SDSS filters in which significant offsets are measured, the number of source pairs used for astrometry in each filter, and whether or not the offset is contained entirely within the fiber.  
\emph{In this first paper of the series, our analysis is focused on the SDSS spectroscopic properties of spatially offset AGN.  For that reason, we only discuss the \cata~sample from here onward, and include the \catb~sample in Paper II}.  The host galaxy environments and spatial offsets of the \cata~offset AGN are shown in Figure \ref{fig:SDSS_offset_catA}.  The \catb~sample is shown in the Appendix. 

\begin{deluxetable*}{ccccccccccc}
\tabletypesize{\footnotesize}
\tablecolumns{11}
\tablecaption{\cata~Spatially Offset AGN: Spectroscopic Properties.}
\tablehead{
\colhead{SDSS Name \vspace*{0.05in}} &
\colhead{\deltaVsigforb} &
\colhead{\deltaVsigbalm} &
\colhead{\NsigHbeta} &
\colhead{\NsigOIII} &
\colhead{\NsigHalphaNII} &
\colhead{\Nsig} &
\colhead{\dvdiff} &
\colhead{\skewHbeta} &
\colhead{\skewOIII} &
\colhead{DP} \\
\colhead{(1)} &
\colhead{(2)} &
\colhead{(3)} &
\colhead{(4)} &
\colhead{(5)} &
\colhead{(6)} &
\colhead{(7)} &
\colhead{(8)} &
\colhead{(9)} &
\colhead{(10)} &
\colhead{(11)}
}
\startdata 
J0813$+$5418 & 0.47 & 0.78 & [0.43] & [-0.45] & [0.90] & [1.27] & [0.22] & [0.41] & [0.16] & [no] \\
J1108$+$0659 & [3.35] & 2.42 & [1.20] & [1.87] & [-0.93] & 6.75 & [0.66] & [-0.31] & -0.59 & yes \\
J1114$+$4036 & [4.07] & 1.18 & [0.23] & [-0.02] & [-1.67] & [-0.09] & 1.98 & -0.55 & [-0.31] & [no] \\
J1115$+$5423 & [4.30] & [3.88] & [-0.18] & [0.26] & [-1.24] & [-0.55] & [0.27] & [-0.21] & [-0.32] & [no] \\
J1234$+$4751 & 2.56 & 2.64 & [0.45] & [-0.50] & [0.76] & [0.14] & [0.07] & 0.58 & [-0.01] & yes \\
J1317$+$4115 & 0.67 & 0.08 & [-0.54] & [-0.90] & [0.53] & [-1.25] & [0.47] & [0.21] & [0.03] & [no] \\
J1344$+$5553 & 2.18 & 1.78 & [0.34] & [-0.07] & [-1.65] & 15.61 & [0.28] & [-0.29] & -0.83 & [no] \\
J1539$+$0324 & 0.92 & 0.55 & [0.50] & [-0.06] & [-0.09] & [1.15] & [0.27] & [-0.03] & [0.33] & [no] \\
J2125$-$0713 & 0.42 & 0.57 & [1.73] & [-0.11] & [0.85] & 8.16 & [0.70] & [-0.21] & -0.62 & [no]
\enddata
\tablecomments{Column 1: abbreviated SDSS galaxy name; Column 2: velocity offset significance of the forbidden emission lines; Column 3: velocity offset significance of the Balmer emission lines; Column 4: H$\beta$ fit quality; Column 5: \oiii~fit quality; Column 6: H$\alpha$/[NII] fit quality; Column 7: continuum fit quality; Column 8: forbidden and Balmer emission line offset difference relative to the error; Column 9: H$\beta$ skewness; Column 10: \oiii~skewness; {\bf Column 11: double-peaked status.}  Values in Columns 2-11 that satisfy the criteria for inclusion in the CG14 spectroscopic candidate offset AGN sample are bracketed.  A source with all ten values bracketed will be in the final sample of CG14.}
 \label{tab:spec_props}
\end{deluxetable*}

\section{Analysis}
\label{sec:analysis}

In this section, we apply the spectroscopic selection parameters of CG14 to our spatially selected sample to determine the overlap of the two techniques (Section \ref{subsec:overlap}).  Then, we use simulations to estimate the physical properties toward which the spatial selection is biased, and these simulated results are compared to our observed results (Section \ref{subsec:simulations}).

\subsection{Overlap of Spatial and Spectroscopic Selections}
\label{subsec:overlap}

Each source in our offset AGN sample comes from the same catalogue of \parentAGNsdsssz~optically-selected AGN used to develop the sample of spectroscopic offset AGN candidates in CG14.  We refer the reader to CG14 for full details of the selection process, but here we summarize the basic spectroscopic requirements used since they are pertinent to our analysis.  The CG14 sample was assembled in a deliberately conservative fashion to include only AGN with 1) robustly modeled emission lines based on the \Nsig~fitting quality assessment parameter (\Nsig~$<3$; see \citealt{Oh:2011}); and 2) significant velocity offsets between the emission lines and stellar absorption lines (\deltaVsig~$>3$).  Furthermore, the selection was designed to exclude spectroscopic features common among gaseous outflows: 3) the forbidden and Balmer emission line velocity offsets should agree to within $1\sigma_{V}$ (\dvdiff~$<1$); and 4) a small skewness ($\gamma$) magnitude in the line profiles of H$\beta$ ($|$\skewHbeta$|<0.5$) and \oiii~($|$\skewOIII$|<0.5$).  Finally, since the goal of the CG14 analysis was to identify candidate \emph{offset} AGN, the selection deliberately omitted sources with double-peaked emission lines (\dpeak$=no$) based on the combined catalogues of \citet{Wang2009}, \citet{Liu2010a}, \citet{Smith:2010} and \citet{Ge:2012} as they represent spectroscopic \emph{dual} AGN candidates.  For reference, the optical fiber spectra and velocity offsets of the \cata~spatially offset AGN are shown in Figure \ref{fig:spectra}.  The host galaxy continuum has been subtracted from each spectrum using \texttt{STARLIGHT} \citep{Fernandes:2005}.

While these criteria are rigorous spectroscopic cuts intended to omit false positives produced by outflows, they are unable to omit cases in which outflows produce large velocity offsets (satisfying criterion 2), do not significantly stratify the NLR (satisfying criterion 3), or produce symmetric emission lines (satisfying criterion 4).  Therefore, the sample is likely to contain a combination of outflows and offset AGN.  For example, two of the spectroscopic candidates from CG14 have been observed with IFU spectroscopy, but with no direct evidence of offset AGN \citep{Allen:2015}.  Furthermore, the strictness of the cuts will omit some true offset AGN which have poor emission line signals (rejected based on criterion 1), small velocity offsets (rejected based on criterion 2), radial gas kinematics which introduce ionization stratifications (rejected based on criterion 3) and emission line asymmetries (rejected based on criterion 4). 

Spatially resolving the location of the AGN relative to the host galaxy is ultimately necessary in all cases to determine if there is actually a secondary active SMBH that is spatially offset from the primary nucleus and producing the kinematic offset in the emission lines.   Our sample of spatially-selected offset AGN show \emph{direct} evidence for offset AGN and therefore avoids the ambiguity introduced by emission line selection.  Table \ref{tab:offset_gen} lists the X- and Y-values of the spatial offsets for each of the \catasz~\cata~offset AGN.  Values in brackets pass the $|\Delta X|\ge3.0$\sigxfinpos~or $|\Delta Y|\ge3.0$\sigyfinpos~criteria.  A source with either value bracketed will be in the spatially offset sample.  By definition, all \cata~sources have at least one value bracketed.    Since the spatial offsets of the \cata~sample are contained entirely within the SDSS fiber (accounting for the relative astrometric uncertainties), they will be detectable as spectroscopic offset AGN if they produce the emission line signatures passing the criteria of CG14.  Therefore, our sample can be used to determine the efficiency of spectroscopic selection.  Specifically, we determine the fractions of spatially offset AGN that are recovered by each of the spectroscopic selection parameters when they are implemented individually and when they are combined.


\begin{figure*} $
\begin{array}{cc}
\hspace*{-0.in} \includegraphics[width=3.3in]{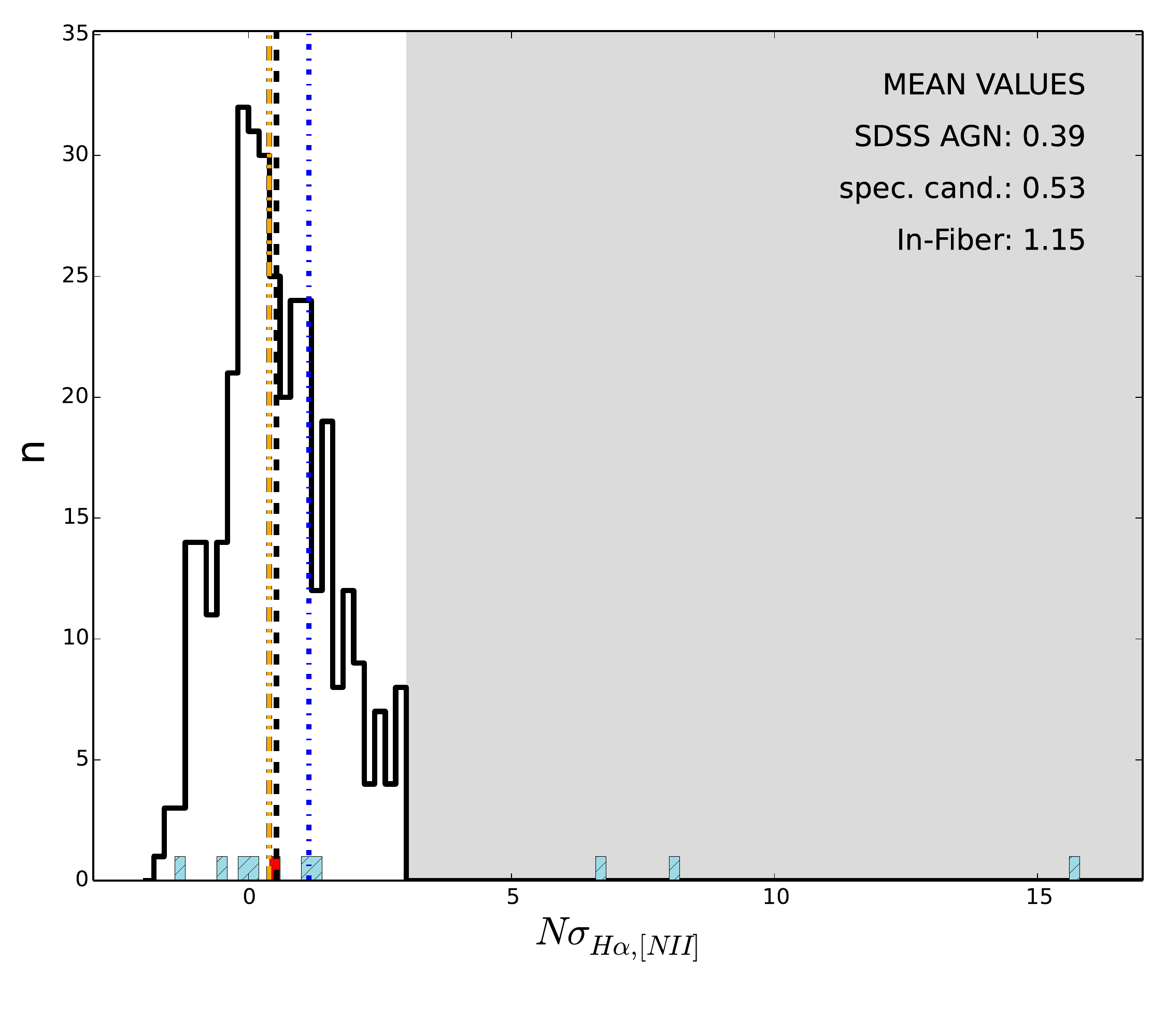} 
\vspace*{-0.22in}
\hspace*{-0.in} \includegraphics[width=3.3in]{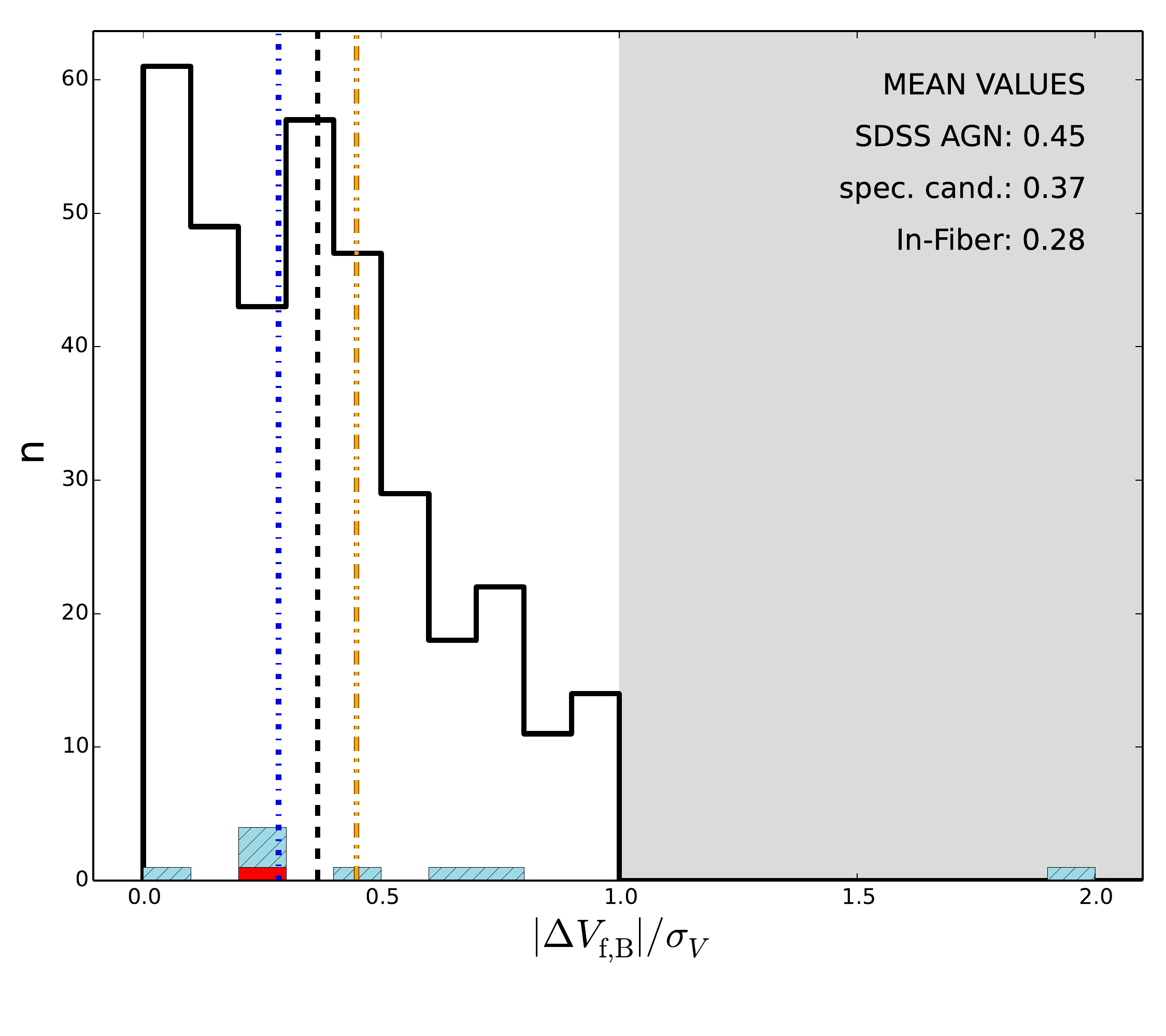} \\
\hspace*{-0.in} \includegraphics[width=3.3in]{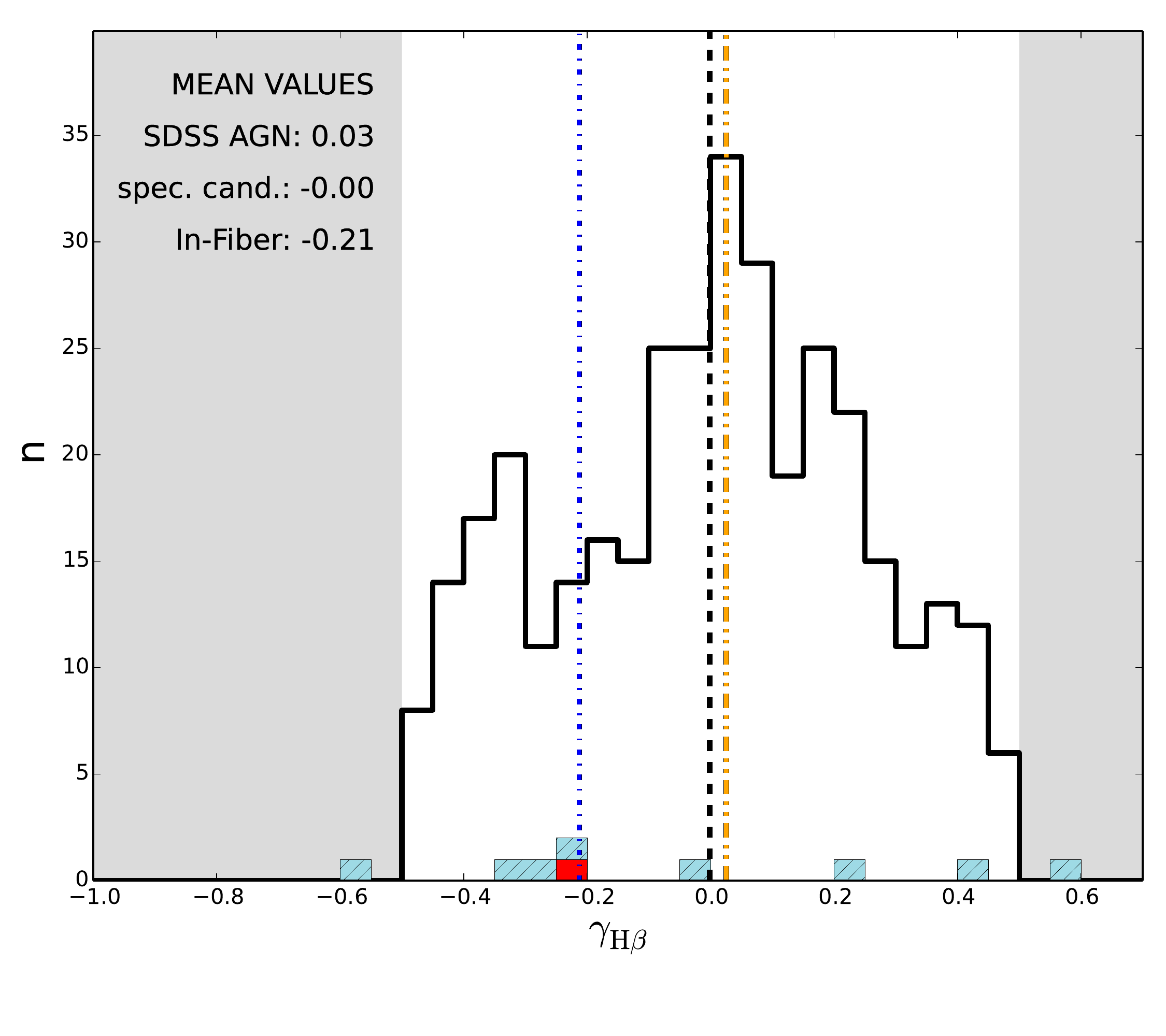} 
\vspace*{-0.22in}
\hspace*{-0.in} \includegraphics[width=3.3in]{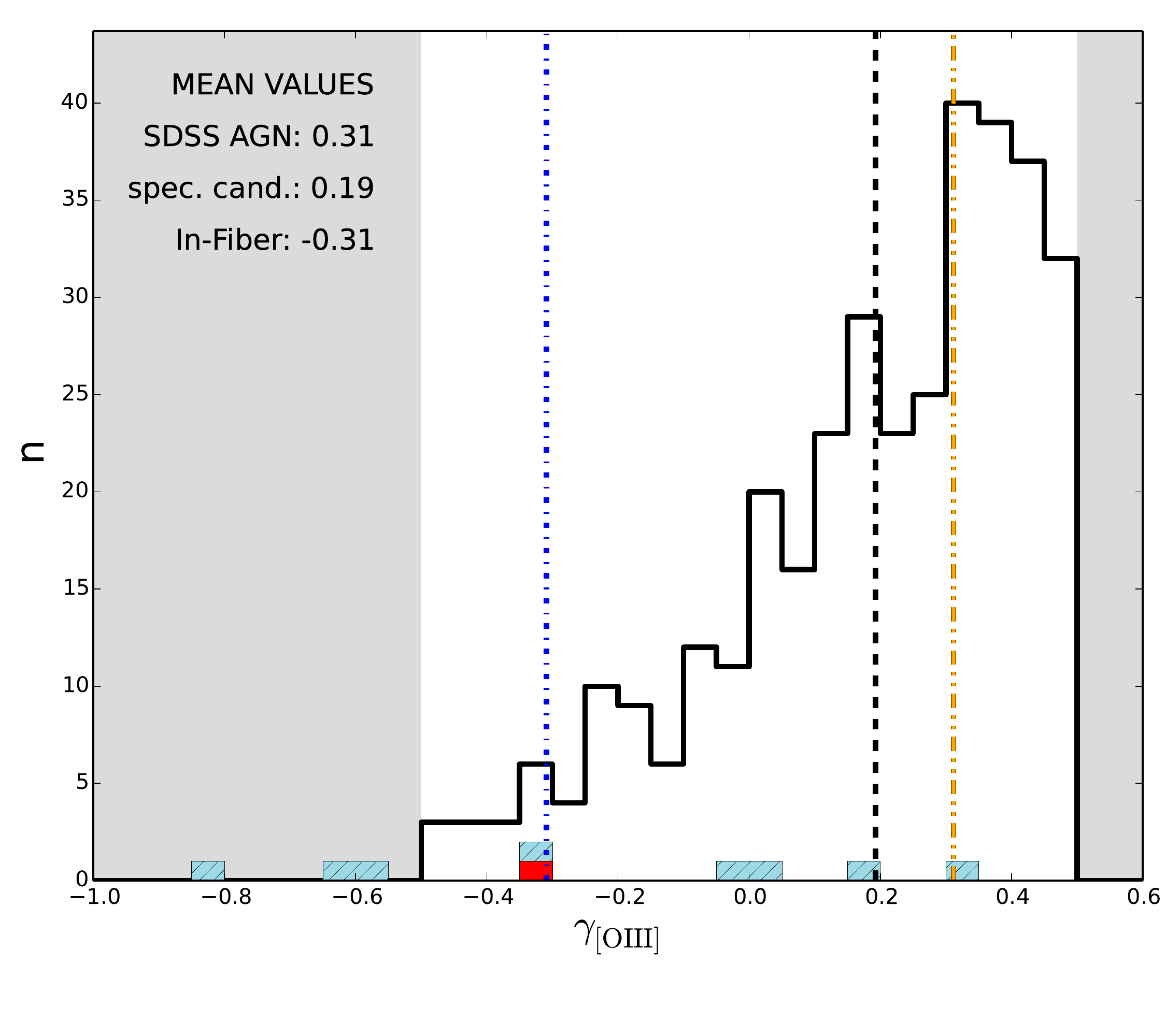} \\
\hspace*{-0.in} \includegraphics[width=3.3in]{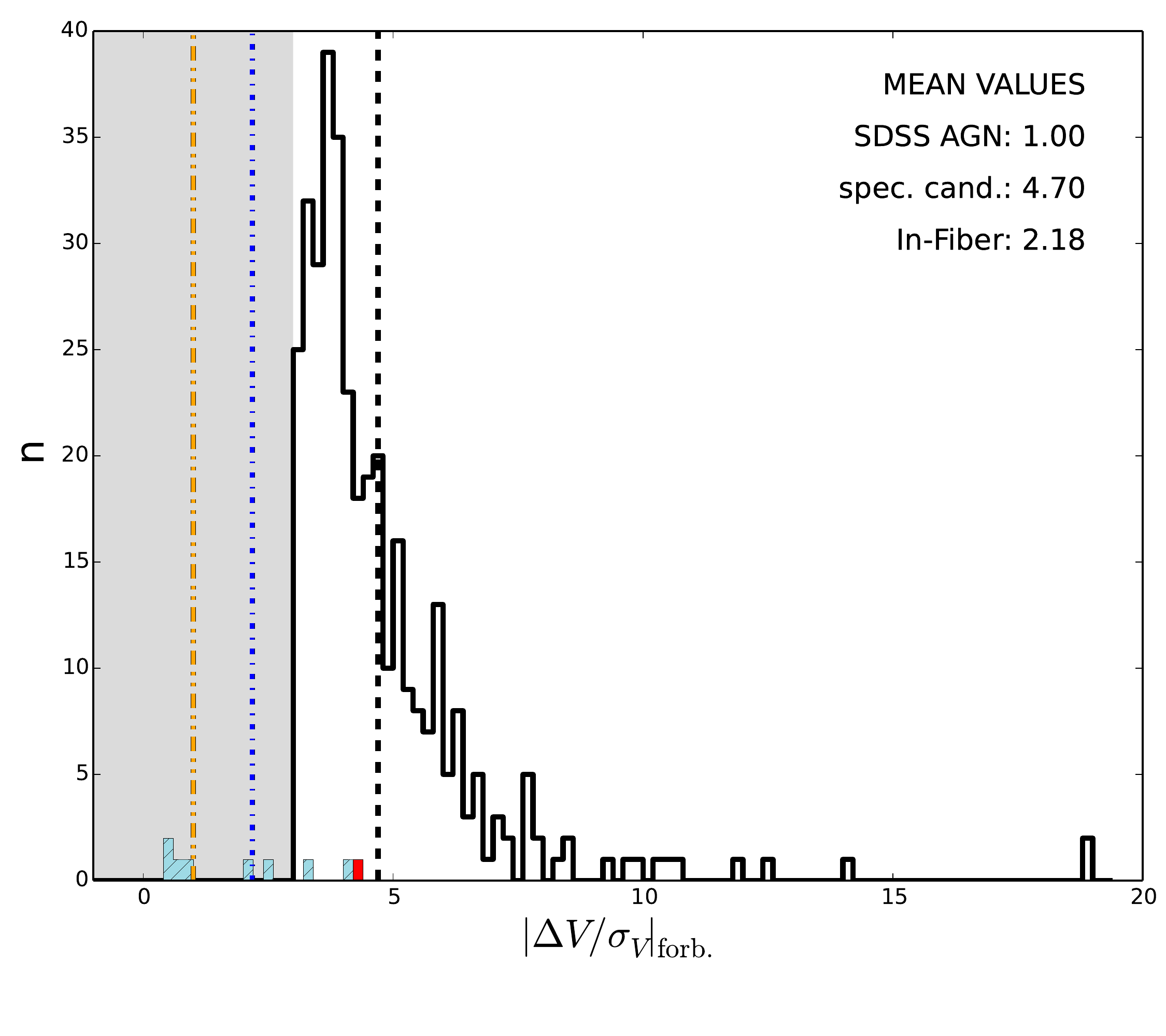}
\hspace*{-0.in} \includegraphics[width=3.3in]{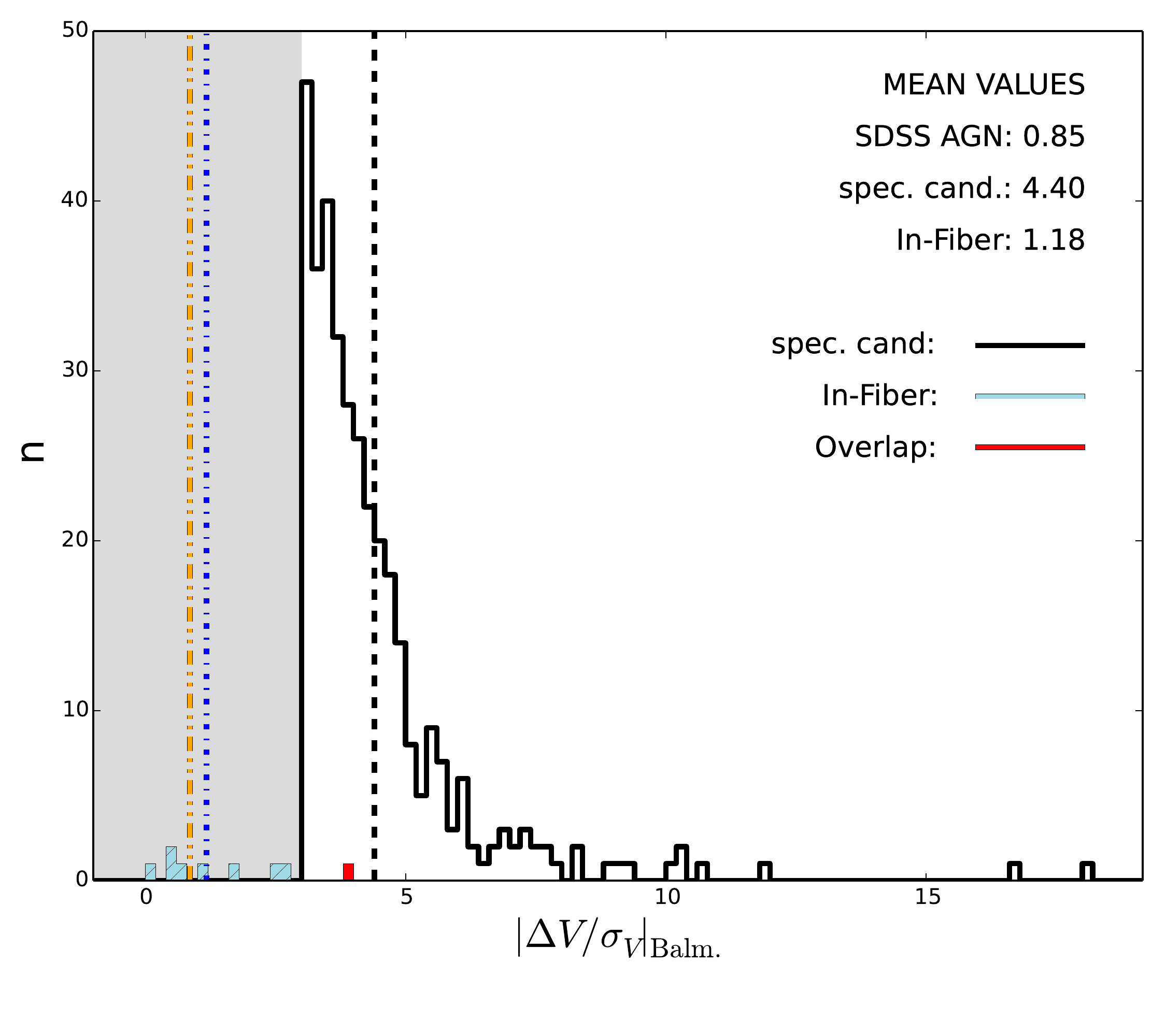}
\end{array} $
\vspace*{-0.22in}
\caption{\footnotesize{Distributions of spectroscopic parameters for the sample of candidate spectroscopic offset AGN (unfilled histogram), the sample of \cata~spatially offset AGN (light blue, filled, hatched histogram), and their overlap (red, filled histogram).  In each panel, the grey-shaded regions indicate values that do not pass the threshold for that particular spectroscopic parameter.  Top left: H$\alpha$/\niia~fit quality; Top right: difference between the forbidden and Balmer emission line velocity offsets relative to their error; Middle left: H$\beta$ skewness; Middle right: \oiii~skewness; Bottom left: velocity offset significance of the forbidden emission lines; Bottom right: velocity offset significance of the Balmer emission lines.  The vertical lines indicate mean values of the initial sample of \parentAGNsdsssz~optically-selected AGN (orange, dot-dot-dashed), the spectroscopic sample of candidates from CG14 (black, dashed) and the \cata~spatially offset AGN (blue, dot-dashed).}
}
\label{fig:spec_ovrlp}
\end{figure*}  

Table \ref{tab:spec_props} lists the spectroscopic parameter values for each of the \catasz~\cata~offset AGN.  Values in brackets pass the threshold for that individual parameter.  All \catasz~of the \cata~offset AGN pass the fitting quality assessment thresholds of the continuum ($|$\NsigCont$|<3$), H$\beta$ ($|$\NsigHbeta$|<3$)~and \oiii~($|$\NsigOIII$|<3$).  However, the fitting quality assessment criterion of H$\alpha$/\nii~($|$\NsigHalphaNII$|<3$) rejects \cataNHalphaNII/\catasz~offset AGN.  The criterion of \dvdiff$<1$ rejects \catadvdiff/\catasz~offset AGN.  The criteria of $|$\skewHbeta$|$$<0.5$ and $|$\skewOIII$|<0.5$ reject \cataskewHbeta/\catasz~and \cataskewOIII/\catasz~offset AGN, respectively.  The single-peak emission line criterion (\dpeak$=no$) rejects \catadp/\catasz~offset AGN.  Finally, the criteria of forbidden emission line velocity offset significance (\deltaVsigforb$>3$) and the Balmer emission line velocity offset significance (\deltaVsigbalm$>3$) reject \catadeltaVforb/\catasz~and \catadeltaVbalm/\catasz~offset AGN, respectively.  Only one offset AGN passes the thresholds for all ten spectroscopic parameters.  Figure \ref{fig:spec_ovrlp} shows the distribution of values for each numerical spectroscopic parameter that rejects at least one \cata~spatially offset AGN.  From Table \ref{tab:spec_props}, we see that the parameters of \NsigCont, \NsigHbeta, \NsigOIII, \NsigHalphaNII, \dvdiff, \skewHbeta, \skewOIII~and \dpeak~individually reject no more than \excludevala$\%$ of \cata~offset AGN, while the velocity offset significance criteria of \deltaVsigforb~and \deltaVsigbalm~are the most exclusive individual criteria.  Table \ref{tab:spec_props} shows that \catadeltaVforb/\catasz$=$\excludevalb$\%$ and \catadeltaVbalm/\catasz$=$\excludevalc$\%$ offset AGN are rejected based on the thresholds of \deltaVsigforb~and \deltaVsigbalm, respectively.

For any AGN spectrum, the parameters of \NsigCont, \NsigHbeta, \NsigOIII, \NsigHalphaNII, \deltaVsigforb~and \deltaVsigbalm~are subject to technical limits such as spectral resolution and/or signal-to-noise ratio that may also exclude true offset AGN, while measurements of the other four parameters (\dvdiff, \skewHbeta, \skewOIII~and \dpeak) will not exclude true offset AGN due to those limits.  We find that 5/\catasz~offset AGN are rejected based on their combined thresholds.  This suggests that emission line signatures in approximately half of the \cata~offset AGN systems are effected by gas kinematics that introduce ionization stratifications or asymmetries.

\subsection{Spatial and Spectroscopic Biases}
\label{subsec:simulations} 

Our selection of offset AGN is inherently conservative in that it includes sources with spatial offsets at a confidence level of $\geq99.73\%$.  Ultimately, this method is designed to construct a \emph{clean} sample of offset AGN free from potentially mis-identified mergers as discussed in Section \ref{sec:intro}.  Thus, our selection is likely to be sensitive to large projected angular offsets between nuclei, \septheta.  Within our physical interpretation of these offset AGN systems (Section \ref{subsec:physical}), the spatial offset selection may be biased toward systems in which the orbit is viewed closer to face-on, resulting in generally larger values of \septheta~but smaller values of projected velocity offsets, \deltaVproj, at all orbital phases.  This may be reflected in the emission line velocity offsets of the \cata~spatially offset AGN (\deltaVforb~and \deltaVbalm), assuming the offset AGN travels with a distinct NLR that is not experiencing complex kinematic disturbances, such as outflows.  However, within the context of the unified model of AGN \citep{Antonucci:1993}, our selection of Type II AGN will introduce a bias toward obscuring features along the line-of-sight.  If obscuration is caused by a dusty torus, this may imply a bias toward edge-on accretion disks as well.  Since the reservoir of material available to form the accretion disk comes from the larger interstellar medium of the host galaxy, a correlation between the host galaxy and accretion disk net angular momentum axes is expected \citep{Hopkins:2012}.  In this scenario, selection of Type II AGN may result in a bias toward edge-on galaxies, and therefore a bias toward edge-on dual SMBH systems, assuming they are orbiting in the galaxy plane following a merger.  Indeed, there is some evidence that Type II AGN  host galaxies show a statistically significant deviation from random orientations, and may deviate toward edge-on systems \citep{Lagos:2011}.

\begin{figure*}[t!] $
\begin{array}{cc}
\hspace*{-0.in} \includegraphics[width=3.5in]{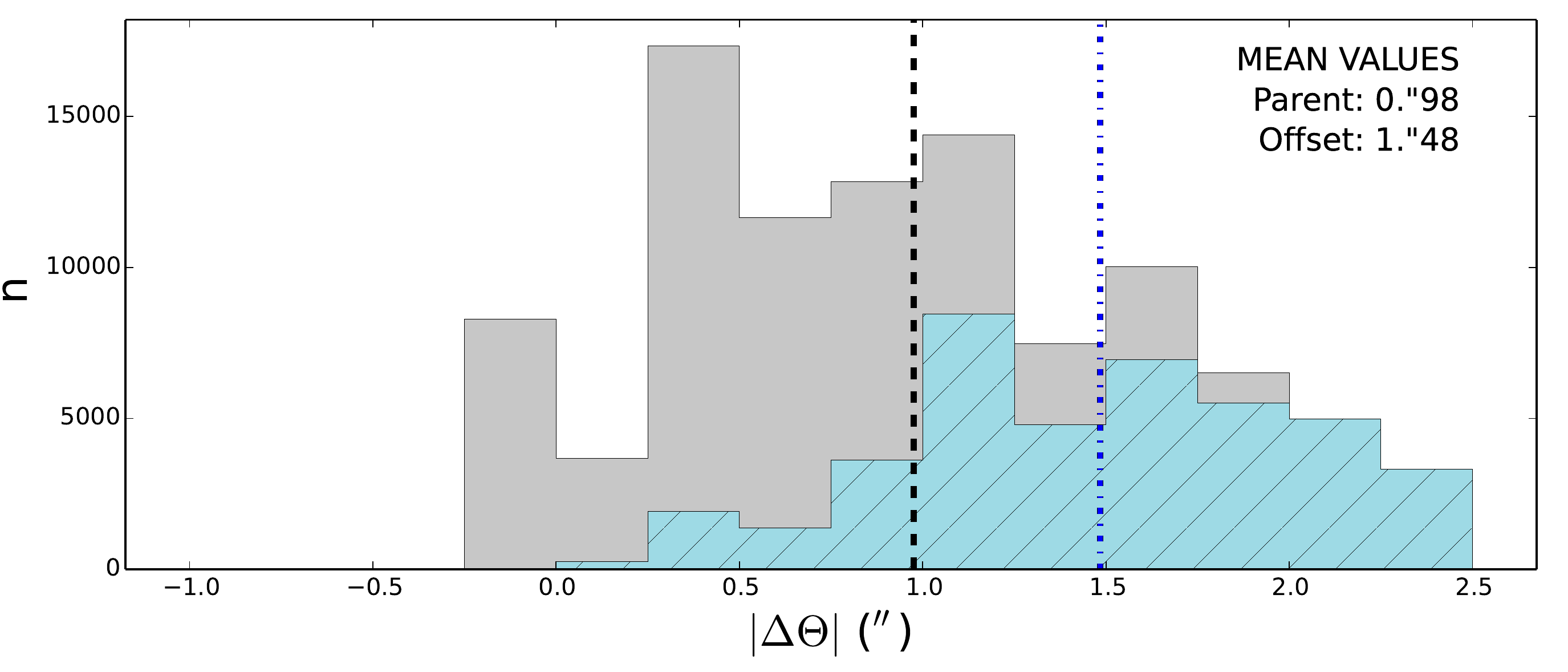}
\hspace*{-0.in} \includegraphics[width=3.5in]{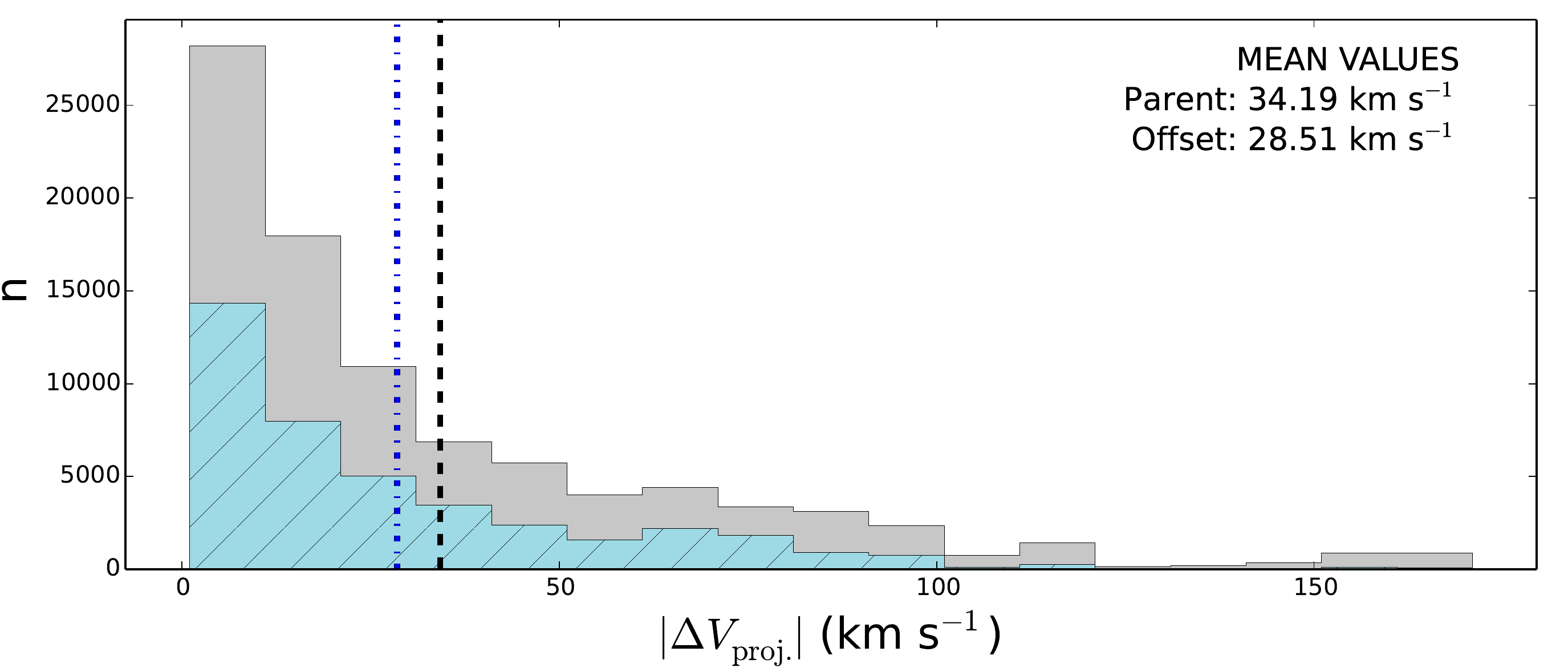} \\
\hspace*{-0.in} \includegraphics[width=3.5in]{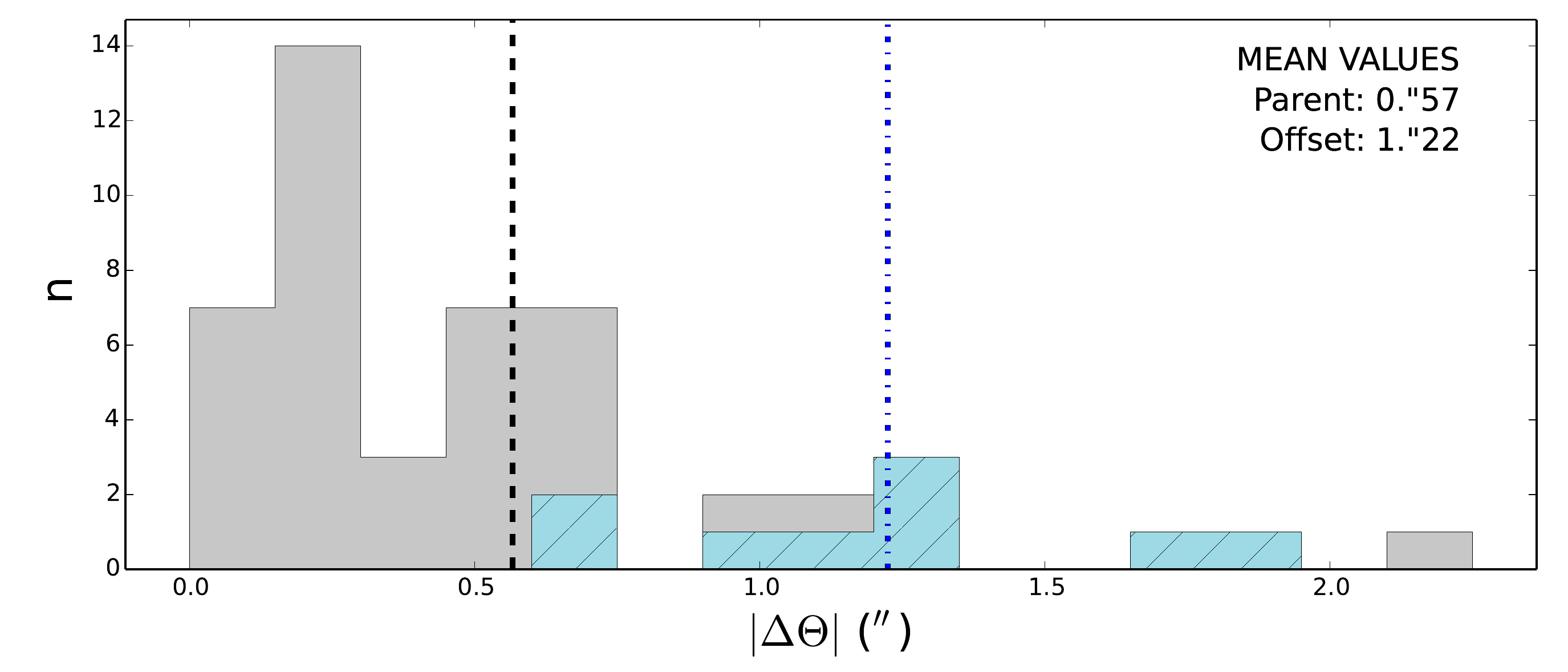}
\hspace*{-0.in} \includegraphics[width=3.5in]{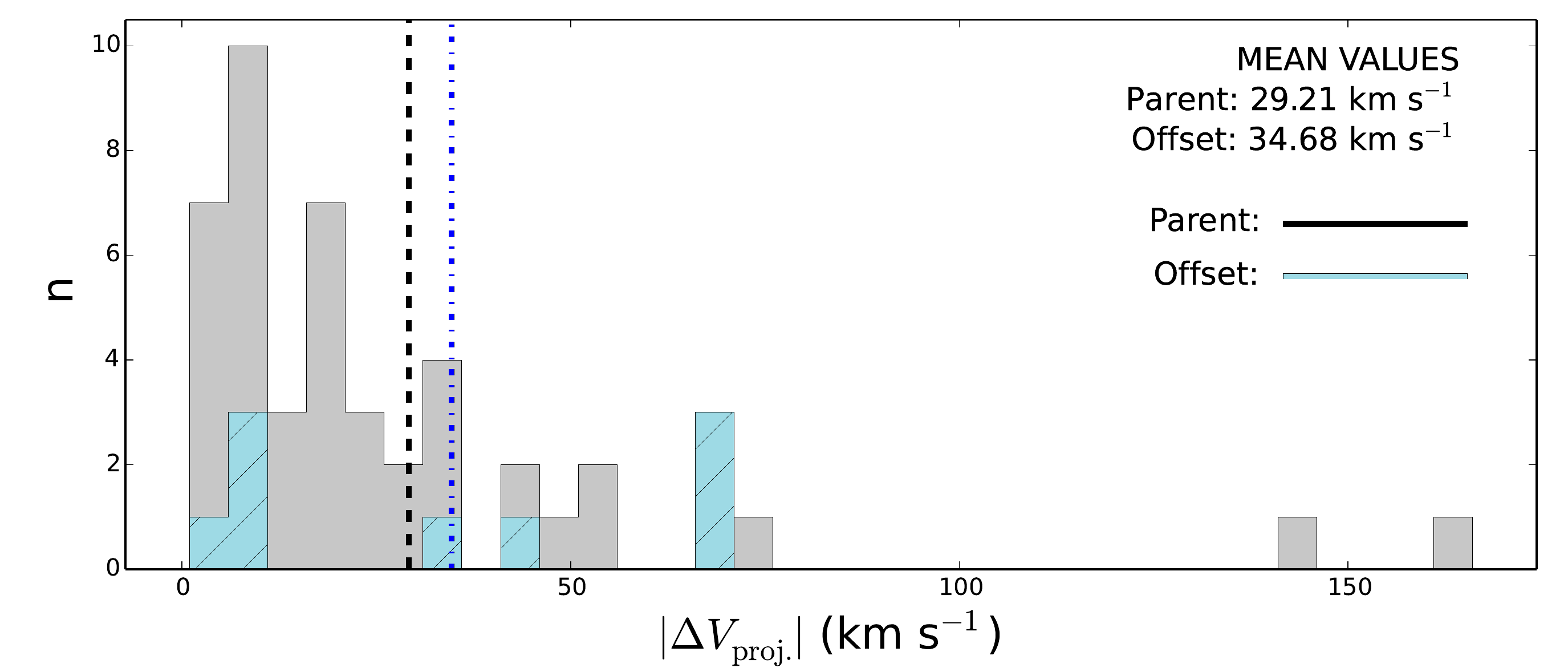}
\end{array} $
\caption{\footnotesize{Left: distribution of angular separation absolute values for the simulated sample (top) and the observed sample (bottom).  Right: distribution of projected velocity offset absolute values for the simulated sample (top) and the observed sample (bottom).  In all panels, the parent sample is shown as the grey, filled histogram, and the offset sample is shown as the light blue, filled, hatched histogram.  The mean values of the parent and offset samples are displayed in the upper right and represented by the black, solid, vertical line and the blue, dot-dashed, vertical line, respectively.
}
}
\label{fig:sim_hist_sep_vel}
\end{figure*}

To gain insight into the sensitivity of our selections to \septheta~and \deltaVproj, we have run simulations of offset nuclei to determine their expected values.  The simulated variables consist of the 3-dimensional physical separation (\sepphys), orbital inclination ($i$), orbital phase ($\Phi$), uncertainty in the separation of the nuclei (\sigfinpos), and host galaxy mass (\mass).  Without any further knowledge of the lifetime of galaxy merger stages, we assume a uniform distribution of \sepphys~from -20 to 20 kpc (i.e. we considered the 40x40 kpc field of each galaxy in the optical image modeling; see Section \ref{subsec:opt_img_mod}).  Values of $\Phi$ are drawn from a uniform distribution, while values of $i$ are drawn from the sample of Type II AGN host galaxy orientations from \citet{Lagos:2011} to account for a potential bias toward edge-on systems.  $i=0^{\circ}$ corresponds to orbital axes parallel to the line-of-sight (face-on orbit), and $\Phi=0^{\circ}$ corresponds to the orbit quadrature.  The distribution of \sigfinpos~is taken directly from our \cata~offset AGN sample.  In the simulation, these parameters are entirely independent of each other and randomly selected.  We iteratively increased the number of simulated offset nuclei until the standard deviation of all values changed by less than $10\%$ (\simitersz~simulated offset nuclei).  We then selected offset nuclei with significant offsets using the same criteria from Section \ref{subsec:offset_agn}.  The results of these simulations are shown in Figure \ref{fig:sim_hist_sep_vel}.

The top left panel displays the distribution of \septheta~for the simulated parent sample (mean value of \sepmeansimpar) compared to the simulated offset nuclei sample (mean value of \sepmeansimoff).  A clear bias toward large \septheta~exists in the offset nuclei sample. 
For comparison, the bottom left panel displays the distribution of \septheta~for the observed parent and \cata~offset AGN from our sample.  In agreement with the simulation results, we observe a bias toward large values of \septheta~reflected in the mean values ($0\farcs57$ for the parent sample compared to $1\farcs22$ for the offset AGN sample).  From a Kolmogorov$-$Smirnov (KS) test, we find a probability of $p=4\times 10^{-4}$ for the null hypothesis that the observed parent and offset AGN values of \septheta~are draw from the same distribution.  

The top right panel displays the distribution of \deltaVproj~for the simulated parent sample (mean value of \velmeansimpar) compared to the simulated offset nuclei sample (mean value of \velmeansimoff).  A bias toward small \deltaVproj~exists in the selected sample. 
For comparison, the bottom right panel displays the distribution of \deltaVproj~for the observed parent and \cata~offset AGN from our sample.  In contrast to the simulation results, we do not observe a bias toward small values of \deltaVproj.  From a KS test, we find a probability of $p=0.41$ for the null hypothesis that the observed parent and offset AGN values of \deltaVproj~are draw from the same distribution.  
The contrast between the simulations and our observed values is also reflected in the mean values, in which we actually see a larger mean value of \deltaVproj~for the \cata~offset AGN sample ($34.68$ km s$^{-1}$) compared to the parent sample  ($29.21$ km s$^{-1}$).

\section{Discussion}
\label{sec:discussion}

In Section \ref{sec:analysis}, we examined the differences inherent between the spatial and spectroscopic selection techniques.  In this section, we discuss the impact of these differences on the number of true offset AGN recovered (Section \ref{subsec:fraction}) and on the physical nature of emission line kinematics in offset AGN systems (Section \ref{subsec:velocity_offsets}).

\subsection{Comparison of Spatial and Spectroscopic Offset AGN Selection Efficiencies}
\label{subsec:fraction}

The analysis in Section \ref{subsec:overlap} shows that the overlap of the spatial and spectroscopic selection techniques is \ovrlp/\catasz.  To determine if this frequency is expected, we first consider that the candidate offset AGN selection rate from CG14 is $351/20,098=0.017$ ($1.7\pm0.09\%$).  Therefore, this same rate would be expected out of a random subsample of the \parentAGNsdsssz~optically-selected AGN.  After removing the \catabiassz~$Chandra$ targeted \cata~offset AGN, we are left with a rate of \ovrlp/\parentnobiassz$=$\ovrlpparenetsz, consistent with the selection rate from CG14 and suggesting that our sample contains the expected number of spectroscopic offset AGN candidates.

However, this means that the spectroscopic selection missed $8/9$ (\excludevalc$\%$) true offset AGN because they do not meet all ten of the spectroscopic criteria.  Since spatially offset AGN are selected independently of these spectroscopic parameters, we expect them to have values consistent with a random sampling of the parent sample (unless spatial selection does introduce a bias toward certain spectroscopic properties, which we discuss in Section \ref{subsec:velocity_offsets}).  From Figure \ref{fig:spec_ovrlp} we see that the thresholds of the \NsigHalphaNII, \dvdiff, \skewHbeta~and \skewOIII~parameters select values typical of the parent AGN distribution and the spatially offset AGN distribution as indicated by their mean values.  On the other hand, the thresholds of the \deltaVsigforb~and \deltaVsigbalm~parameters do not select values typical of the parent AGN distribution or the spatially offset AGN distribution, and this is reflected in the small fraction of recovered spatially offset AGN.  Specifically, the \deltaVsigforb~and \deltaVsigbalm~criteria individually exclude \excludevalb$\%$ and \excludevalc$\%$ of spatially offset AGN, respectively, versus the maximum \excludevala$\%$ excluded by any of the other eight individual parameters.  

Indeed, we expect large line-of-sight velocity offsets are rare and produced only if there is a powerful outflow or an offset AGN with a favorable orientation.  For example, CG14 ran simulations of offset AGN orbital motion to estimate the true number of offset AGN in the parent sample, finding that $\sim50\%$ of offset AGN are missed due to the \deltaVsig~limit.  This fraction is smaller but similar to the lower bound of our estimate (\excludevalb~$-$ \excludevalc$\%$), suggesting that projection effects play in role in missing true offset AGN if a significant velocity offset is required.  

To examine the spectroscopic recovery of offset AGN free of statistical effects, in Section \ref{subsec:overlap} we combined only parameters that do not exclude AGN based on measurement uncertainties, finding that about half of true offset AGN are still rejected.  Since these rejections are not based on measurement limitations, they reflect real ionization stratifications and emission line blueshifts or redshifts typically associated with AGN outflows.  

\subsection{Origin of Emission Line Kinematics in Spatially Offset AGN}
\label{subsec:velocity_offsets} 


The analysis in Section \ref{subsec:simulations} shows that a simple orbital model of double nuclei results in a bias toward large angular offsets with the selection of spatially offset nuclei, and this is consistent with our observations.  This also has the effect of introducing a bias toward face-on orbits and thus small projected velocity offsets (Figure \ref{fig:sim_hist_sep_vel}).  However, the observed values of \deltaVproj~show no evidence for the same bias and are not consistent with this prediction.  We now consider several conditions under which this bias would not be present: \\ \\ 
\noindent 1) The offset X-ray AGN is in a \emph{face-on} orbit about the primary optical nucleus of the host galaxy and traveling with a physically distinct NLR that produces the kinematically offset emission line.  In this case, the relative space-velocity of the SMBHs is so large that even a small radial projection corresponds to a large emission line velocity offset.  
We note that, while this scenario can explain large velocity offsets for individual AGN, it does not explain the statistical enhancement in the observed velocity offsets compared to predictions from our simulations. \\ \\
\noindent 2) The offset X-ray AGN is in an \emph{edge-on} orbit about the primary optical nucleus of the host galaxy and traveling with a physically distinct NLR that produces the kinematically offset emission line.  In this case, the spatial offset is greater than the threshold because we are viewing the orbit near the quadrature phase when the angular separation would be largest.  However, as with scenario 1, this scenario is not a general solution.\\ \\
\noindent 3)  An X-ray AGN is physically offset from the optical galaxy core, while the offsets of the emission lines are due to gas kinematics, including outflows.  In this case, the bulk gas kinematics would not necessarily be confined to the orbital plane and could therefore produce large projected velocities, regardless of the orbital inclination.  This scenario may also be expected because these systems are galaxy mergers, and such events are known to induce kinematic disturbances in NLRs or trigger enhanced AGN accretion that can drive NLR outflows. \\

From this discussion, we have seen that the spatially offset AGN sample do not show a bias toward small projected emission line velocity offsets as predicted by our simulations.  We hypothesize that this may be explained by a combination of large space velocity offsets and favorable orientation/phase combinations, though these scenarios do not provide a statistical solution.  A more general explanation is one in which face-on orbits are preferred for selecting offset AGN but the NLR gas kinematics do not necessarily trace the dual SMBH orbit, and instead have a significant radial component due possibly to outflows.  This explanation is further supported by the presence of spectroscopic signatures suggestive of outflows in half of the sample.


\section{Conclusions}
\label{sec:conclusions}

We have constructed a sample of \catabAGNsz~AGN that are spatially offset from a nearby galaxy stellar core: \emph{spatially offset AGN}.  While the initial sample consisted of Type II AGN chosen from the SDSS galaxy catalogue based on emission line ratios, the sources in our final sample are uniformly classified as AGN based on X-ray properties, with the X-ray detections being spatially within the SDSS fiber.  They represent the first systematically developed sample of X-ray AGN with detections that are nearby to, but spatially isolated from, the nuclear region of a galaxy (hosting a SMBH) and hence provide a clean sample of AGN fueling in galaxy mergers.  The conclusions regarding our selection process are as follows:  

\begin{itemize}

\item[$\circ$] \hspace*{-0.05in} 
The combination of our registration procedure and the spatial resolution of $Chandra$ is such that we measure significant projected separations down to $0\farcs6$ or 0.8 kpc. We also imposed a projected physical separation upper bound of 20 kpc to ensure that the systems are interacting, limiting the maximum separations to $17\farcs4$ or 19.4 kpc.  With this range of separations, our spatially offset AGN sample can be used to study the properties of AGN in mergers from early-stage interactions (\sepproj$\sim20$ kpc) to later merger stages than previous systematic studies (\sepproj$<1$ kpc).  

\item[$\circ$] \hspace*{-0.05in} Due to the fiber radius of 1\farcs5, in \catasz~cases the spatial offset is contained entirely within the SDSS fiber (\cata~offset AGN) such that the emission line kinematics induced by the merger will be reflected in the spectroscopic signatures.  On the other hand, in \catbsz~cases the spatially offset stellar core is outside of the fiber (\catb~offset AGN).  In \catabovrlpsz~cases, the X-ray AGN detection is offset from a stellar core within the fiber and from a stellar core outside of the fiber.

\end{itemize}

Sources in our sample represent \emph{direct} detections of offset AGN.  Since they are selected independent of kinematic properties, we have used emission lines from the SDSS fiber spectra to study the efficiency of spectroscopic selection and the dynamics of AGN photo-ionized gas in galaxy mergers.  For this analysis, we only used the \cata~sample as the offsets are contained entirely within the fiber.  The analysis of kinematic properties produced the following results:

\begin{itemize}

\item[$\circ$] \hspace*{-0.05in} We find that our sample of spatially offset AGN contains the expected number of spectroscopic offset AGN candidates when applying the strict spectroscopic criteria of CG14.  However, this also suggests that spectroscopic selection may miss up to \excludevalc$\%$~of true offset AGN.  

\item[$\circ$] \hspace*{-0.05in} From an examination of the spectroscopic parameters individually, the forbidden and Balmer emission line velocity offset significance criteria (\deltaVsigforb$>3$ and \deltaVsigbalm$>3$, respectively) are responsible for rejecting most true offset AGN (\catadeltaVforb/\catasz~and \catadeltaVbalm/\catasz, respectively).  
This result is consistent with the prediction that most true offset AGN have small line-of-sight emission line velocity shifts relative to the host galaxy systemic velocity due to projection effects.  

\item[$\circ$] \hspace*{-0.05in} The magnitudes of the observed velocity offsets are larger than expected from simulations that predict a bias toward face-on orbits in spatially-selected offset AGN.  This may point toward a weak correlation between SMBH orbital motion and narrow emission line offsets, suggesting complex kinematics in the extended NLR gas.  It is also suggested by the fact that half of the \cata~offset AGN display emission line ionization stratifications and asymmetries, suggestive of outflows. \\ 

\end{itemize}

In Paper II of this series, we will use our sample of spatially offset AGN to investigate the parameters that effect the level of AGN fueling in galaxy mergers.  In Paper III of this series, we will present new and archival $HST$ imaging for a subset of our offset AGN sample. \\

We would like to thank an anonymous referee for detailed and insightful comments that greatly improved the quality of the paper.  The analysis in this paper was also helped by constructive discussions with Rebecca Nevin and Francisco M{\"u}ller-S{\'a}nchez.  Support for this work was provided by NASA through Chandra Award Number AR5-16010A issued by the \emph{Chandra X-ray Observatory Center}, which is operated by the Smithsonian Astrophysical Observatory for and on behalf of NASA under contract NAS8-03060.  The scientific results reported in this article are based in part on observations made by the \emph{Chandra X-ray Observatory}, and this research has made use of software provided by the Chandra X-ray Center in the application packages CIAO and Sherpa.  We also acknowledge use of the the Sloan Digital Sky Survey and NASA's Astrophysics Data System. 

\emph{Facilities}: \emph{CXO} (ACIS)


\section*{Appendix}
\label{sec:appendix}

The host galaxy environments and spatial offsets of the \catb~spatially offset AGN are shown in Figure \ref{fig:SDSS_offset_catB}.  These spatially offset AGN will be combined with the \cata~spatially offset AGN in Paper II.  Notes on individual \catb~spatially offset AGN are in Section \ref{subsec:notes}.


\begin{figure*}[b!]
\subfloat {\hspace*{0.10in} \includegraphics[width=6.8in]{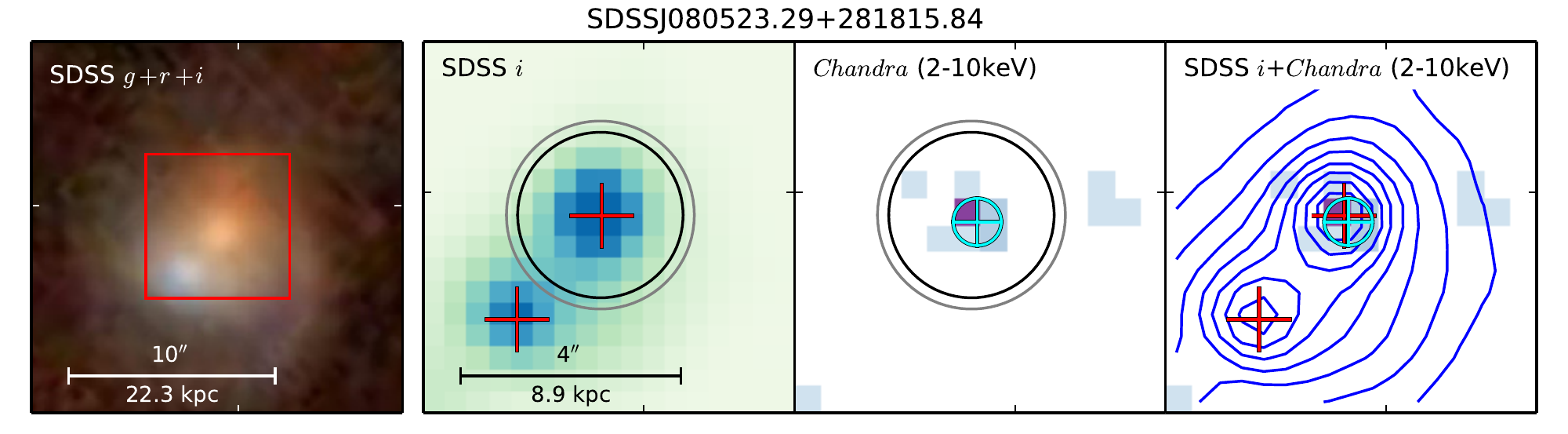}}
\vspace*{-0.20in}
\subfloat {\hspace*{0.10in} \includegraphics[width=6.8in]{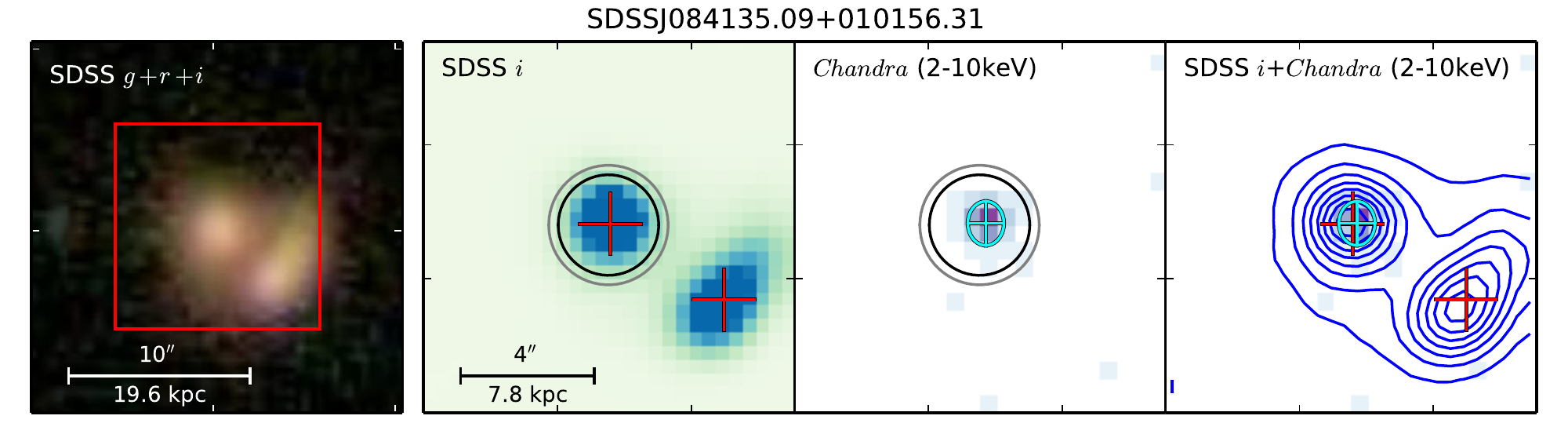}}
\vspace*{-0.20in}
\subfloat {\hspace*{0.10in} \includegraphics[width=6.8in]{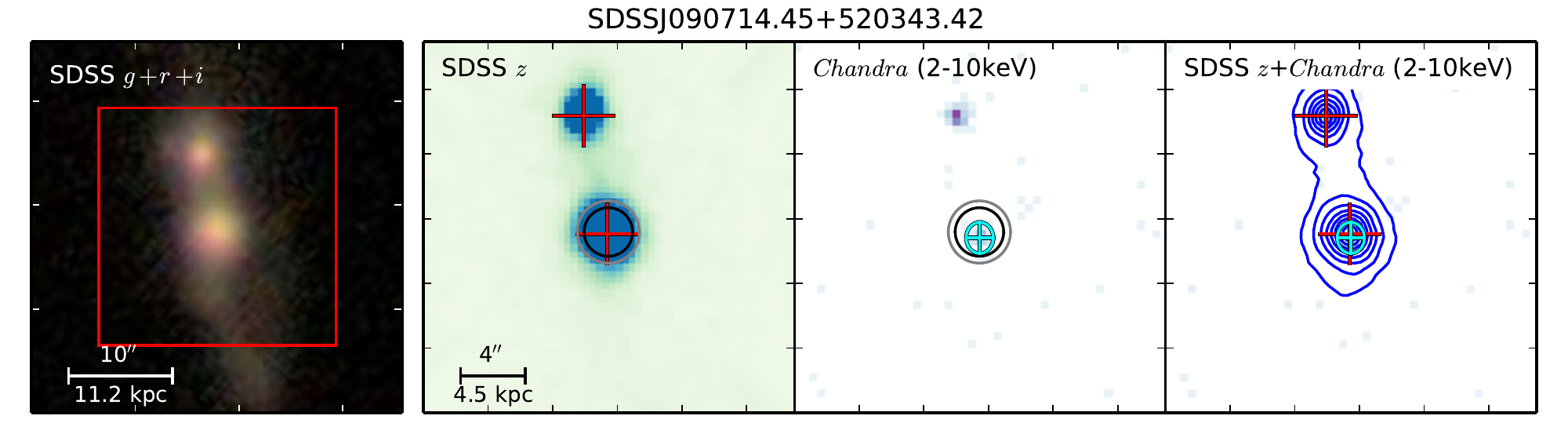}}
\vspace*{-0.20in}
\subfloat {\hspace*{0.10in} \includegraphics[width=6.8in]{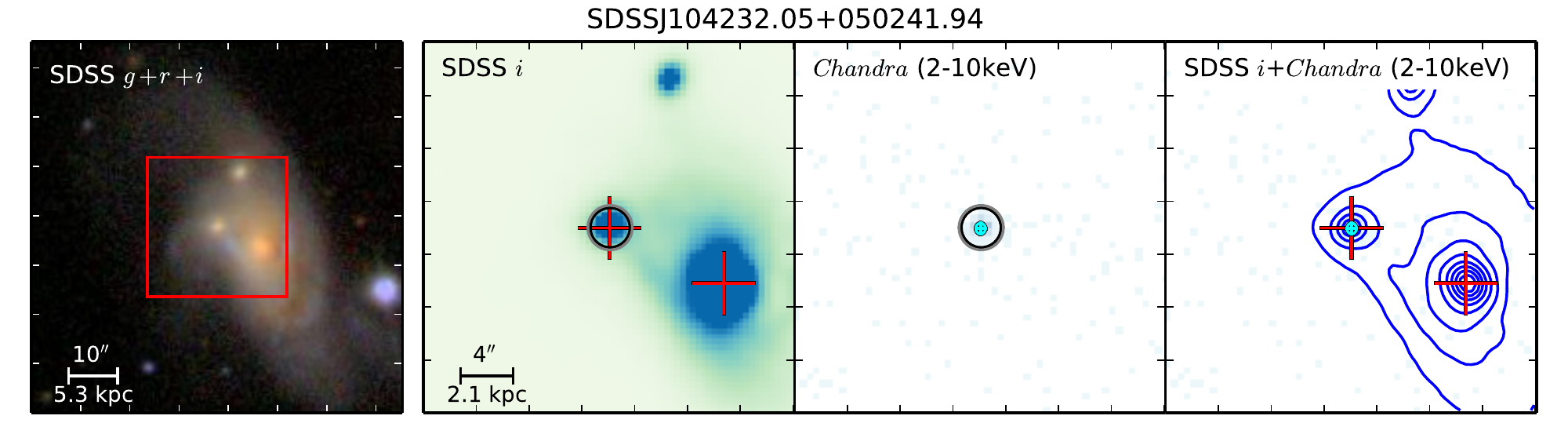}}
\caption{Same as Figure \ref{fig:SDSS_offset_catA} but for the sample of \catb~offset AGN.}
\label{fig:SDSS_offset_catB}
\end{figure*}
\begin{figure*}
\ContinuedFloat
\subfloat {\hspace*{0.10in} \includegraphics[width=6.8in]{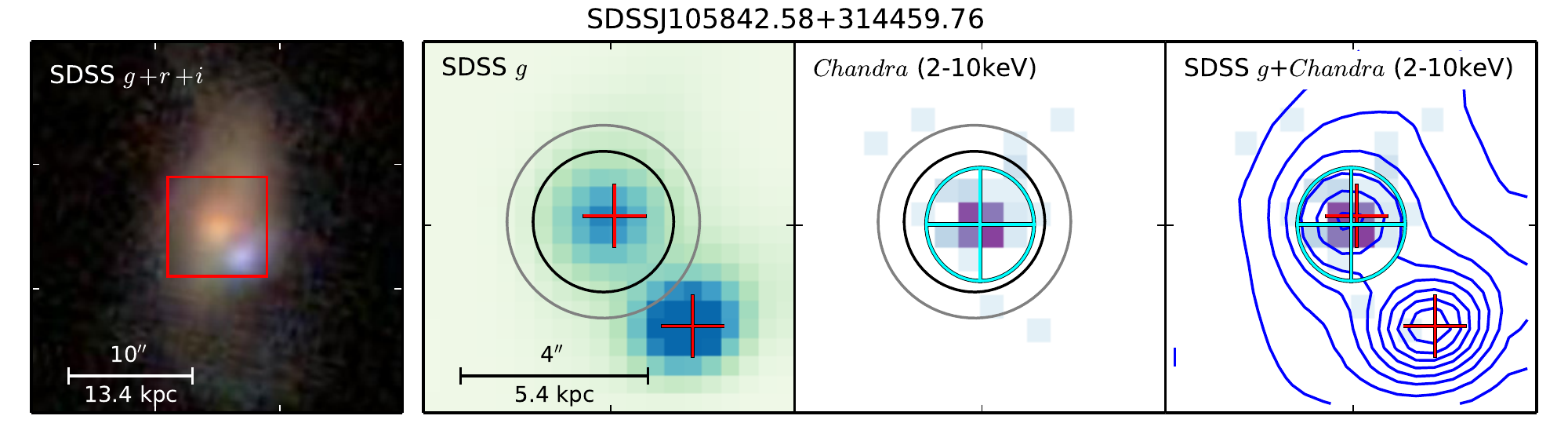}}
\vspace*{-0.20in}
\subfloat {\hspace*{0.10in} \includegraphics[width=6.8in]{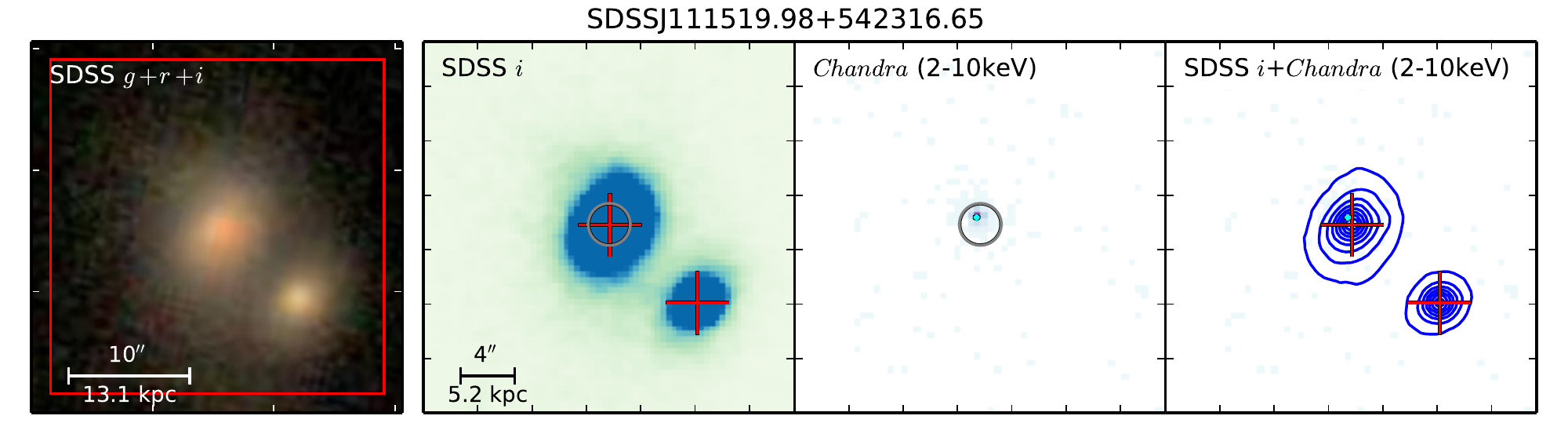}}
\vspace*{-0.20in}
\subfloat {\hspace*{0.10in} \includegraphics[width=6.8in]{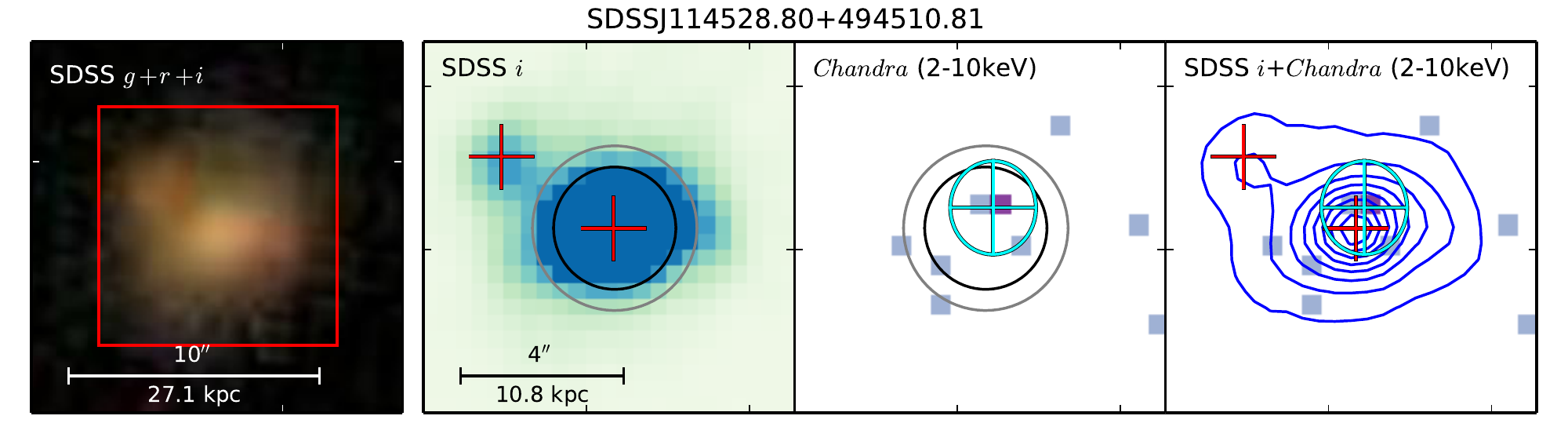}}
\vspace*{-0.20in}
\subfloat {\hspace*{0.10in} \includegraphics[width=6.8in]{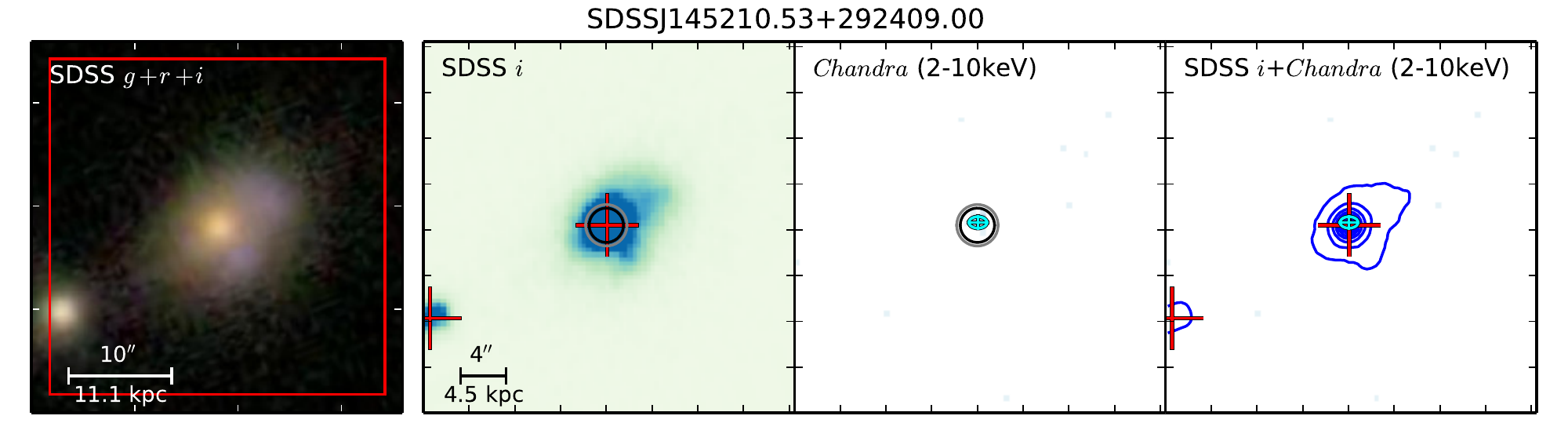}}
\vspace*{-0.20in}
\subfloat {\hspace*{0.10in} \includegraphics[width=6.8in]{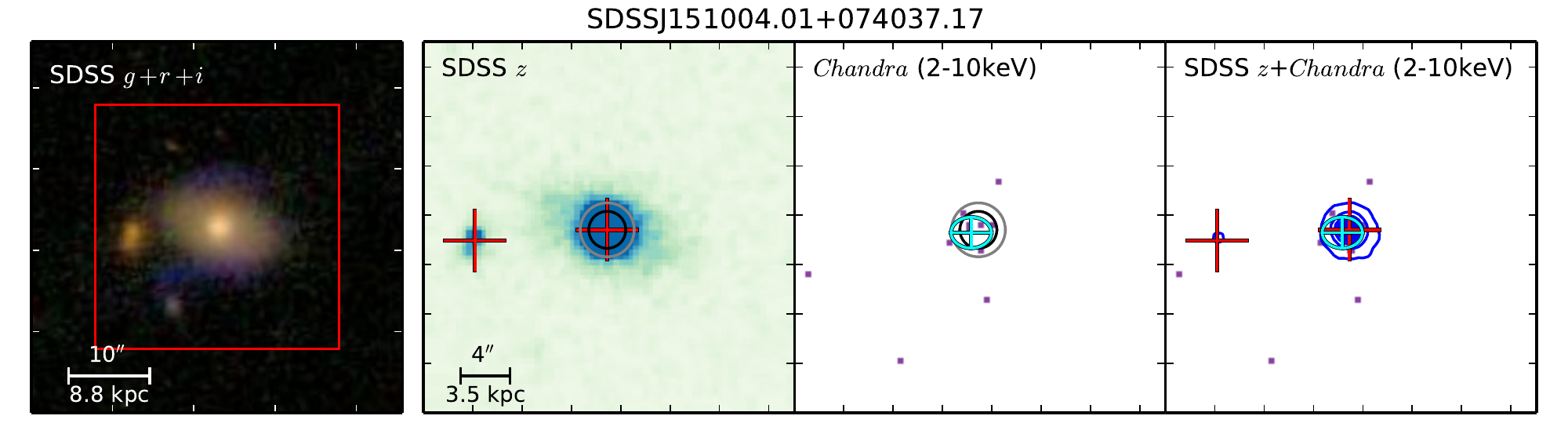}}
\caption{continued}
\end{figure*}
\begin{figure*}
\ContinuedFloat
\subfloat {\hspace*{0.10in} \includegraphics[width=6.8in]{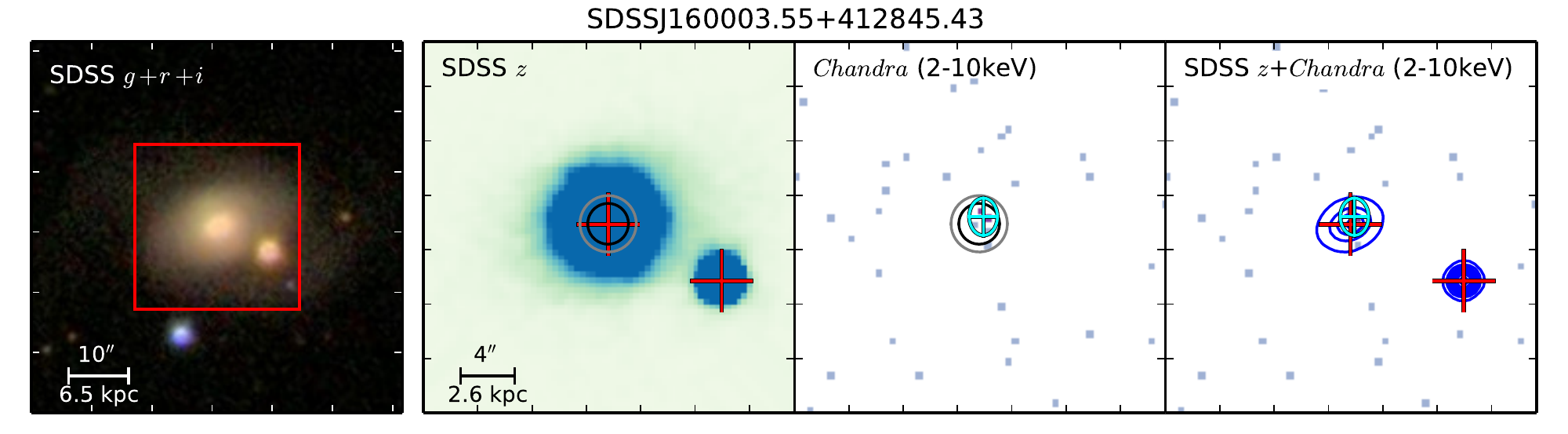}}
\vspace*{-0.20in}
\subfloat {\hspace*{0.10in} \includegraphics[width=6.8in]{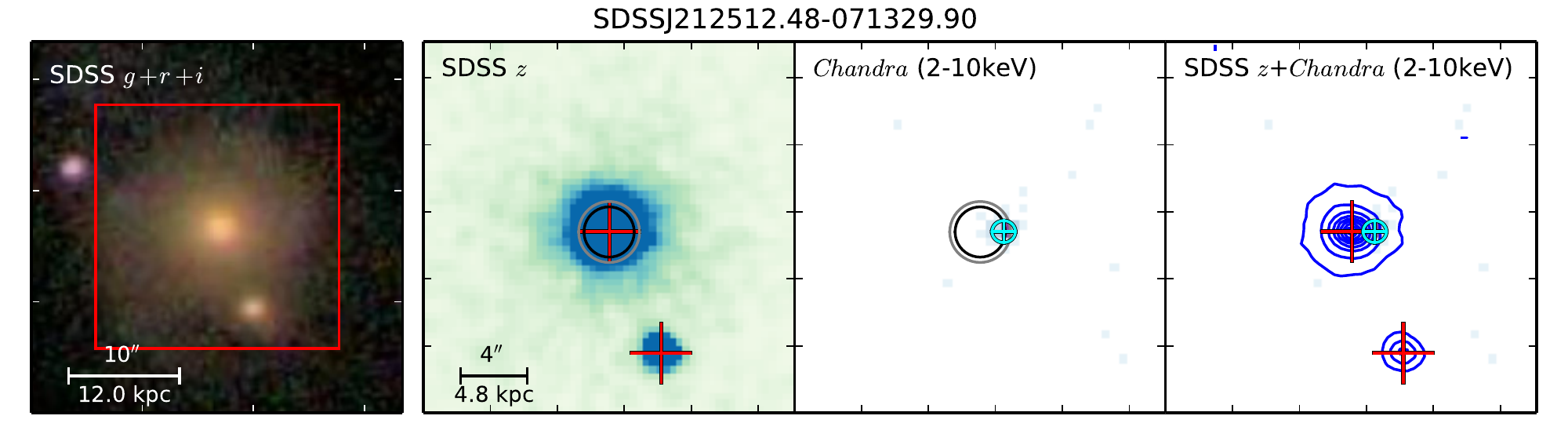}}
\caption{continued}
\end{figure*}


\begin{thebibliography}{106}
\expandafter\ifx\csname natexlab\endcsname\relax\def\natexlab#1{#1}\fi

\bibitem[{{Abazajian} {et~al.}(2009){Abazajian}, {Adelman-McCarthy},
  {Ag{\"u}eros}, {Allam}, {Allende Prieto}, {An}, {Anderson}, {Anderson},
  {Annis}, {Bahcall}, {Bailer-Jones}, {Barentine}, {Bassett}, {Becker},
  {Beers}, {Bell}, {Belokurov}, {Berlind}, {Berman}, {Bernardi}, {Bickerton},
  {Bizyaev}, {Blakeslee}, {Blanton}, {Bochanski}, {Boroski}, {Brewington},
  {Brinchmann}, {Brinkmann}, {Brunner}, {Budav{\'a}ri}, {Carey}, {Carliles},
  {Carr}, {Castander}, {Cinabro}, {Connolly}, {Csabai}, {Cunha}, {Czarapata},
  {Davenport}, {de Haas}, {Dilday}, {Doi}, {Eisenstein}, {Evans}, {Evans},
  {Fan}, {Friedman}, {Frieman}, {Fukugita}, {G{\"a}nsicke}, {Gates},
  {Gillespie}, {Gilmore}, {Gonzalez}, {Gonzalez}, {Grebel}, {Gunn},
  {Gy{\"o}ry}, {Hall}, {Harding}, {Harris}, {Harvanek}, {Hawley}, {Hayes},
  {Heckman}, {Hendry}, {Hennessy}, {Hindsley}, {Hoblitt}, {Hogan}, {Hogg},
  {Holtzman}, {Hyde}, {Ichikawa}, {Ichikawa}, {Im}, {Ivezi{\'c}}, {Jester},
  {Jiang}, {Johnson}, {Jorgensen}, {Juri{\'c}}, {Kent}, {Kessler}, {Kleinman},
  {Knapp}, {Konishi}, {Kron}, {Krzesinski}, {Kuropatkin}, {Lampeitl},
  {Lebedeva}, {Lee}, {Lee}, {Leger}, {L{\'e}pine}, {Li}, {Lima}, {Lin}, {Long},
  {Loomis}, {Loveday}, {Lupton}, {Magnier}, {Malanushenko}, {Malanushenko},
  {Mandelbaum}, {Margon}, {Marriner}, {Mart{\'{\i}}nez-Delgado}, {Matsubara},
  {McGehee}, {McKay}, {Meiksin}, {Morrison}, {Mullally}, {Munn}, {Murphy},
  {Nash}, {Nebot}, {Neilsen}, {Newberg}, {Newman}, {Nichol}, {Nicinski},
  {Nieto-Santisteban}, {Nitta}, {Okamura}, {Oravetz}, {Ostriker}, {Owen},
  {Padmanabhan}, {Pan}, {Park}, {Pauls}, {Peoples}, {Percival}, {Pier}, {Pope},
  {Pourbaix}, {Price}, {Purger}, {Quinn}, {Raddick}, {Fiorentin}, {Richards},
  {Richmond}, {Riess}, {Rix}, {Rockosi}, {Sako}, {Schlegel}, {Schneider},
  {Scholz}, {Schreiber}, {Schwope}, {Seljak}, {Sesar}, {Sheldon}, {Shimasaku},
  {Sibley}, {Simmons}, {Sivarani}, {Smith}, {Smith}, {Smol{\v c}i{\'c}},
  {Snedden}, {Stebbins}, {Steinmetz}, {Stoughton}, {Strauss}, {Subba Rao},
  {Suto}, {Szalay}, {Szapudi}, {Szkody}, {Tanaka}, {Tegmark}, {Teodoro},
  {Thakar}, {Tremonti}, {Tucker}, {Uomoto}, {Vanden Berk}, {Vandenberg},
  {Vidrih}, {Vogeley}, {Voges}, {Vogt}, {Wadadekar}, {Watters}, {Weinberg},
  {West}, {White}, {Wilhite}, {Wonders}, {Yanny}, {Yocum}, {York}, {Zehavi},
  {Zibetti}, \& {Zucker}}]{Abazajian09}
{Abazajian}, K.~N., {et~al.} 2009, ApJ, 182, 543

\bibitem[{{Alexander} {et~al.}(2003){Alexander}, {Bauer}, {Brandt},
  {Schneider}, {Hornschemeier}, {Vignali}, {Barger}, {Broos}, {Cowie},
  {Garmire}, {Townsley}, {Bautz}, {Chartas}, \& {Sargent}}]{Alexander:2003}
{Alexander}, D.~M., {et~al.} 2003, \aj, 126, 539

\bibitem[{{Allen} {et~al.}(2015){Allen}, {Schaefer}, {Scott}, {Fogarty}, {Ho},
  {Medling}, {Leslie}, {Bland-Hawthorn}, {Bryant}, {Croom}, {Goodwin}, {Green},
  {Konstantopoulos}, {Lawrence}, {Owers}, {Richards}, \& {Sharp}}]{Allen:2015}
{Allen}, J.~T., {et~al.} 2015, \mnras, 451, 2780

\bibitem[{{Anderson} {et~al.}(2011){Anderson}, {Mossman}, {Kim}, {Allen},
  {Glotfelty}, \& {Fabbiano}}]{Anderson:2011}
{Anderson}, C.~S., {Mossman}, A.~E., {Kim}, D.-W., {Allen}, G.~E., {Glotfelty},
  K.~J., \& {Fabbiano}, G. 2011, in Astronomical Society of the Pacific
  Conference Series, Vol. 442, Astronomical Data Analysis Software and Systems
  XX, ed. I.~N. {Evans}, A.~{Accomazzi}, D.~J. {Mink}, \& A.~H. {Rots}, 139

\bibitem[{{Antonucci}(1993)}]{Antonucci:1993}
{Antonucci}, R. 1993, \araa, 31, 473

\bibitem[{{Arias} {et~al.}(1995){Arias}, {Charlot}, {Feissel}, \&
  {Lestrade}}]{Arias:1995}
{Arias}, E.~F., {Charlot}, P., {Feissel}, M., \& {Lestrade}, J.-F. 1995, \aap,
  303, 604

\bibitem[{{Bachetti} {et~al.}(2014){Bachetti}, {Harrison}, {Walton},
  {Grefenstette}, {Chakrabarty}, {F{\"u}rst}, {Barret}, {Beloborodov}, {Boggs},
  {Christensen}, {Craig}, {Fabian}, {Hailey}, {Hornschemeier}, {Kaspi},
  {Kulkarni}, {Maccarone}, {Miller}, {Rana}, {Stern}, {Tendulkar}, {Tomsick},
  {Webb}, \& {Zhang}}]{Bachetti:2014}
{Bachetti}, M., {et~al.} 2014, \nat, 514, 202

\bibitem[{{Baldwin} {et~al.}(1981){Baldwin}, {Phillips}, \&
  {Terlevich}}]{Baldwin1981}
{Baldwin}, J.~A., {Phillips}, M.~M., \& {Terlevich}, R. 1981, \pasp, 93, 5

\bibitem[{{Barnes} \& {Hernquist}(1991)}]{Barnes:1991}
{Barnes}, J.~E., \& {Hernquist}, L.~E. 1991, \apjl, 370, L65

\bibitem[{{Barrows} {et~al.}(2013){Barrows}, {Sandberg Lacy}, {Kennefick},
  {Comerford}, {Kennefick}, \& {Berrier}}]{Barrows:2013}
{Barrows}, R.~S., {Sandberg Lacy}, C.~H., {Kennefick}, J., {Comerford}, J.~M.,
  {Kennefick}, D., \& {Berrier}, J.~C. 2013, \apj, 769, 95

\bibitem[{{Barrows} {et~al.}(2012){Barrows}, {Stern}, {Madsen}, {Harrison},
  {Assef}, {Comerford}, {Cushing}, {Fassnacht}, {Gonzalez}, {Griffith},
  {Hickox}, {Kirkpatrick}, \& {Lagattuta}}]{Barrows:2012}
{Barrows}, R.~S., {et~al.} 2012, \apj, 744, 7

\bibitem[{{Barth} {et~al.}(2008){Barth}, {Bentz}, {Greene}, \&
  {Ho}}]{Barth:2008}
{Barth}, A.~J., {Bentz}, M.~C., {Greene}, J.~E., \& {Ho}, L.~C. 2008, \apjl,
  683, L119

\bibitem[{Bentz {et~al.}(2009)Bentz, Peterson, Pogge, \&
  Vestergaard}]{Bentz:2009c}
Bentz, M.~C., Peterson, B.~M., Pogge, R.~W., \& Vestergaard, M. 2009, The
  Astrophysical Journal Letters, 694, L166

\bibitem[{{Bertin} \& {Arnouts}(1996)}]{Bertin:Arnouts:1996}
{Bertin}, E., \& {Arnouts}, S. 1996, \aaps, 117, 393

\bibitem[{{Bianchi} {et~al.}(2008){Bianchi}, {Chiaberge}, {Piconcelli},
  {Guainazzi}, \& {Matt}}]{Bianchi2008}
{Bianchi}, S., {Chiaberge}, M., {Piconcelli}, E., {Guainazzi}, M., \& {Matt},
  G. 2008, \mnras, 386, 105

\bibitem[{{Blecha} {et~al.}(2016){Blecha}, {Sijacki}, {Kelley}, {Torrey},
  {Vogelsberger}, {Nelson}, {Springel}, {Snyder}, \& {Hernquist}}]{Blecha:2016}
{Blecha}, L., {et~al.} 2016, \mnras, 456, 961

\bibitem[{Bondi {et~al.}(2016)Bondi, P{\'e}rez-Torres, Piconcelli, \&
  Fu}]{Bondi:2016}
Bondi, M., P{\'e}rez-Torres, M., Piconcelli, E., \& Fu, H. 2016, arXiv preprint
  arXiv:1603.01452

\bibitem[{{Brinchmann} {et~al.}(2004){Brinchmann}, {Charlot}, {White},
  {Tremonti}, {Kauffmann}, {Heckman}, \& {Brinkmann}}]{Brinchmann:2004}
{Brinchmann}, J., {Charlot}, S., {White}, S.~D.~M., {Tremonti}, C.,
  {Kauffmann}, G., {Heckman}, T., \& {Brinkmann}, J. 2004, \mnras, 351, 1151

\bibitem[{{Calabretta} \& {Greisen}(2002)}]{Calabretta:2002}
{Calabretta}, M.~R., \& {Greisen}, E.~W. 2002, \aap, 395, 1077

\bibitem[{{Comerford} {et~al.}(2012){Comerford}, {Gerke}, {Stern}, {Cooper},
  {Weiner}, {Newman}, {Madsen}, \& {Barrows}}]{Comerford:2012}
{Comerford}, J.~M., {Gerke}, B.~F., {Stern}, D., {Cooper}, M.~C., {Weiner},
  B.~J., {Newman}, J.~A., {Madsen}, K., \& {Barrows}, R.~S. 2012, \apj, 753, 42

\bibitem[{{Comerford} \& {Greene}(2014)}]{Comerford:Greene:2014}
{Comerford}, J.~M., \& {Greene}, J.~E. 2014, \apj, 789, 112

\bibitem[{{Comerford} {et~al.}(2015){Comerford}, {Pooley}, {Barrows}, {Greene},
  {Zakamska}, {Madejski}, \& {Cooper}}]{Comerford:2015}
{Comerford}, J.~M., {Pooley}, D., {Barrows}, R.~S., {Greene}, J.~E.,
  {Zakamska}, N.~L., {Madejski}, G.~M., \& {Cooper}, M.~C. 2015, \apj, 806, 219

\bibitem[{{Comerford} {et~al.}(2013){Comerford}, {Schluns}, {Greene}, \&
  {Cool}}]{Comerford:2013}
{Comerford}, J.~M., {Schluns}, K., {Greene}, J.~E., \& {Cool}, R.~J. 2013,
  \apj, 777, 64

\bibitem[{{Comerford} {et~al.}(2009){Comerford}, {Gerke}, {Newman}, {Davis},
  {Yan}, {Cooper}, {Faber}, {Koo}, {Coil}, {Rosario}, \&
  {Dutton}}]{Comerford2009a}
{Comerford}, J.~M., {et~al.} 2009, \apj, 698, 956

\bibitem[{{Conselice} {et~al.}(2003){Conselice}, {Bershady}, {Dickinson}, \&
  {Papovich}}]{Conselice:2003}
{Conselice}, C.~J., {Bershady}, M.~A., {Dickinson}, M., \& {Papovich}, C. 2003,
  \aj, 126, 1183

\bibitem[{{Feng} \& {Kaaret}(2009)}]{Feng:Kaaret:2009}
{Feng}, H., \& {Kaaret}, P. 2009, \apj, 696, 1712

\bibitem[{Fernandes {et~al.}(2005)Fernandes, Mateus, Sodr{\'e}, Stasi{\'n}ska,
  \& Gomes}]{Fernandes:2005}
Fernandes, R.~C., Mateus, A., Sodr{\'e}, L., Stasi{\'n}ska, G., \& Gomes, J.~M.
  2005, Monthly Notices of the Royal Astronomical Society, 358, 363

\bibitem[{{Ferrarese} \& {Merritt}(2000)}]{Ferrarese2000}
{Ferrarese}, L., \& {Merritt}, D. 2000, \apjl, 539, L9

\bibitem[{{Fu} {et~al.}(2011{\natexlab{a}}){Fu}, {Myers}, {Djorgovski}, \&
  {Yan}}]{Fu:2011a}
{Fu}, H., {Myers}, A.~D., {Djorgovski}, S.~G., \& {Yan}, L. 2011{\natexlab{a}},
  \apj, 733, 103

\bibitem[{{Fu} {et~al.}(2015{\natexlab{a}}){Fu}, {Myers}, {Djorgovski}, {Yan},
  {Wrobel}, \& {Stockton}}]{Fu:2015a}
{Fu}, H., {Myers}, A.~D., {Djorgovski}, S.~G., {Yan}, L., {Wrobel}, J.~M., \&
  {Stockton}, A. 2015{\natexlab{a}}, \apj, 799, 72

\bibitem[{{Fu} {et~al.}(2015{\natexlab{b}}){Fu}, {Wrobel}, {Myers},
  {Djorgovski}, \& {Yan}}]{Fu:2015b}
{Fu}, H., {Wrobel}, J.~M., {Myers}, A.~D., {Djorgovski}, S.~G., \& {Yan}, L.
  2015{\natexlab{b}}, \apjl, 815, L6

\bibitem[{{Fu} {et~al.}(2012){Fu}, {Yan}, {Myers}, {Stockton}, {Djorgovski},
  {Aldering}, \& {Rich}}]{Fu:2012}
{Fu}, H., {Yan}, L., {Myers}, A.~D., {Stockton}, A., {Djorgovski}, S.~G.,
  {Aldering}, G., \& {Rich}, J.~A. 2012, \apj, 745, 67

\bibitem[{{Fu} {et~al.}(2011{\natexlab{b}}){Fu}, {Zhang}, {Assef}, {Stockton},
  {Myers}, {Yan}, {Djorgovski}, {Wrobel}, \& {Riechers}}]{Fu:2011}
{Fu}, H., {et~al.} 2011{\natexlab{b}}, \apjl, 740, L44

\bibitem[{Gaetz \& Jerius(2004)}]{Gaetz:2004}
Gaetz, T., \& Jerius, D. 2004

\bibitem[{{Ge} {et~al.}(2012){Ge}, {Hu}, {Wang}, {Bai}, \& {Zhang}}]{Ge:2012}
{Ge}, J.-Q., {Hu}, C., {Wang}, J.-M., {Bai}, J.-M., \& {Zhang}, S. 2012, \apjs,
  201, 31

\bibitem[{{Gebhardt} {et~al.}(2000){Gebhardt}, {Bender}, {Bower}, {Dressler},
  {Faber}, {Filippenko}, {Green}, {Grillmair}, {Ho}, {Kormendy}, {Lauer},
  {Magorrian}, {Pinkney}, {Richstone}, \& {Tremaine}}]{Gebhardt00}
{Gebhardt}, K., {et~al.} 2000, ApJ, 539, L13

\bibitem[{{Graham} \& {Driver}(2005)}]{Graham:2005}
{Graham}, A.~W., \& {Driver}, S.~P. 2005, PASA, 22, 118

\bibitem[{{Greene} {et~al.}(2011){Greene}, {Zakamska}, {Ho}, \&
  {Barth}}]{Greene:2011}
{Greene}, J.~E., {Zakamska}, N.~L., {Ho}, L.~C., \& {Barth}, A.~J. 2011, \apj,
  732, 9

\bibitem[{{Guainazzi} {et~al.}(2005){Guainazzi}, {Piconcelli},
  {Jim{\'e}nez-Bail{\'o}n}, \& {Matt}}]{Guainazzi2005}
{Guainazzi}, M., {Piconcelli}, E., {Jim{\'e}nez-Bail{\'o}n}, E., \& {Matt}, G.
  2005, \aap, 429, L9

\bibitem[{{G{\"u}ltekin} {et~al.}(2009){G{\"u}ltekin}, {Richstone}, {Gebhardt},
  {Lauer}, {Tremaine}, {Aller}, {Bender}, {Dressler}, {Faber}, {Filippenko},
  {Green}, {Ho}, {Kormendy}, {Magorrian}, {Pinkney}, \&
  {Siopis}}]{Gultekin:2009}
{G{\"u}ltekin}, K., {et~al.} 2009, \apj, 698, 198

\bibitem[{{H{\"a}ring} \& {Rix}(2004)}]{Haring:2004}
{H{\"a}ring}, N., \& {Rix}, H. 2004, \apjl, 604, L89

\bibitem[{{Harris} {et~al.}(2004){Harris}, {Mossman}, \&
  {Walker}}]{Harris:2004}
{Harris}, D.~E., {Mossman}, A.~E., \& {Walker}, R.~C. 2004, \apj, 615, 161

\bibitem[{{H{\"a}ussler} {et~al.}(2007){H{\"a}ussler}, {McIntosh}, {Barden},
  {Bell}, {Rix}, {Borch}, {Beckwith}, {Caldwell}, {Heymans}, {Jahnke}, {Jogee},
  {Koposov}, {Meisenheimer}, {S{\'a}nchez}, {Somerville}, {Wisotzki}, \&
  {Wolf}}]{Haussler:2007}
{H{\"a}ussler}, B., {et~al.} 2007, \apjs, 172, 615

\bibitem[{Hog {et~al.}(1994)Hog, Makarov, \& Pederson}]{Hog:1994}
Hog, E., Makarov, V., \& Pederson, H. 1994, in Galactic and solar system
  optical astrometry: proceedings of the Royal Greenwich Observatory and the
  Institute of Astronomy workshop held in Cambridge, June 21-24, 1993,
  Cambridge Univ Pr, 71

\bibitem[{{H{\o}g} {et~al.}(2000){H{\o}g}, {Fabricius}, {Makarov}, {Urban},
  {Corbin}, {Wycoff}, {Bastian}, {Schwekendiek}, \& {Wicenec}}]{Hog:2000}
{H{\o}g}, E., {et~al.} 2000, \aap, 355, L27

\bibitem[{{Hopkins} {et~al.}(2005){Hopkins}, {Hernquist}, {Cox}, {Di Matteo},
  {Martini}, {Robertson}, \& {Springel}}]{Hopkins05}
{Hopkins}, P.~F., {Hernquist}, L., {Cox}, T.~J., {Di Matteo}, T., {Martini},
  P., {Robertson}, B., \& {Springel}, V. 2005, ApJ, 630, 705

\bibitem[{Hopkins {et~al.}(2012)Hopkins, Hernquist, Hayward, \&
  Narayanan}]{Hopkins:2012}
Hopkins, P.~F., Hernquist, L., Hayward, C.~C., \& Narayanan, D. 2012, Monthly
  Notices of the Royal Astronomical Society, 425, 1121

\bibitem[{{Hudson} {et~al.}(2006){Hudson}, {Reiprich}, {Clarke}, \&
  {Sarazin}}]{Hudson2006}
{Hudson}, D.~S., {Reiprich}, T.~H., {Clarke}, T.~E., \& {Sarazin}, C.~L. 2006,
  \aap, 453, 433

\bibitem[{Ishibashi \& Courvoisier(2010)}]{Ishibashi:2010}
Ishibashi, W., \& Courvoisier, T.-L. 2010, Astronomy \& Astrophysics, 512, A58

\bibitem[{{Jonker} {et~al.}(2010){Jonker}, {Torres}, {Fabian}, {Heida},
  {Miniutti}, \& {Pooley}}]{Jonker:2010}
{Jonker}, P.~G., {Torres}, M.~A.~P., {Fabian}, A.~C., {Heida}, M., {Miniutti},
  G., \& {Pooley}, D. 2010, \mnras, 407, 645

\bibitem[{{Juneau} {et~al.}(2011){Juneau}, {Dickinson}, {Alexander}, \&
  {Salim}}]{Juneau:2011}
{Juneau}, S., {Dickinson}, M., {Alexander}, D.~M., \& {Salim}, S. 2011, \apj,
  736, 104

\bibitem[{{Kauffmann} {et~al.}(2003){Kauffmann}, {Heckman}, {Tremonti},
  {Brinchmann}, {Charlot}, {White}, {Ridgway}, {Brinkmann}, {Fukugita}, {Hall},
  {Ivezi{\'c}}, {Richards}, \& {Schneider}}]{Kauffmann:2003}
{Kauffmann}, G., {et~al.} 2003, \mnras, 346, 1055

\bibitem[{{Kewley} {et~al.}(2001){Kewley}, {Dopita}, {Sutherland}, {Heisler},
  \& {Trevena}}]{Kewley:2001}
{Kewley}, L.~J., {Dopita}, M.~A., {Sutherland}, R.~S., {Heisler}, C.~A., \&
  {Trevena}, J. 2001, \apj, 556, 121

\bibitem[{{King} \& {Dehnen}(2005)}]{King:2005}
{King}, A.~R., \& {Dehnen}, W. 2005, \mnras, 357, 275

\bibitem[{{Komatsu} {et~al.}(2011){Komatsu}, {Smith}, {Dunkley}, {Bennett},
  {Gold}, {Hinshaw}, {Jarosik}, {Larson}, {Nolta}, {Page}, {Spergel},
  {Halpern}, {Hill}, {Kogut}, {Limon}, {Meyer}, {Odegard}, {Tucker}, {Weiland},
  {Wollack}, \& {Wright}}]{Komatsu:2011}
{Komatsu}, E., {et~al.} 2011, \apjs, 192, 18

\bibitem[{{Komossa} {et~al.}(2003){Komossa}, {Burwitz}, {Hasinger}, {Predehl},
  {Kaastra}, \& {Ikebe}}]{Komossa2003}
{Komossa}, S., {Burwitz}, V., {Hasinger}, G., {Predehl}, P., {Kaastra}, J.~S.,
  \& {Ikebe}, Y. 2003, \apjl, 582, L15

\bibitem[{{Komossa} \& {Merritt}(2008)}]{Komossa:2008}
{Komossa}, S., \& {Merritt}, D. 2008, \apjl, 689, L89

\bibitem[{{Koss} {et~al.}(2011){Koss}, {Mushotzky}, {Treister}, {Veilleux},
  {Vasudevan}, {Miller}, {Sanders}, {Schawinski}, \& {Trippe}}]{Koss:2011}
{Koss}, M., {et~al.} 2011, \apjl, 735, L42+

\bibitem[{{Lackner} {et~al.}(2014){Lackner}, {Silverman}, {Salvato},
  {Kampczyk}, {Kartaltepe}, {Sanders}, {Capak}, {Civano}, {Halliday}, {Ilbert},
  {Jahnke}, {Koekemoer}, {Lee}, {Le F{\`e}vre}, {Liu}, {Scoville}, {Sheth}, \&
  {Toft}}]{Lackner:2014}
{Lackner}, C.~N., {et~al.} 2014, \aj, 148, 137

\bibitem[{Lagarias {et~al.}(1998)Lagarias, Reeds, Wright, \&
  Wright}]{Lagarias:1998}
Lagarias, J.~C., Reeds, J.~A., Wright, M.~H., \& Wright, P.~E. 1998, SIAM
  Journal on optimization, 9, 112

\bibitem[{Lagos {et~al.}(2011)Lagos, Padilla, Strauss, Cora, \&
  Hao}]{Lagos:2011}
Lagos, C. d.~P., Padilla, N.~D., Strauss, M.~A., Cora, S.~A., \& Hao, L. 2011,
  Monthly Notices of the Royal Astronomical Society, 414, 2148

\bibitem[{{Lehmer} {et~al.}(2005){Lehmer}, {Brandt}, {Alexander}, {Bauer},
  {Schneider}, {Tozzi}, {Bergeron}, {Garmire}, {Giacconi}, {Gilli}, {Hasinger},
  {Hornschemeier}, {Koekemoer}, {Mainieri}, {Miyaji}, {Nonino}, {Rosati},
  {Silverman}, {Szokoly}, \& {Vignali}}]{Lehmer:2005}
{Lehmer}, B.~D., {et~al.} 2005, \apjs, 161, 21

\bibitem[{{Liu} {et~al.}(2013){Liu}, {Civano}, {Shen}, {Green}, {Greene}, \&
  {Strauss}}]{Liu:2013}
{Liu}, X., {Civano}, F., {Shen}, Y., {Green}, P., {Greene}, J.~E., \&
  {Strauss}, M.~A. 2013, \apj, 762, 110

\bibitem[{{Liu} {et~al.}(2010{\natexlab{a}}){Liu}, {Greene}, {Shen}, \&
  {Strauss}}]{Liu2010b}
{Liu}, X., {Greene}, J.~E., {Shen}, Y., \& {Strauss}, M.~A. 2010{\natexlab{a}},
  \apjl, 715, L30

\bibitem[{{Liu} {et~al.}(2010{\natexlab{b}}){Liu}, {Shen}, {Strauss}, \&
  {Greene}}]{Liu2010a}
{Liu}, X., {Shen}, Y., {Strauss}, M.~A., \& {Greene}, J.~E. 2010{\natexlab{b}},
  \apj, 708, 427

\bibitem[{{Liu} {et~al.}(2011){Liu}, {Shen}, {Strauss}, \& {Hao}}]{Liu:2011}
{Liu}, X., {Shen}, Y., {Strauss}, M.~A., \& {Hao}, L. 2011, \apj, 737, 101

\bibitem[{{Lotz} {et~al.}(2011){Lotz}, {Jonsson}, {Cox}, {Croton}, {Primack},
  {Somerville}, \& {Stewart}}]{Lotz:2011}
{Lotz}, J.~M., {Jonsson}, P., {Cox}, T.~J., {Croton}, D., {Primack}, J.~R.,
  {Somerville}, R.~S., \& {Stewart}, K. 2011, \apj, 742, 103

\bibitem[{Lupton(1993)}]{Lupton:1993}
Lupton, R. 1993, Journal of the British Astronomical Association, 103, 320

\bibitem[{{Maiolino} {et~al.}(2001){Maiolino}, {Marconi}, {Salvati},
  {Risaliti}, {Severgnini}, {Oliva}, {La Franca}, \& {Vanzi}}]{Maiolino:2001}
{Maiolino}, R., {Marconi}, A., {Salvati}, M., {Risaliti}, G., {Severgnini}, P.,
  {Oliva}, E., {La Franca}, F., \& {Vanzi}, L. 2001, \aap, 365, 28

\bibitem[{{Marconi} \& {Hunt}(2003)}]{Marconi:Hunt:2003}
{Marconi}, A., \& {Hunt}, L.~K. 2003, \apjl, 589, L21

\bibitem[{{Mazzarella} {et~al.}(2012){Mazzarella}, {Iwasawa}, {Vavilkin},
  {Armus}, {Kim}, {Bothun}, {Evans}, {Spoon}, {Haan}, {Howell}, {Lord},
  {Marshall}, {Ishida}, {Xu}, {Petric}, {Sanders}, {Surace}, {Appleton},
  {Chan}, {Frayer}, {Inami}, {Khachikian}, {Madore}, {Privon}, {Sturm}, {U}, \&
  {Veilleux}}]{Mazzarella:2012}
{Mazzarella}, J.~M., {et~al.} 2012, ArXiv e-prints

\bibitem[{{McGurk} {et~al.}(2015){McGurk}, {Max}, {Medling}, {Shields}, \&
  {Comerford}}]{McGurk:2015}
{McGurk}, R.~C., {Max}, C.~E., {Medling}, A.~M., {Shields}, G.~A., \&
  {Comerford}, J.~M. 2015, \apj, 811, 14

\bibitem[{{McGurk} {et~al.}(2011){McGurk}, {Max}, {Rosario}, {Shields},
  {Smith}, \& {Wright}}]{McGurk:2011}
{McGurk}, R.~C., {Max}, C.~E., {Rosario}, D.~J., {Shields}, G.~A., {Smith},
  K.~L., \& {Wright}, S.~A. 2011, \apjl, 738, L2

\bibitem[{McLure \& Dunlop(2002)}]{McLure:2002}
McLure, R., \& Dunlop, J. 2002, Monthly Notices of the Royal Astronomical
  Society, 331, 795

\bibitem[{{Middleton} {et~al.}(2008){Middleton}, {Done}, \&
  {Schurch}}]{Middleton:2008}
{Middleton}, M., {Done}, C., \& {Schurch}, N. 2008, \mnras, 383, 1501

\bibitem[{{Mihos} \& {Hernquist}(1996)}]{Mihos:1996}
{Mihos}, J.~C., \& {Hernquist}, L. 1996, \apj, 464, 641

\bibitem[{{M{\"u}ller-Sanchez} {et~al.}(2015){M{\"u}ller-Sanchez}, {Comerford},
  {Nevin}, {Barrows}, {Cooper}, \& {Greene}}]{Mueller-Sanchez:2015}
{M{\"u}ller-Sanchez}, F., {Comerford}, J.~M., {Nevin}, R., {Barrows}, R.~S.,
  {Cooper}, M.~C., \& {Greene}, J.~E. 2015, ArXiv e-prints

\bibitem[{{Nandra} \& {Pounds}(1994)}]{Nandra:1994}
{Nandra}, K., \& {Pounds}, K.~A. 1994, \mnras, 268, 405

\bibitem[{{Oh} {et~al.}(2011){Oh}, {Sarzi}, {Schawinski}, \& {Yi}}]{Oh:2011}
{Oh}, K., {Sarzi}, M., {Schawinski}, K., \& {Yi}, S.~K. 2011, \apjs, 195, 13

\bibitem[{{Peng} {et~al.}(2010){Peng}, {Ho}, {Impey}, \& {Rix}}]{Peng:2010}
{Peng}, C.~Y., {Ho}, L.~C., {Impey}, C.~D., \& {Rix}, H.-W. 2010, \aj, 139,
  2097

\bibitem[{Perryman {et~al.}(1997)Perryman, Lindegren, Kovalevsky, Hoeg,
  Bastian, Bernacca, Cr{\'e}z{\'e}, Donati, Grenon, Grewing,
  {et~al.}}]{Perryman:1997}
Perryman, M.~A., {et~al.} 1997, Astronomy and Astrophysics, 323, L49

\bibitem[{{Piconcelli} {et~al.}(2005){Piconcelli}, {Jimenez-Bail{\'o}n},
  {Guainazzi}, {Schartel}, {Rodr{\'{\i}}guez-Pascual}, \&
  {Santos-Lle{\'o}}}]{Piconcelli:2005}
{Piconcelli}, E., {Jimenez-Bail{\'o}n}, E., {Guainazzi}, M., {Schartel}, N.,
  {Rodr{\'{\i}}guez-Pascual}, P.~M., \& {Santos-Lle{\'o}}, M. 2005, \aap, 432,
  15

\bibitem[{{Piconcelli} {et~al.}(2010){Piconcelli}, {Vignali}, {Bianchi},
  {Mathur}, {Fiore}, {Guainazzi}, {Lanzuisi}, {Maiolino}, \&
  {Nicastro}}]{Piconcelli2010}
{Piconcelli}, E., {et~al.} 2010, ArXiv e-prints

\bibitem[{{Pier} {et~al.}(2003){Pier}, {Munn}, {Hindsley}, {Hennessy}, {Kent},
  {Lupton}, \& {Ivezi{\'c}}}]{Pier:2003}
{Pier}, J.~R., {Munn}, J.~A., {Hindsley}, R.~B., {Hennessy}, G.~S., {Kent},
  S.~M., {Lupton}, R.~H., \& {Ivezi{\'c}}, {\v Z}. 2003, \aj, 125, 1559

\bibitem[{{Pooley} {et~al.}(2009){Pooley}, {Rappaport}, {Blackburne},
  {Schechter}, {Schwab}, \& {Wambsganss}}]{Pooley:2009}
{Pooley}, D., {Rappaport}, S., {Blackburne}, J., {Schechter}, P.~L., {Schwab},
  J., \& {Wambsganss}, J. 2009, \apj, 697, 1892

\bibitem[{{Pushkarev} {et~al.}(2012){Pushkarev}, {Hovatta}, {Kovalev},
  {Lister}, {Lobanov}, {Savolainen}, \& {Zensus}}]{Pushkarev:2012}
{Pushkarev}, A.~B., {Hovatta}, T., {Kovalev}, Y.~Y., {Lister}, M.~L.,
  {Lobanov}, A.~P., {Savolainen}, T., \& {Zensus}, J.~A. 2012, \aap, 545, A113

\bibitem[{{Readhead} {et~al.}(1979){Readhead}, {Pearson}, {Cohen}, {Ewing}, \&
  {Moffet}}]{Readhead:1979}
{Readhead}, A.~C.~S., {Pearson}, T.~J., {Cohen}, M.~H., {Ewing}, M.~S., \&
  {Moffet}, A.~T. 1979, \apj, 231, 299

\bibitem[{{Reeves} \& {Turner}(2000)}]{Reeves:2000}
{Reeves}, J.~N., \& {Turner}, M.~J.~L. 2000, \mnras, 316, 234

\bibitem[{{Reyes} {et~al.}(2008){Reyes}, {Zakamska}, {Strauss}, {Green},
  {Krolik}, {Shen}, {Richards}, {Anderson}, \& {Schneider}}]{Reyes:2008}
{Reyes}, R., {et~al.} 2008, \aj, 136, 2373

\bibitem[{{Rosario} {et~al.}(2011){Rosario}, {McGurk}, {Max}, {Shields},
  {Smith}, \& {Ammons}}]{Rosario:2011}
{Rosario}, D.~J., {McGurk}, R.~C., {Max}, C.~E., {Shields}, G.~A., {Smith},
  K.~L., \& {Ammons}, S.~M. 2011, \apj, 739, 44

\bibitem[{{Rosario} {et~al.}(2010){Rosario}, {Shields}, {Taylor}, {Salviander},
  \& {Smith}}]{Rosario:2010}
{Rosario}, D.~J., {Shields}, G.~A., {Taylor}, G.~B., {Salviander}, S., \&
  {Smith}, K.~L. 2010, \apj, 716, 131

\bibitem[{{Schlegel} {et~al.}(1998){Schlegel}, {Finkbeiner}, \&
  {Davis}}]{Schlegel98}
{Schlegel}, D.~J., {Finkbeiner}, D.~P., \& {Davis}, M. 1998, ApJ, 500, 525

\bibitem[{{Shen} {et~al.}(2011){Shen}, {Liu}, {Greene}, \&
  {Strauss}}]{Shen:2011b}
{Shen}, Y., {Liu}, X., {Greene}, J.~E., \& {Strauss}, M.~A. 2011, \apj, 735, 48

\bibitem[{{Silverman} {et~al.}(2005){Silverman}, {Green}, {Barkhouse}, {Kim},
  {Aldcroft}, {Cameron}, {Wilkes}, {Mossman}, {Ghosh}, {Tananbaum}, {Smith},
  {Smith}, {Smith}, {Foltz}, {Wik}, \& {Jannuzi}}]{Silverman:2005}
{Silverman}, J.~D., {et~al.} 2005, \apj, 618, 123

\bibitem[{{Smith} {et~al.}(2010){Smith}, {Shields}, {Bonning}, {McMullen},
  {Rosario}, \& {Salviander}}]{Smith:2010}
{Smith}, K.~L., {Shields}, G.~A., {Bonning}, E.~W., {McMullen}, C.~C.,
  {Rosario}, D.~J., \& {Salviander}, S. 2010, \apj, 716, 866

\bibitem[{Souchay {et~al.}(2009)Souchay, Andrei, Barache, Bouquillon, Gontier,
  Lambert, Le~Poncin-Lafitte, Taris, Arias, Suchet, {et~al.}}]{Souchay:2009}
Souchay, J., {et~al.} 2009, Astronomy \& Astrophysics, 494, 799

\bibitem[{{Springel} {et~al.}(2005){Springel}, {Di Matteo}, \&
  {Hernquist}}]{Springel2005}
{Springel}, V., {Di Matteo}, T., \& {Hernquist}, L. 2005, \apjl, 620, L79

\bibitem[{{Steinborn} {et~al.}(2015){Steinborn}, {Dolag}, {Comerford},
  {Hirschmann}, {Remus}, \& {Teklu}}]{Steinborn:2015}
{Steinborn}, L.~K., {Dolag}, K., {Comerford}, J.~M., {Hirschmann}, M., {Remus},
  R.-S., \& {Teklu}, A.~F. 2015, ArXiv e-prints

\bibitem[{{Stern} {et~al.}(2002){Stern}, {Tozzi}, {Stanford}, {Rosati},
  {Holden}, {Eisenhardt}, {Elston}, {Wu}, {Connolly}, {Spinrad}, {Dawson},
  {Dey}, \& {Chaffee}}]{Stern:2002}
{Stern}, D., {et~al.} 2002, \aj, 123, 2223

\bibitem[{{U} {et~al.}(2013){U}, {Medling}, {Sanders}, {Max}, {Armus},
  {Iwasawa}, {Evans}, {Kewley}, \& {Fazio}}]{U:2013}
{U}, V., {et~al.} 2013, \apj, 775, 115

\bibitem[{{Van Wassenhove} {et~al.}(2012){Van Wassenhove}, {Volonteri},
  {Mayer}, {Dotti}, {Bellovary}, \& {Callegari}}]{Van_Wassenhove:2012}
{Van Wassenhove}, S., {Volonteri}, M., {Mayer}, L., {Dotti}, M., {Bellovary},
  J., \& {Callegari}, S. 2012, \apjl, 748, L7

\bibitem[{{Wang} {et~al.}(2009){Wang}, {Chen}, {Hu}, {Mao}, {Zhang}, \&
  {Bian}}]{Wang2009}
{Wang}, J., {Chen}, Y., {Hu}, C., {Mao}, W., {Zhang}, S., \& {Bian}, W. 2009,
  \apjl, 705, L76

\bibitem[{{Wilkes} {et~al.}(2005){Wilkes}, {Pounds}, {Schmidt}, {Smith},
  {Cutri}, {Ghosh}, {Nelson}, \& {Hines}}]{Wilkes:2005}
{Wilkes}, B.~J., {Pounds}, K.~A., {Schmidt}, G.~D., {Smith}, P.~S., {Cutri},
  R.~M., {Ghosh}, H., {Nelson}, B., \& {Hines}, D.~C. 2005, \apj, 634, 183

\bibitem[{{Wolter} {et~al.}(2006){Wolter}, {Trinchieri}, \&
  {Colpi}}]{Wolter:2006}
{Wolter}, A., {Trinchieri}, G., \& {Colpi}, M. 2006, \mnras, 373, 1627

\bibitem[{{Xue} {et~al.}(2011){Xue}, {Luo}, {Brandt}, {Bauer}, {Lehmer},
  {Broos}, {Schneider}, {Alexander}, {Brusa}, {Comastri}, {Fabian}, {Gilli},
  {Hasinger}, {Hornschemeier}, {Koekemoer}, {Liu}, {Mainieri}, {Paolillo},
  {Rafferty}, {Rosati}, {Shemmer}, {Silverman}, {Smail}, {Tozzi}, \&
  {Vignali}}]{Xue:2011}
{Xue}, Y.~Q., {et~al.} 2011, \apjs, 195, 10

\bibitem[{{Zacharias} {et~al.}(2000){Zacharias}, {Urban}, {Zacharias}, {Hall},
  {Wycoff}, {Rafferty}, {Germain}, {Holdenried}, {Pohlman}, {Gauss}, {Monet},
  \& {Winter}}]{Zacharias:2000}
{Zacharias}, N., {et~al.} 2000, \aj, 120, 2131

\end{thebibliography}
\end{document}